\documentclass[iop]{emulateapj}

\usepackage{color}

\usepackage{graphicx}
\usepackage{natbib}
\usepackage{txfonts}

\shorttitle{Formation of the penumbra and start of the Evershed flow}
\shortauthors{Murabito et al.}

\begin{document}


\title{Formation of the penumbra and start of the Evershed flow}


\author{M. Murabito\altaffilmark{1}, P. Romano\altaffilmark{2}, S.L. Guglielmino\altaffilmark{1},  F. Zuccarello\altaffilmark{1} and S. K. Solanki\altaffilmark{3,}\altaffilmark{4}}
\email{mmurabito@oact.inaf.it}




\altaffiltext{1}{Dipartimento di Fisica e Astronomia - Sezione Astrofisica, Universit\`{a} degli Studi di Catania,
			 Via S. Sofia 78, 95123 Catania, Italy}
\altaffiltext{2}{INAF - Osservatorio Astrofisico di Catania,
              Via S. Sofia 78, 95123 Catania, Italy.}
\altaffiltext{3}{Max-Planck-Institut f\"{u}r Sonnensystemforschung
Justus-von-Liebig-Weg 3, 37077 G\"{o}ttingen, Germany}
\altaffiltext{4}{School of Space Research, Kyung Hee University, Yongin, Gyeonggi-Do, 446-701, Republic of Korea}		 			 


\begin{abstract}

We studied the variations of line-of-sight photospheric plasma flows during the formation phase of the penumbra around a pore in Active Region NOAA 11490. We used a high spatial, spectral, and temporal resolution data set acquired by the Interferometric BIdimensional Spectrometer (IBIS) operating at the NSO/Dunn Solar Telescope as well as data taken by the Helioseismic and Magnetic Imager onboard the Solar Dynamics Observatory satellite (SDO/HMI). 
Before the penumbra formed we observed a redshift of the spectral line in the inner part of the annular zone surrounding the pore as well as a blueshift of material associated with opposite magnetic polarity further away from the pore.
We found that the onset of the classical Evershed flow occurs in a very short time scale -- 1-3 hours -- while the penumbra is forming.
During the same time interval we found changes in the magnetic field inclination in the penumbra, with the vertical field actually changing sign near the penumbral edge, while the total magnetic field showed a significant increase, about 400 G.
To explain these and other observations related to the formation of the penumbra and the onset of the Evershed flow we propose a scenario in which the penumbra is formed by magnetic flux dragged down from the canopy surrounding the initial pore. The Evershed flow starts when the sinking magnetic field dips below the solar surface and magnetoconvection sets in. 

\end{abstract}




\section{Introduction}

During their lifetimes, sunspots show several dynamic phenomena whose underlying physical processes can be clarified only by additional observational and theoretical studies \citep{Sol03}. In particular, the mechanisms responsible for penumbra formation remain unclear because their study requires time series observations of sunspots with high temporal, spatial, and spectral resolution carried out from their first appearance as a pore \citep{Thom04}.

Currently, there are two main explanations for penumbra formation. \citet{Lek98} suggested that emerging, horizontal field lines could be trapped and form a penumbra rather than continuing to rise to higher layers, due to the presence of the overlying magnetic canopy in the emerging region. In contrast, \citet{Shi12} and \citet{Rom13, Rom14} showed that some signatures of penumbra formation around pores in the chromosphere appear earlier than in the photosphere. Their findings suggest that the field lines of the magnetic canopy, already existing at a higher level of the solar atmosphere and overlying the pore, may be responsible for the formation of the penumbra if they sink down into the photosphere and below the solar surface. Thus, \citet{Shi12}, using images in the Ca II H 396.8 nm line acquired with the Hinode~Solar Optical Telescope, showed that in Active Region (AR) NOAA 11039 a 3\arcsec- 5\arcsec\/ wide annular zone surrounding a pore already existed in the chromosphere some hours before the penumbra became visible in the photosphere. Using spectro-polarimetric scans through the \ion{Fe}{1} 630.25 nm line, \citet{Rom13, Rom14} detected the presence of several patches at the edge of the annular zone around a pore of the AR NOAA 11490, with a typical size of about 1\arcsec\/, that were characterized by a rather vertical magnetic field with polarity opposite to that of the pore. These patches showed radially outward displacements with horizontal velocities of about 2 km s$^{-1}$, that have been interpreted as due to portions of the pore's magnetic field returning beneath the photosphere, being progressively stretched and pushed down by the overlying magnetic fields.
 
Other studies assessed the presence of a critical value of some physical parameters above which penumbra formation takes place.
\citet{Lek98} found a threshold of 1-1.5 $\times$ 10$^{20}$ Mx, above which a pore can develop a penumbra. From the analysis of a data set taken at the German Vacuum Tower Telescope, \citet{Rez12} studied the formation of a sunspot penumbra in AR NOAA 11024 and proposed a critical magnetic field strength B$_{crit}\leq1.6$ kG and a critical inclination angle of the magnetic field $\alpha\geq60^{\circ}$ with respect to the normal to the photosphere, above which the penumbra begins to form. \citet{Jur11} investigated nine stable sunspots and concluded that the umbra-penumbra (UP) boundary, traditionally defined by an intensity threshold, is also characterized by a [critical] value of the vertical component of the magnetic field, $B_{ver}^{stable}=1860$ $(\pm 190)$ G. \citet{Jur15} confirmed this result: extending the analysis to cover the phase of penumbra formation, they also deduced that the UP boundary migrates toward the umbra and $B_{ver}$ increases. Therefore, during penumbra formation, the pore is partially converted into penumbra. To explain this critical value of $B_{ver}$, they propose that there are two modes of magneto-convection. The penumbral mode takes over in areas with $B_{ver}<B^{stable}_{ver}$, while the umbral mode prevails instead in areas with $B_{ver}>B^{stable}_{ver}$. Moreover, through the study of the AR NOAA 11024, \citet{Sch10b} found that the penumbra forms in segments and that, initially, it cannot settle down on the side towards the opposite polarity where flux emergence is still occurring.

Another important issue that needs to be clarified in the formation of the penumbra is the initiation of the Evershed flow \citep{Eve09}. It consists of a nearly horizontal outflow along the penumbral filaments, mainly manifested as red and blue wavelength shifts in the photospheric absorption lines at the limb side and disc-center side of the penumbra, respectively. This flow, with typical, spatially averaged speeds of 1-2 km s$^{-1}$, is confined in nearly horizontal magnetic field channels (i.e., the so-called interspines, \citealp{Bor08}). In one set of models \citep{Mey68} it is ascribed to the difference in the magnetic field strength between the two footpoints of a penumbral filament. This causes a difference of gas pressure and drives a flow towards the footpoint with higher magnetic field strength, i.e., the footpoint further away from the umbra \citep{Bor11} although often still located inside the sunspot, since most of the flow returns back inside the Sun within the penumbra (see Solanki et al. 1994). Alternatively, \citet{Scha08}, \citet{Spr06}, \citet{Scha06} proposed that the Evershed effect is produced by convection (i.e., that the Evershed effect is mainly driven by gas pressure gradients produced by horizontal gradients of temperature).

However, before the formation of the penumbra a line-of-sight (LOS) velocity of opposite sign with respect to that displayed by the typical Evershed flow was observed at some azimuths \citep{Schl12}. This flow seemed to be associated with the early stages of penumbra formation and reversed its sign as the penumbra formed. In fact, also \citet{Rom14} found persistent photospheric plasma upflow before the formation of the penumbra at the locations of the patches at the outer edge of the annular zone, and downflows in the inner part of the annular zone, which
were interpreted as the signature of an inflow towards the pore. They interpreted this plasma motion as a counter-Evershed flow.

It is clear that the comprehension of the counter-Evershed flow during the early stages of the penumbra formation may be useful to explain the processes of energy transport in the formed penumbra where, in principle, the presence of a rather strong (1500 G) and horizontal (40$^{\circ}$-80$^{\circ}$) magnetic field should inhibit the convective motions. 

Various models have been proposed to explain the presence of convective motions in the penumbra: the {\em hot rising flux-tube model} \citep{Sch02}, the {\em azimuthal convection model} \citep{Spr06} and more sophisticated geometries of elongated convection cells \citep{Remp09}. The former predicts the presence of upflows at the inner footpoints of the flux tubes near the umbra and downflows at their outer footpoints at the edge of the sunspot. In this case, the convection is radial with respect to the sunspot barycenter, with convective flows occurring along the penumbral filaments. The latter model provides a very efficient heat transport mechanism: the convective motions are present over the entire length of the bright penumbral filaments, with upflows at the center of the filaments and downflows at the filament edges. Observations suggest that some combination of both types of flows is acting, with both, a flow directed along the penumbral filaments and a flow directed perpendicularly to them playing a role \citep{Jos11, Scha11, RuizRamos13, Scha13, Tiw13, EstPo15}. 

In this scenario, understanding the presence of the counter-Evershed flow before the penumbra formation could be useful to shed light on the dynamics of the penumbral region. For this reason, we present new results obtained from the study of the formation of the penumbra in a sunspot already studied by \citet{Rom13, Rom14}. In this Paper, we analyze a new data set, consisting of spectro-polarimetric scans of the \ion{Fe}{1} 630.25 nm line acquired after the formation of the penumbra, as well as using HMI observations. In particular, we analyze the plasma motions inside the annular zone, providing new constraints for modeling the formation phase of the sunspot penumbra. We focus on the onset of the classical Evershed flow, which is observed to occur during penumbra formation.\\ In the next Section we describe the whole data set and its analysis. In Section 3 we present the results. Finally, in Section 4 we summarize the main conclusions.

\section{Observations and analysis}

We study AR NOAA 11490 using high temporal, spatial, and spectral resolution data acquired by the Interferometric Bidimensional Spectrometer \citep[IBIS;][]{Cav06} operating at the NSO/Dunn Solar Telescope (DST). The observations were carried out on 2012 May 28 from 13:39 UT to 14:12 UT and on May 29 from 13:49 UT to 14:32 UT when the AR was characterized by a cosine of the heliocentric angle $\mu=0.95$ and $\mu=0.97$, respectively.

The data set, whose relevant characteristics were already described by detail in \citet{Rom13}, consists of 30 scans for each day of observation through the \ion{Fe}{1} 630.25 nm line, with 67 s cadence. The line was sampled with a spectral profile having a FWHM of 2 pm, an average wavelength step of 2 pm and an integration time of 60 ms. The \ion{Fe}{1} 630.25 nm line was sampled in spectro-polarimetric mode with 30 spectral points. The field of view (FOV) was 500 $\times$ 1000 pixels with a pixel scale of 0\farcs09. 

For each spectral frame, a simultaneous broad-band (at $633.32 \pm 5$ nm) frame, imaging the same FOV with the same exposure time, was acquired. To reduce the seeing degradation, the images were restored using the Multi-Frame Blind Deconvolution \citep[MFBD;][]{Lof02} technique (see details in \citealp{Rom13}).

To determine the evolution of the LOS plasma velocity, magnetic field strength, inclination and azimuth angles, we performed a single-component inversion of the Stokes profiles for all the available scans of the \ion{Fe}{1} 630.25 nm line using the SIR code \citep{RuizIniesta92}. We used a different procedure to invert the Stokes profiles of the data set acquired after the penumbra formation, with respect to the procedure used in \citet{Rom13}. The spectra were normalized to the quiet sun continuum, $I{_c}$. More precisely, we divided the FOV into three regions, identified by different thresholds in the continuum intensity $I{_c}$ to account for the different physical conditions: quiet Sun ($I{_c}>0.9$), penumbra ($0.7<I{_c}<0.9$), umbra ($I{_c}<0.7$). For the quiet Sun model we used as an initial guess the temperature stratification of the Harvard-Smithsonian Reference Atmosphere \citep[HSRA,][]{HSRA} and a value of 0.1 km s$^{-1}$ for the line-of.sight (LOS) velocity. In the penumbral model, we changed the initial guess of the temperature (T) and the electron pressure ($p_{e^{-}}$) according to the penumbral stratification provided by \citet{Iniesta94}, and we used an initial value for the magnetic field strength of 1000 G and 1 km s$^{-1}$ for the LOS velocity. For the umbral model we changed the initial T and $p_{e^{-}}$ using the values provided by \citet{Col94}, (an umbral model for a small spot), and we also started from a value of 2000 G for the magnetic field strength.\\
The temperature stratification of each component was modified with three nodes, although all other quantities were assumed to be height independent. We modelled the stray-light contamination by averaging over all Stokes I spectra in the 64 pixels characterized by the lowest polarization degree. A magnetic filling factor was introduced as a free parameter of the inversion, which described the weight being assigned to the local atmosphere relative to the stray-light. The spectral point-spread function of IBIS \citep{Rea08} was used to take into account the finite spectral resolution of the instrument.
Once we obtained the magnetic field strength, the inclination and azimuth angles, we solved the 180$^{\circ}$-azimuth ambiguity and transformed the components of the vector magnetic field into the local solar frame using the Non-Potential Field Calculation code \citep{Geo05}.
 
In general, the results obtained  by the SIR inversion code appear reasonable, but in a few places in the penumbra, anomalous velocities were obtained (small patches of strong upflows, with jumps in the velocity at their edges). Various tests, such as changing the number of nodes, starting from different initial values of the free parameters, etc. did not improve the situation. Since the velocity is the most central variable of this study, we also measured the LOS plasma velocity using Gaussian fits to the line profiles, i.e., we reconstructed the profiles of the \ion{Fe}{1} line in each spatial pixel by fitting the corresponding Stokes I component with a linear background and a Gaussian shaped line. The values of LOS velocity were deduced from the Doppler shift of the centroid of the line profiles in each spatial point. We estimated the uncertainty affecting the velocity measurements considering the standard deviation of the centroids of the line profiles estimated in all points of the whole FOV. Thus, the estimated relative error in the velocity is $\pm$0.2 km s$^{-1}$.

The temperature in the umbra is low enough to allow for the formation of molecules, in particular blending with the 630.25 nm line. Therefore, all umbral profiles with $I_{c}<0.7 I_{qs}$, were excluded from the calculation of the line shift, and the Doppler velocity in the umbra were set to zero. The reference for the local frame of rest was calibrated by imposing that the plasma in a quiet Sun region has on average a convective blueshift \citep{Drav81} for the \ion{Fe}{1} 630.25 line, equal to -124 m s $^{-1}$ following \citet{Balth88}.

We also analyzed both Space-weather HMI Active Region Patches
\citep[SHARPs,][]{Bobra14} continuum filtegrams and Dopplergrams taken by the Helioseismic and Magnetic Imager (HMI) on the Solar Dynamics Observatory (SDO) \citep{Sch12} satellite in the Fe I 617.3 nm with a resolution of 1\arcsec\/ to study the evolution of the velocity field in the forming penumbra. These data cover one day of observation, starting from 2012 May 28 at 14:58:25 UT until May 29 at 14:58:25 UT. The cadence of these data is 12 minutes. To calibrate SDO/HMI Dopplergrams we choose the same calibration method used for IBIS velocity maps, with convective blueshift equal to -95 m s$^{-1}$ \citep{Balth88}. Moreover, to study the variation of the magnetic field of the active region we analyzed the components $B_{r}$, $B_{\theta}$, $B_{\phi}$ of the vector magnetic field \textit{B} deduced from SDO/HMI SHARPs data. The uncertainties in the field strength and in the inclinations are $\pm$240 G and $\pm$ 20$^{\circ}$, respectively.

IBIS and SDO/HMI observations were co-aligned using the first spectral image in the continuum of the \ion{Fe}{1} 630.25 nm line in the sequence of IBIS data taken at 13:39 UT and 13:59 UT, for 2012 May 28 and 29, respectively, and a SDO/HMI continuum image closest in time (13:36 UT and 13:58 UT, for 2012 May 28 and 29, respectively). We used the IDL \textit{SolarSoft} mapping routines to take into account the different pixel sizes.

To analyze the evolution of the plasma flow in the forming penumbra, we aligned the SDO/HMI images from 19:00 UT to 24:00 UT taking as reference image the first of these images. Our aim was to overlay the images of the pore exactly on top of each other, so that the evolution of individual parts of the pore (and hence of the forming penumbra) can be followed. The displacement between the reference image and the other images was obtained with cross-correlation techniques. The rapid evolution and motion of the forming sunspot limits the  precision of the alignment, which is of the order of the pixel size of SDO/HMI, i.e., 0\farcs5. From the SDO/HMI observations, we extracted sub-arrays for further analysis, as shown in Figure 1.

\section{Results}

\begin{figure*}[!t]
	\centering
	\includegraphics[scale=0.40, clip, trim=20 120 50 280]{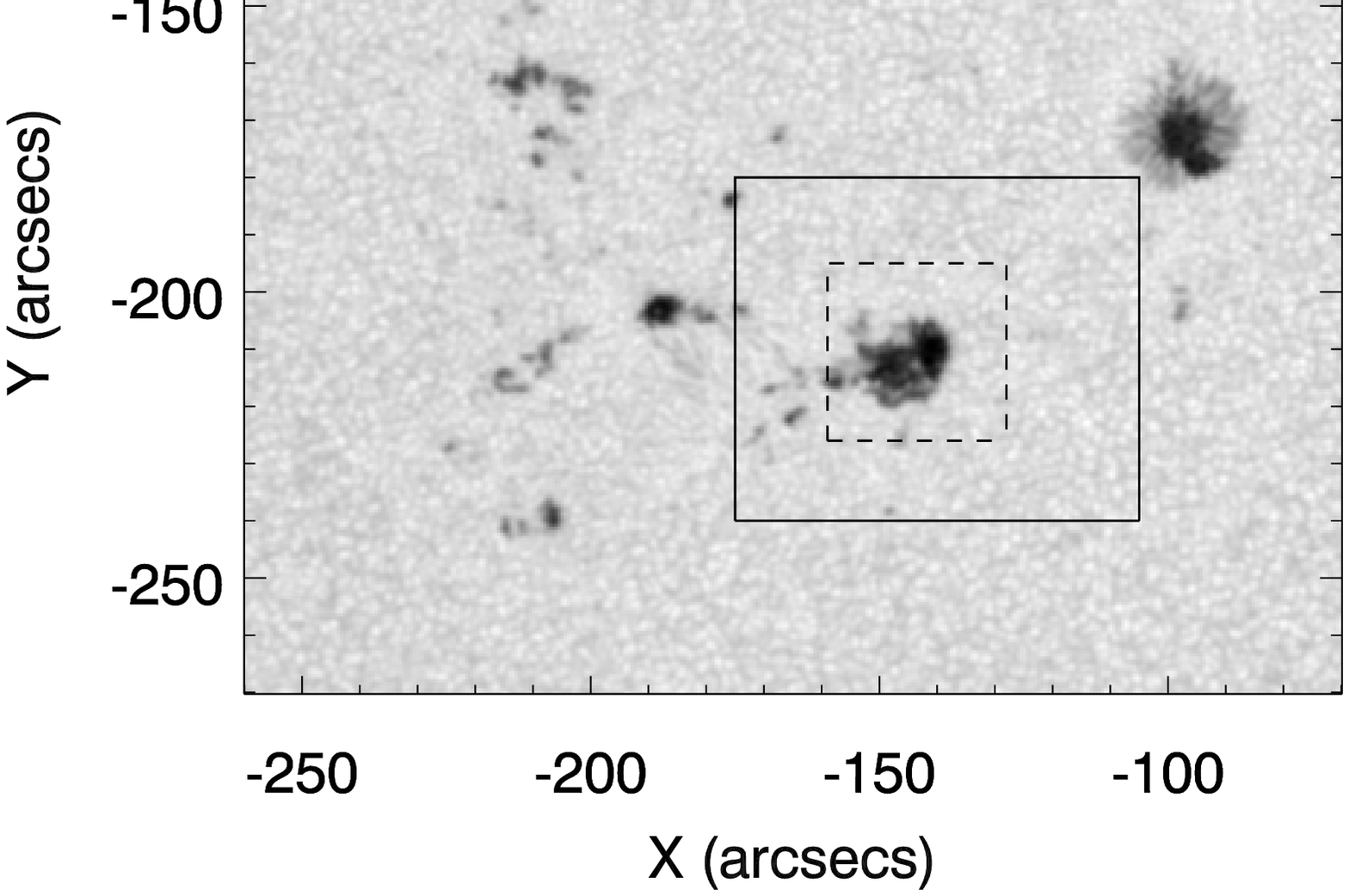}
	\includegraphics[scale=0.40, clip, trim=70 120 50 280]{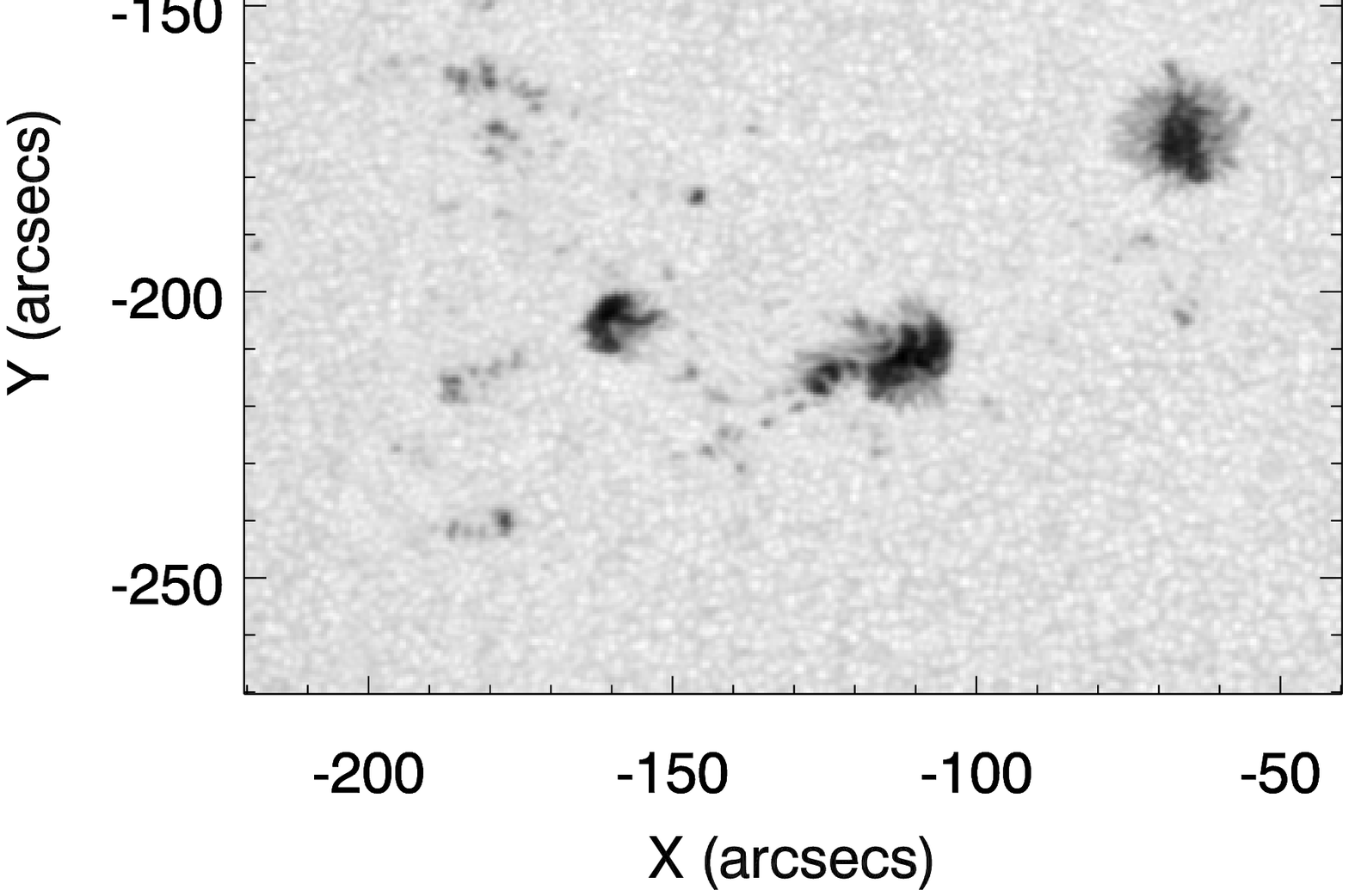}\\
	\includegraphics[scale=0.40, clip, trim=20 120 50 280]{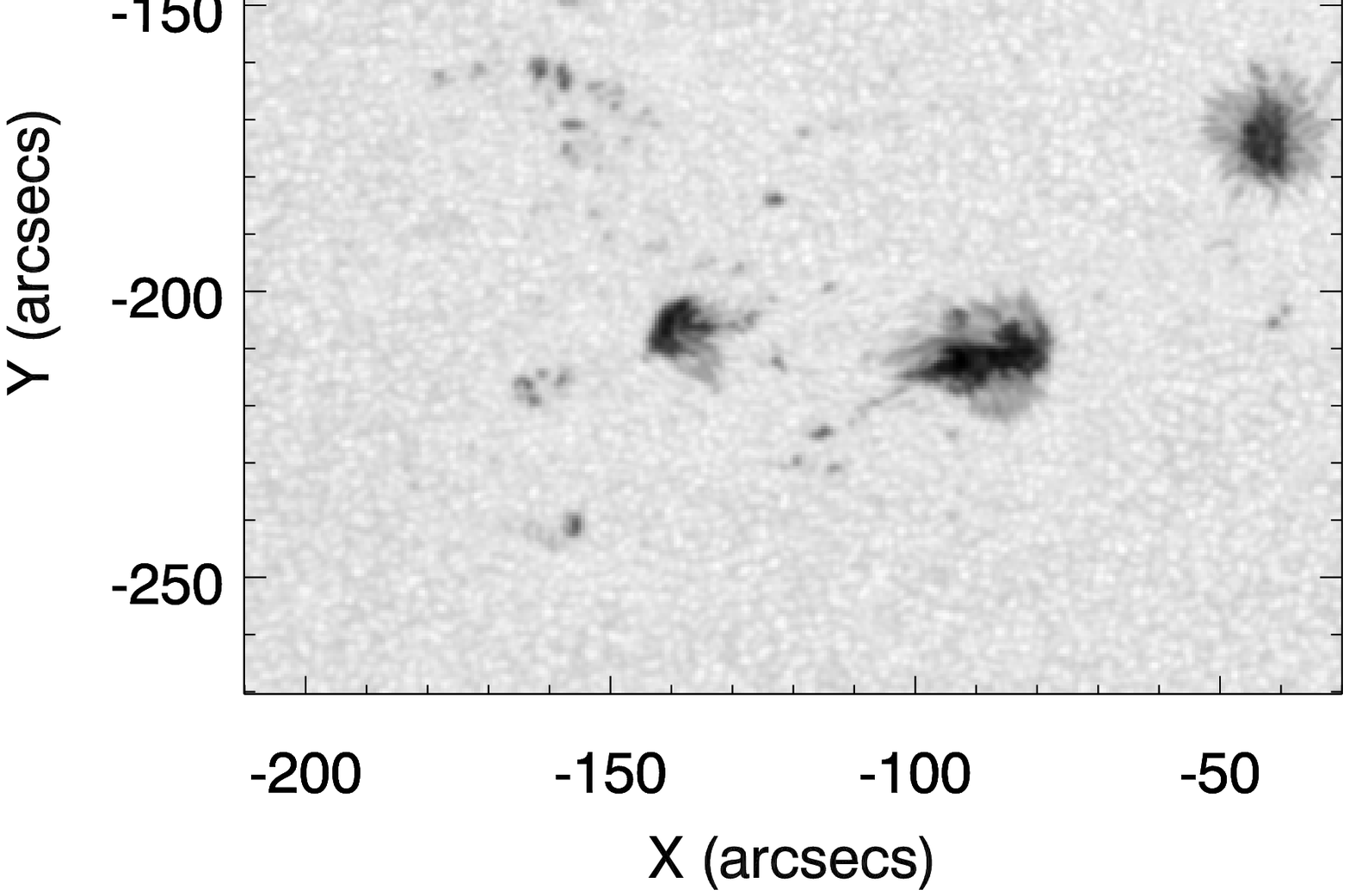}
	\includegraphics[scale=0.40, clip, trim=70 120 50 280]{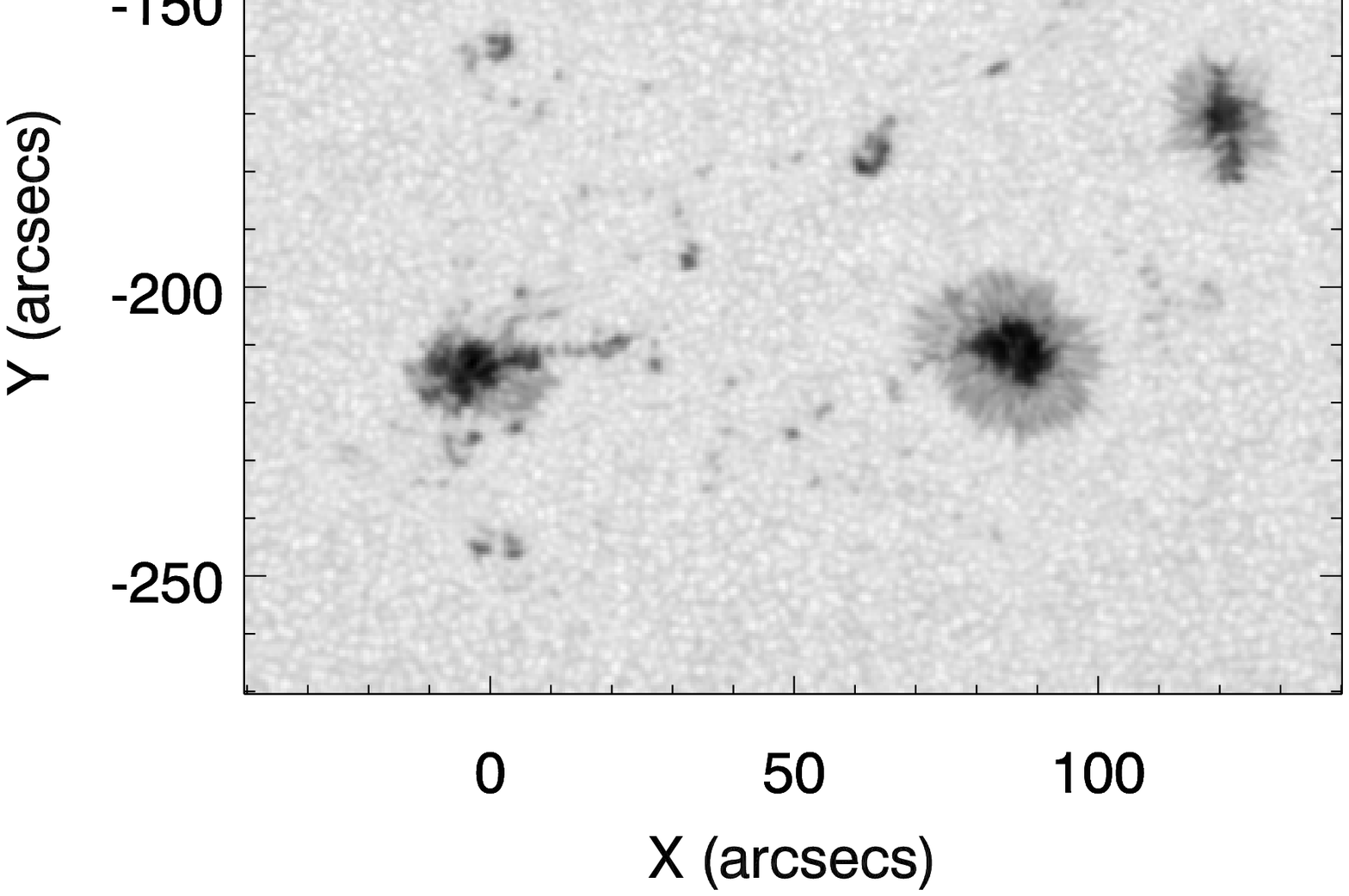}\\
	\includegraphics[scale=0.40, clip, trim=20 90 50 280]{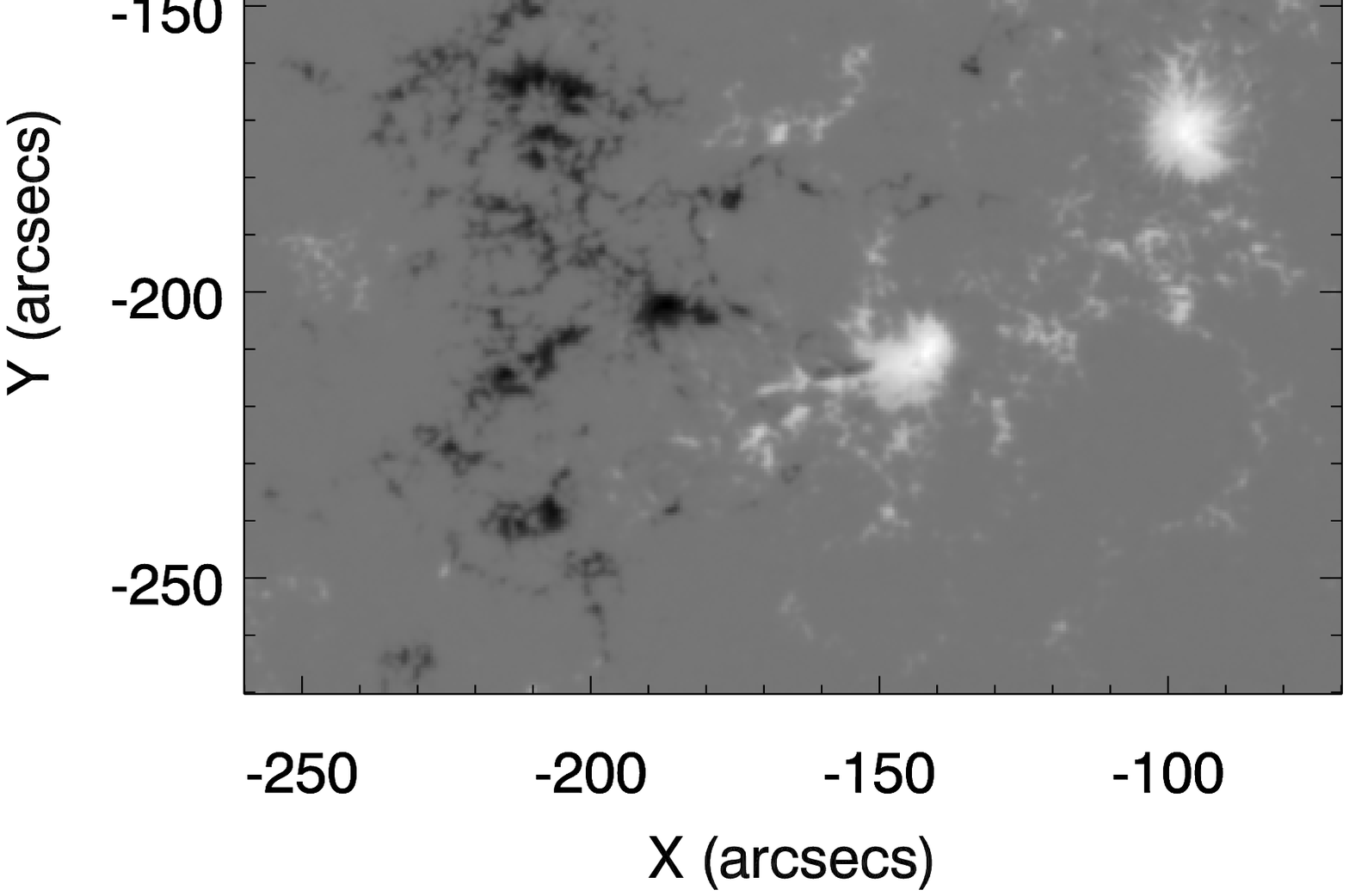}
	\includegraphics[scale=0.40, clip, trim=70 90 50 280]{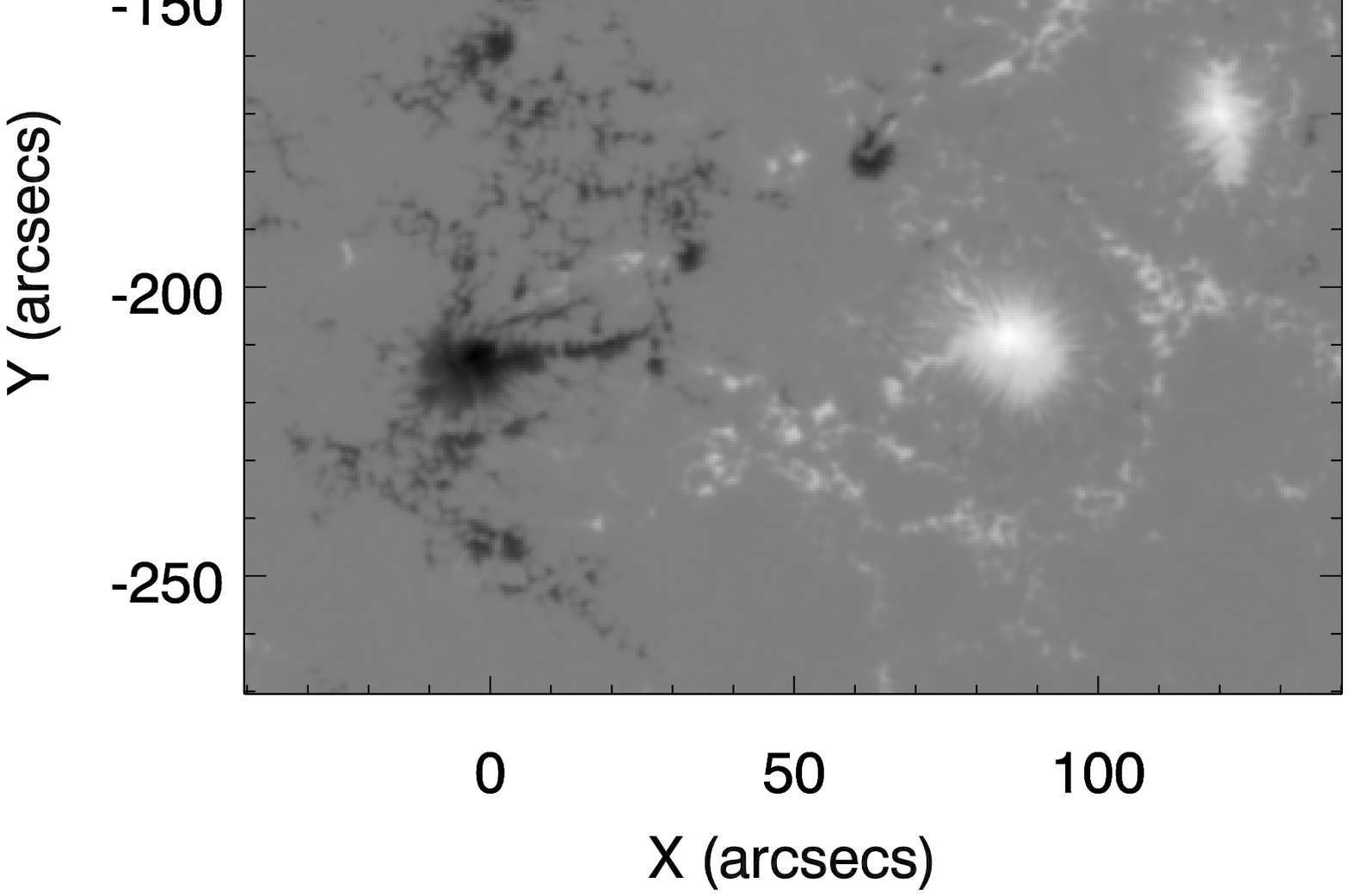}\\
	\caption{Continuum filtergrams (\textit{first and second rows}) and LOS magnetograms (\textit{third row}) taken by SDO/HMI at the times given at the top of each panel, showing the evolution of NOAA AR 11490. In these and in the following images North is at the top, West is to the right. In the \textit{top left panel} the boxes indicate the FoVs of IBIS (dashed line) and SDO/HMI (solid line) displayed in Figure 2 and 3, respectively. The axes give the distance from solar disc center in arcsec. The arrow points to the disc center. \label{fig1}}
\end{figure*}

We can see in the sequence of continuum filtegrams shown in Figure 1 the evolution of AR NOAA 11490 from May 28 at 13:58 UT to May 29 at 13:58 UT. We note that the pore in the boxes in Figure 1 (\textit{top left panel}), characterized by positive polarity (Figure 1, \textit{bottom left panel}), forms its penumbra in the course of the depicted 24 hours. In particular, the pore changes its initial shape, as shown in the \textit{top right panel} of Figure 1 (17:22 UT), and the penumbra initially develops only on the north and south part of the pore. Later, the penumbra develops in the western part of the pore, and only at 23:58 UT does it also develop in the part towards the opposite polarity (see Figure 1, \textit{middle right panel}, showing the situation at 13:58 UT on 29 May). This development is in agreement with the findings of \citet{Sch10a} that the penumbra forms later in the direction of the opposite polarity of an active region, where flux is still emerging. Therefore, we estimated that the pore  becomes surrounded by its penumbra in about 10 hr, i.e., from 13:58 UT to 23:58 UT on 28 May.

Figure 2 shows maps of the continuum intensity, magnetic field strength, and inclination angle on May 28 at 14:00 UT (\textit{left panels, first, second, and third rows}), before the penumbra formed, as obtained from the SIR inversion of the Stokes profiles of the \ion{Fe}{1} 630.25 nm line. These maps reveal that the pore is characterized by an umbra which does not appear to be homogeneous (see also \citealp{Rom13}). The magnetic field strength in the pore is about 1.5 kG. Around the pore we distinguish an annular zone where the magnetic field exceeds 500 G (shown by the yellow contour in Figure 2, \textit{left panel second row}). In this zone the inclination is not constant but there are a number of sectors with different magnetic inclination. One can imagine this as an (upside down) ballerina skirt structure of the magnetic field on a large azimuthal scale (Figure 2, \textit{left panel, third row}). There are also patches, characterized by an inclination of about 180$^{\circ}$, corresponding to the polarity opposite to that of the sunspot. They are located only in some sectors of the annular zone and $\sim$ 3\arcsec\/- 4\arcsec\/ from the pore.

In Figure 2 (\textit{right panels}) we show the continuum intensity, the magnetic field strength, and the inclination angle after the formation of the penumbral region on May 29. The magnetic field in the penumbra gradually decreases from about 1.5 kG at the edge of the umbra to about 500 G at the external border of the penumbra. The inclination angle in the penumbra increases gradually from about on average 40$^{\circ}$-50$^{\circ}$ in the inner most penumbra to about 80$^{\circ}$-90$^{\circ}$ at its outer boundary. We also note that the patches of polarity opposite to that of the sunspot, are now more numerous and are located $\sim$ 10\arcsec\/ from the edge of the umbra, i.e., farther than the previous day. In this case they are visible all around the sunspot.

The \textit{bottom panels} of Figure 2 show the LOS velocity measured by Doppler shift of the centroid of the \ion{Fe}{1} 630.25 nm line. The saturation level chosen for these maps is $\pm$0.8 km s$^{-1}$ to better display the velocities in the annular zone and along the penumbral filaments. Before the penumbra is formed the annular zone is characterized by downflows larger than 1 km s$^{-1}$ in its inner part (Figure 2, \textit{left bottom panels}).  These flows are particularly evident in the north-western sector of the annular zone, where upflows slightly larger than the granular pattern are also visible at greater distance from the pore, but close to the downflows. This region is also characterized by elongated ``cells'' in intensity and an inhomogeneous field strength, with elongated structures (marked by red squares in the \textit{left panels} of Figure 2). 

After the penumbra had formed, on May 29, the LOS velocity map is dominated by the classic Evershed flow all around the spot (Figure 2, \textit{bottom right panel}), characterized by flow towards the observer of about -0.5 km s$^{-1}$ in the north-eastern part of the penumbra, and by flow away from the observer of 0.6-0.7 km s$^{-1}$ in the south-western part.
   
To further study the evolution of the plasma flow in the forming penumbra, we analysed the SDO/HMI data, which allowed us to follow the evolution of the spot over a longer time span, although with lower spatial resolution. Figure 3 shows the evolution of the continuum intensity (\textit{first column}), LOS velocity (\textit{second column}), strength and inclination angle of the magnetic field (\textit{third and fourth column}) from May 28 at 13:58 UT to May 29 at 14:58 UT, as deduced by SDO/HMI SHARP data. On May 28 13:58 UT we identify three sectors characterized by different values of inclination (see the arrows in Figure 3, \textit{fourth column}). In particular, we can see in the north-western part of the annular zone a sector (indicated by label 1 in the\textit{ top rightmost panel}) where the inclination is between 90$^{\circ}$ and 110$^{\circ}$; in the south-eastern part and in the north-eastern part two sectors (indicated by labels 2 and 3) where the inclination is between 30$^{\circ}$ and 60$^{\circ}$. This configuration was identified earlier in the IBIS observations, and described as an (upside down) ballerina skirt structure. In the subsequent 24 hours the region 1, characterized by horizontal field, surrounds the pore in the outer part of the penumbra, while the inner part shows an inclination of about 60$^{\circ}$-80$^{\circ}$.

In Figure 3 (\textit{second column}) we show the LOS velocity deduced from SDO/HMI data. In these maps we can see that before the formation of the penumbra, on May 28, similarly to the IBIS observations, in the north-western (center discward) part of the pore there is a significant redshift corresponding to velocities around 0.4-0.6 km s$^{-1}$. This line shift is opposite to that of the expected Evershed flow. Furthermore, the sequence of the LOS velocity maps show that, while the penumbra is forming, a different velocity pattern appears and a flow of opposite sign, in agreement with the Evershed flow, becomes more and more extended.

\begin{figure*}[tb]
	\centering
	\includegraphics[scale=0.40, clip, trim=30 250 100 250]{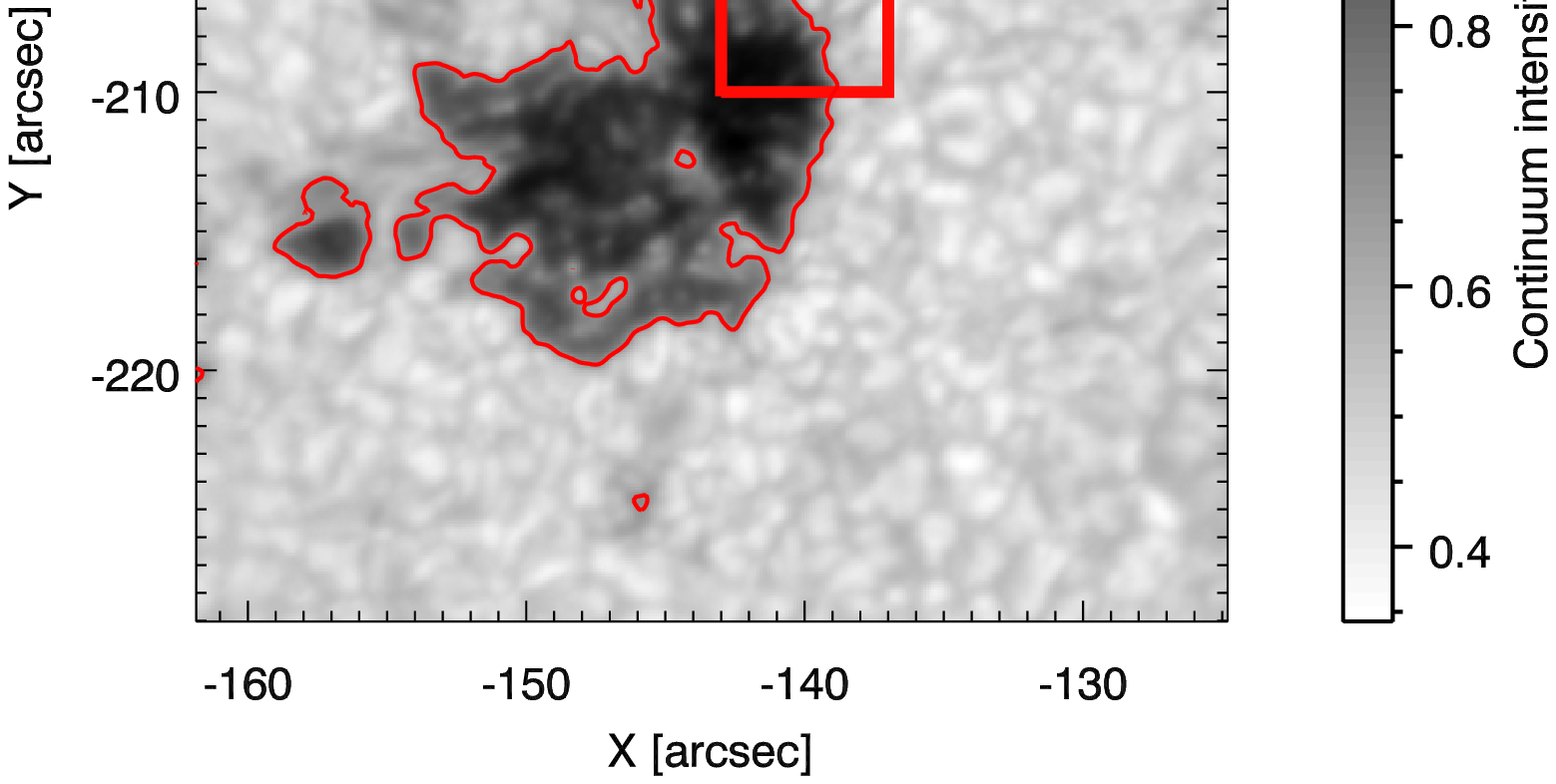}
	\includegraphics[scale=0.40, clip, trim=85 250 100 250]{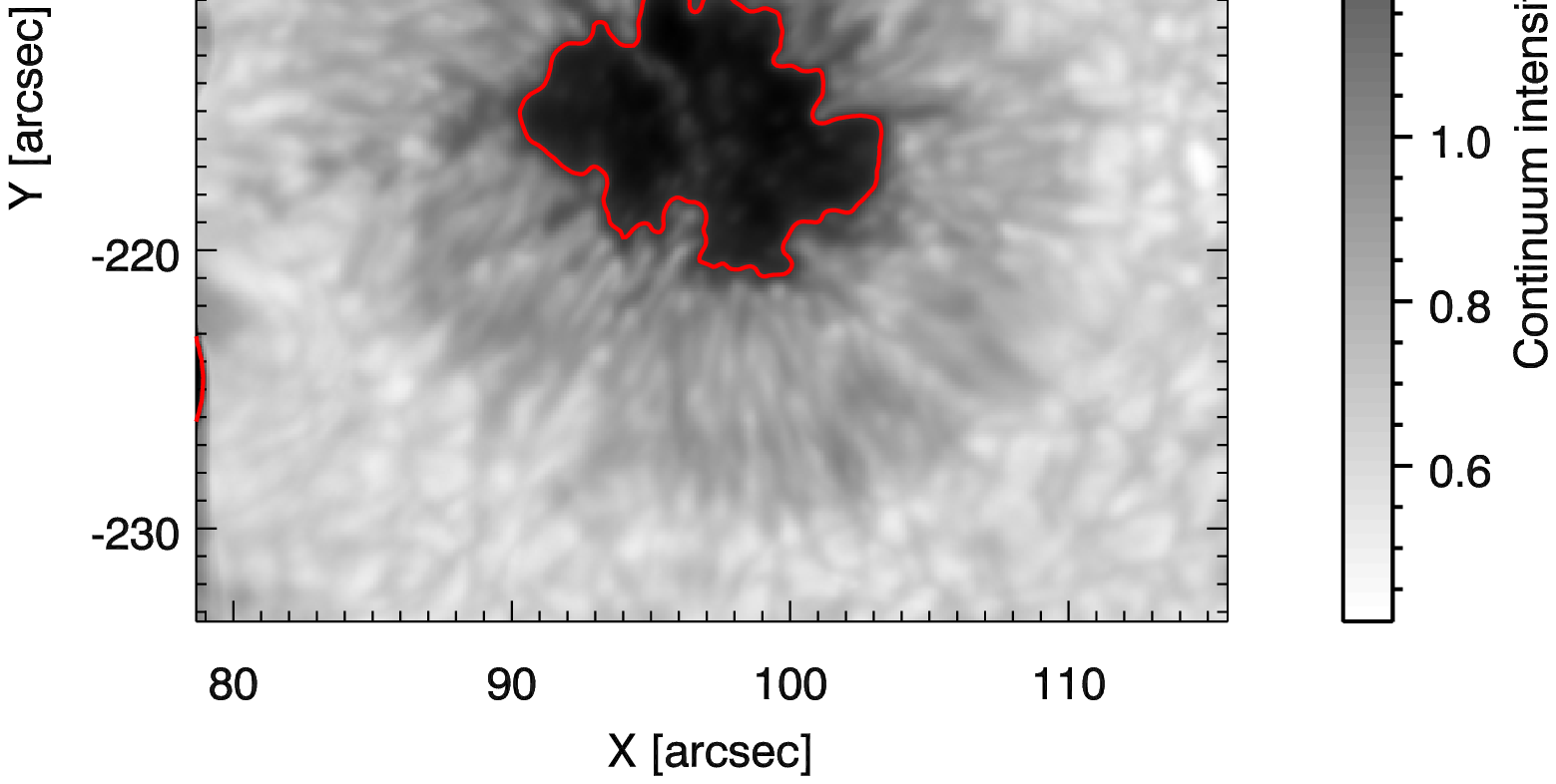}\\
	\includegraphics[scale=0.40, clip, trim=30 250 100 250]{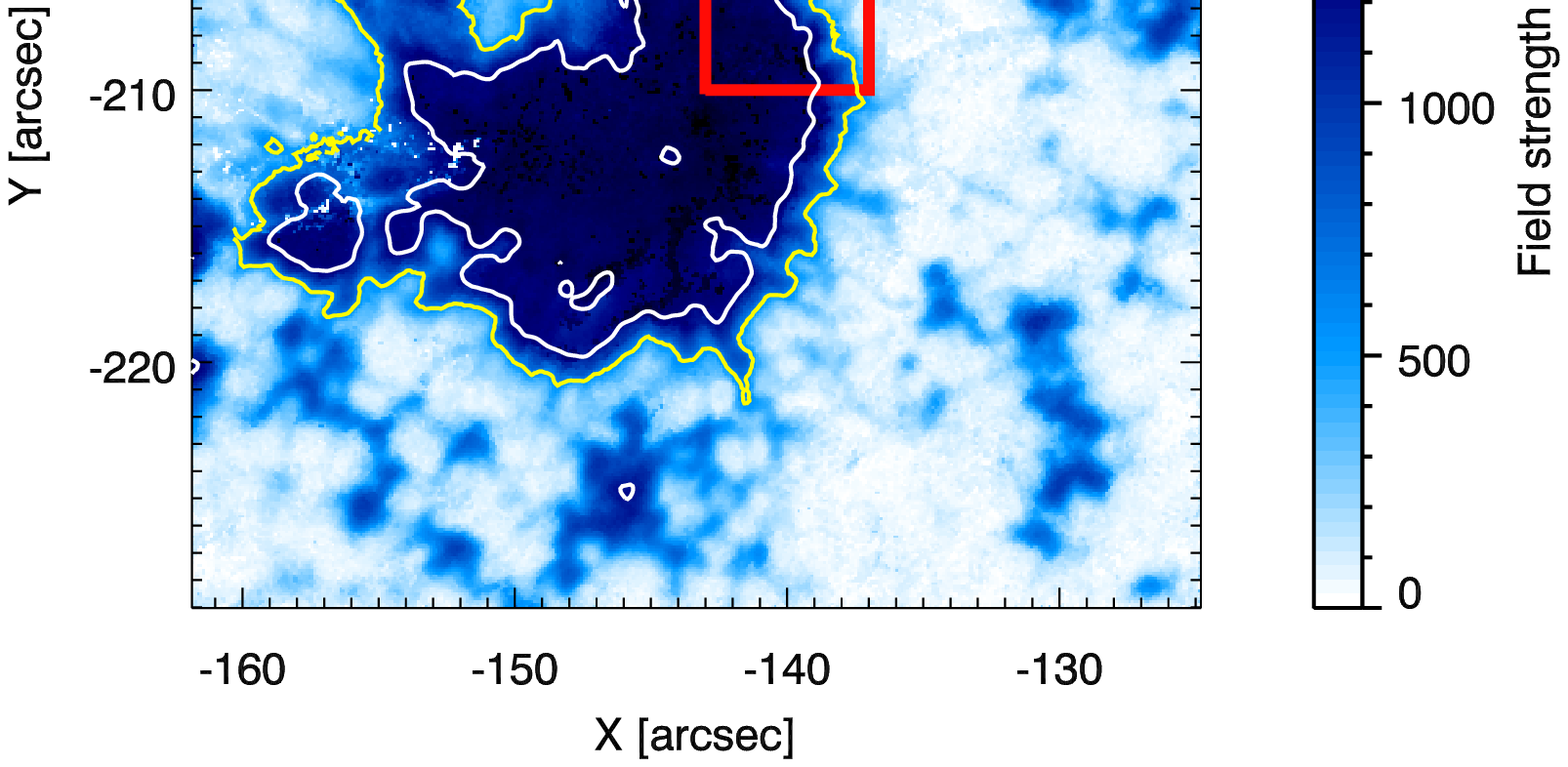}
	\includegraphics[scale=0.40, clip, trim=85 250 100 250]{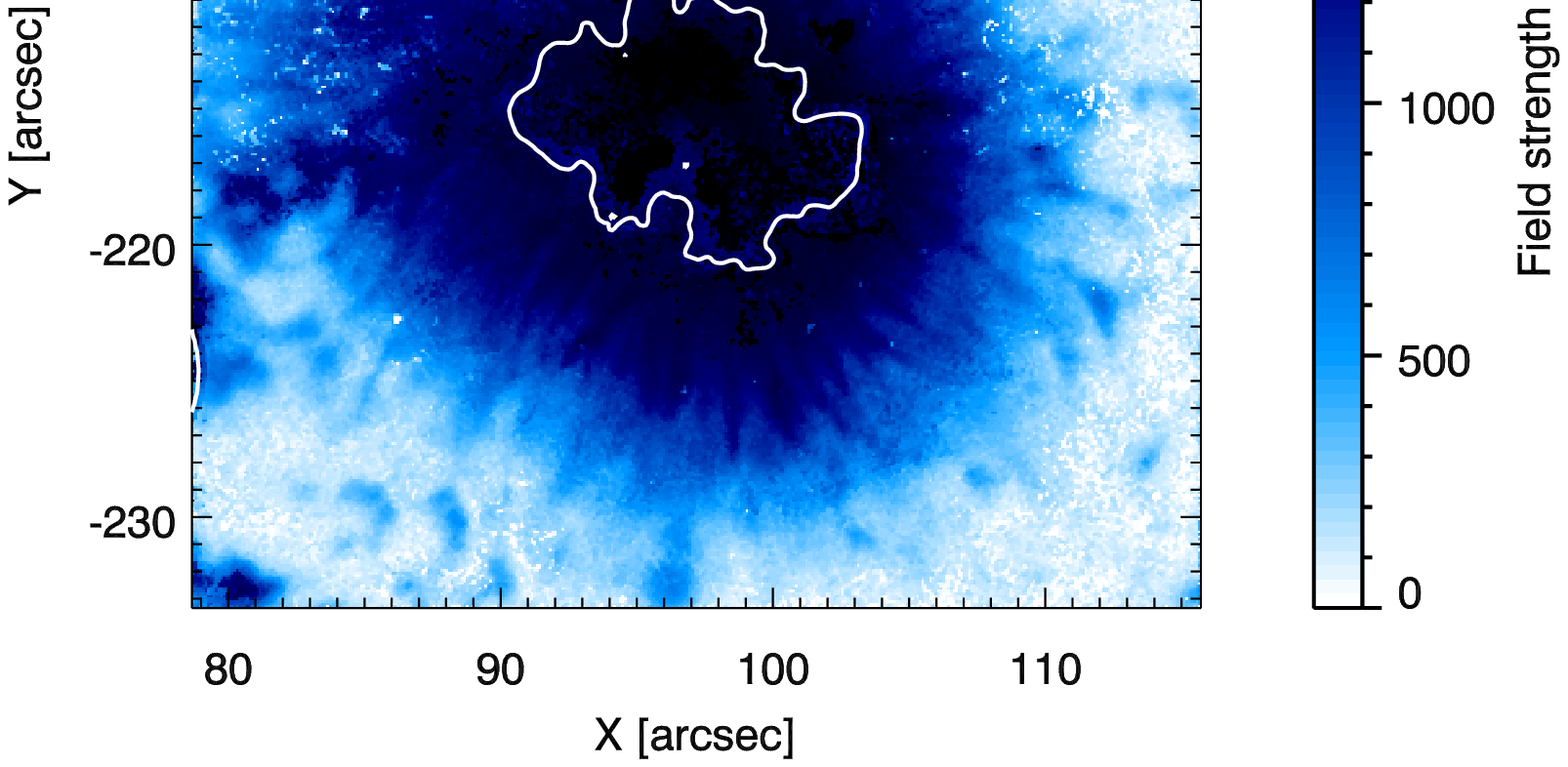}\\
	\includegraphics[scale=0.40, clip, trim=30 250 100 250]{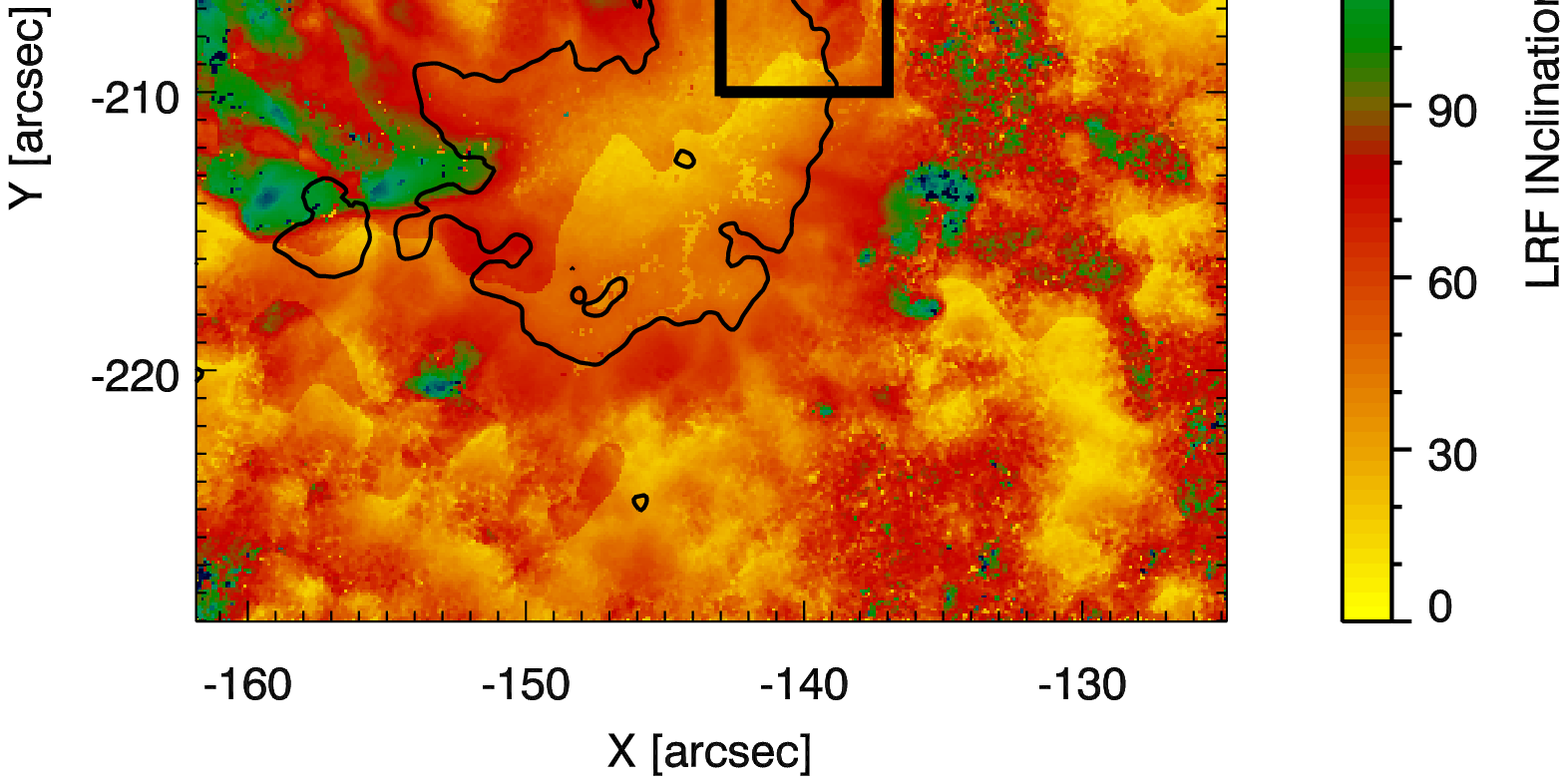}
	\includegraphics[scale=0.40, clip, trim=85 250 100 250]{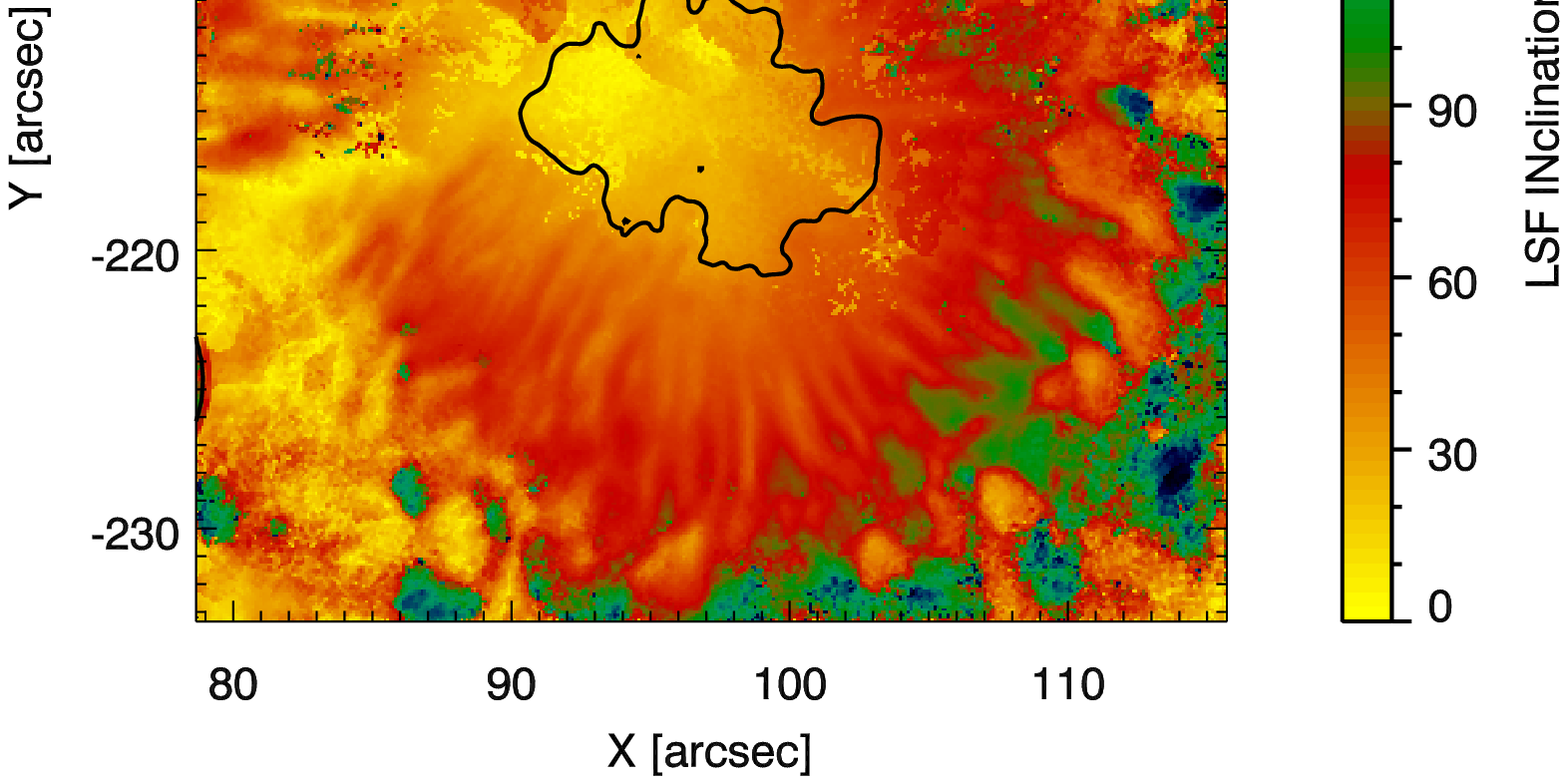}\\
	\includegraphics[scale=0.40, clip, trim=30 200 100 250]{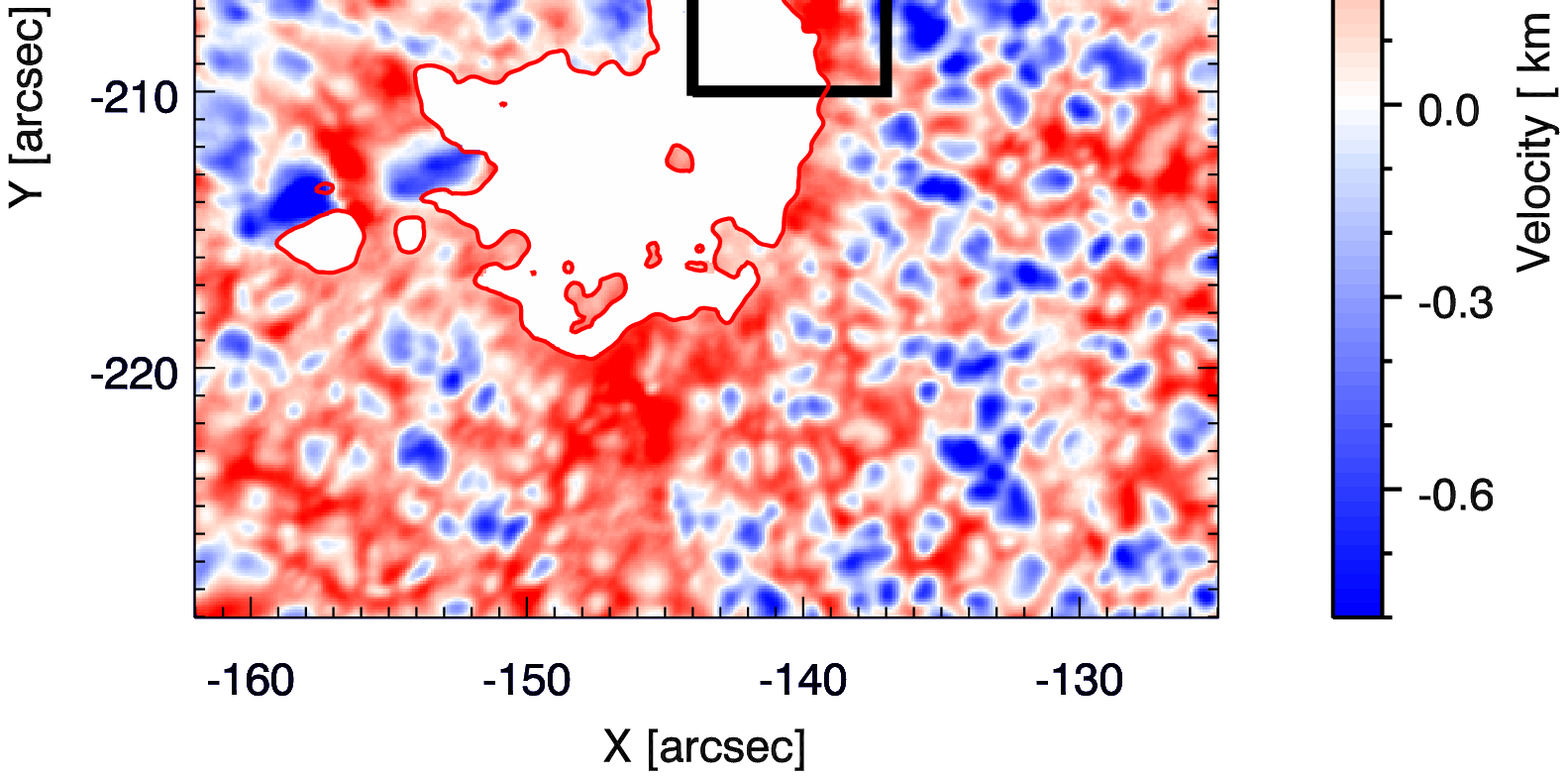}
	\includegraphics[scale=0.40, clip, trim=85 200 100 250]{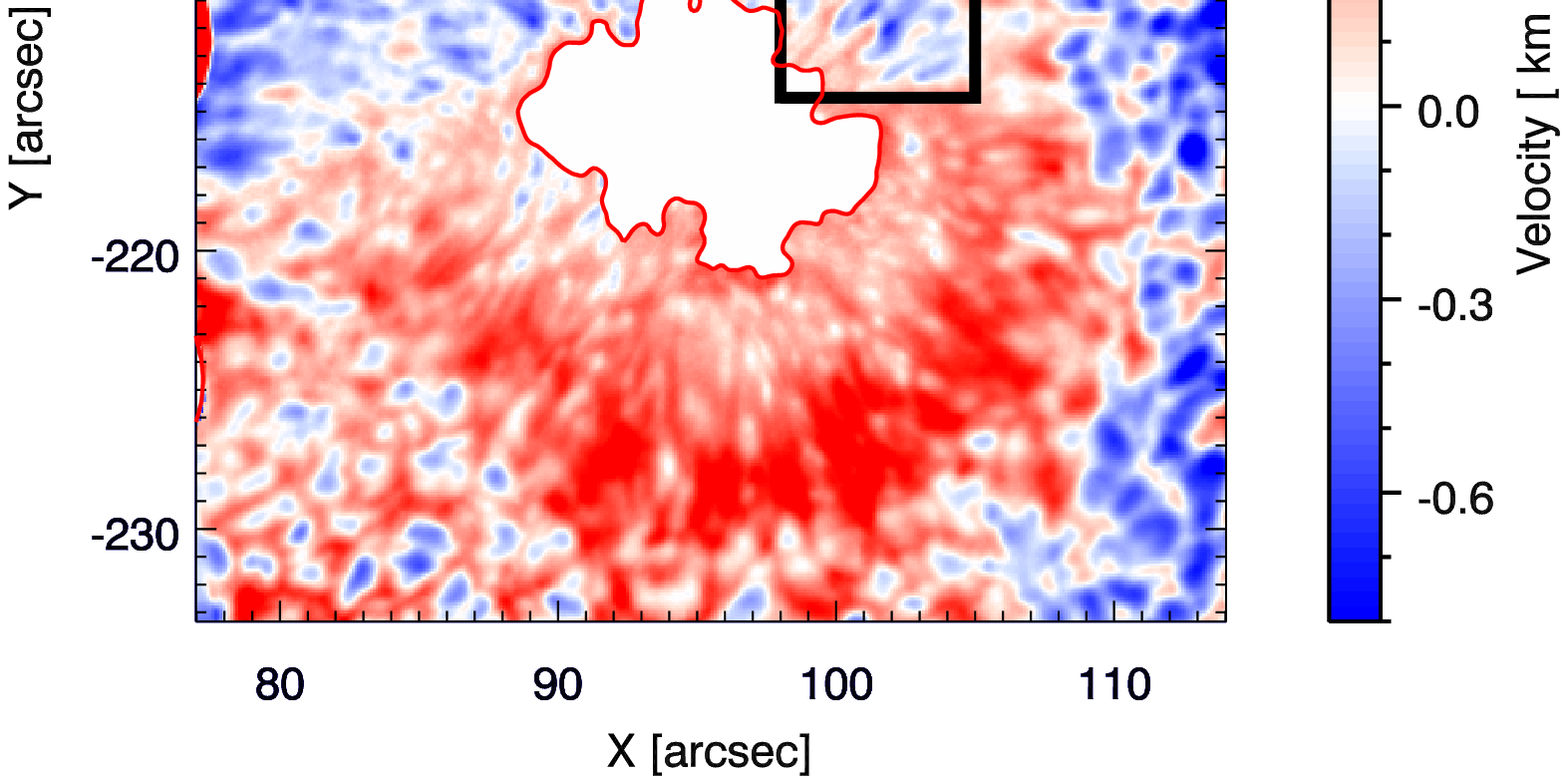}
	\caption{Maps of intensity, magnetic field strength and inclination angle (first, second, and third row) on 2012 May 28 at 14:00 UT (left, before penumbra formation) and on 2012 May 29 at 14:31 UT (right, after penumbra formation), obtained from the SIR inversion of the Stokes profiles of the \ion{Fe}{1} 630.25 nm line acquired by IBIS. The red or black contours indicate the edge of the pore and of the umbra as seen in the continuum intensity image. In the left panel of the second row, the contours indicate the edge of the pore as seen in the continuum intensity image (red contour) and the annular zone as seen in the magnetic field image (yellow contour), respectively. LOS velocity maps (bottom panels) on 2012 May 28 at 14:00 UT and 2012 May 29 at 14:31 UT, are deduced from the Doppler shift of the centroid of the \ion{Fe}{1} 630.25 nm line profile (see the main text for details). Downflow and upflow correspond to positive and negative velocities, respectively. The red or black square encloses a region of particular interest (see main text for details). The arrow points to the disc center. \label{fig2}}
\end{figure*}

One striking feature of Figure 3 is seen in the inclination images. At the beginning of the time series there are a number of small magnetic features surrounding the pore with polarities opposite to the pore. As time goes by more of these appear, forming a nearly complete ring around the sunspot. This opposite polarity ring is itself surrounded (on the outside) by a partial ring with the same polarity as the sunspot. As it forms, this ring moves away from the spot with time, presumably driven by the moat flow. Such features can also be seen in the third right panel of Figure 2.

In order to investigate the conditions that lead to the establishment of the classical Evershed flow, we analyze the evolution of the continuum intensity, LOS velocity, inclination and strength of the magnetic field in the 2-pixel wide (and 25-pixels long) segment A overplotted in the \textit{second row} of Figure 3 and in all frames of Figure 4, which shows the evolution of the continuum intensity and the LOS velocity from 21:12 UT to 21:58 UT on May 28. During this time interval the selected segment lies in a sector where the penumbra is forming. As time passes the blueshifted region covers a larger range of azimuths around the growing spot in the upper right of these images, while the azimuth coverage of the redshifted region decreases. In Figure 5 we can see the evolution of the continuum intensity and the LOS velocity along the segment on May 28 from 19:00 UT to 24:00 UT. In the \textit{top left panel} of Figure 5 we indicated the positions of the umbra-quiet sun boundary before the penumbra formation at 19:00 UT (the black vertical bar) and the umbra-penumbra boundary at 24:00 UT (the vertical orange bar). These positions  have been determined by computing the maxima in the derivative of the continuum intensity signal along the selected segment.  Analyzing the \textit{bottom panels}, in the inner part of the selected segments we note a clear evolution from redshift with a maximum of about 500 m s$^{-1}$ (see the curves taken at 19:00, 20:00 and 21:00 UT) to blueshift, whose maximum velocity of about 700 m s$^{-1}$ is reached at 22:00 UT. Figure 5, (\textit{bottom right panel}) shows in more detail the transition from redshift to blueshift that occurred between 21:00 UT and 22:00 UT on May 28.

\begin{figure*}[htbp]
	\centering
	\includegraphics[scale=0.245, clip, trim=0  200 76 230]{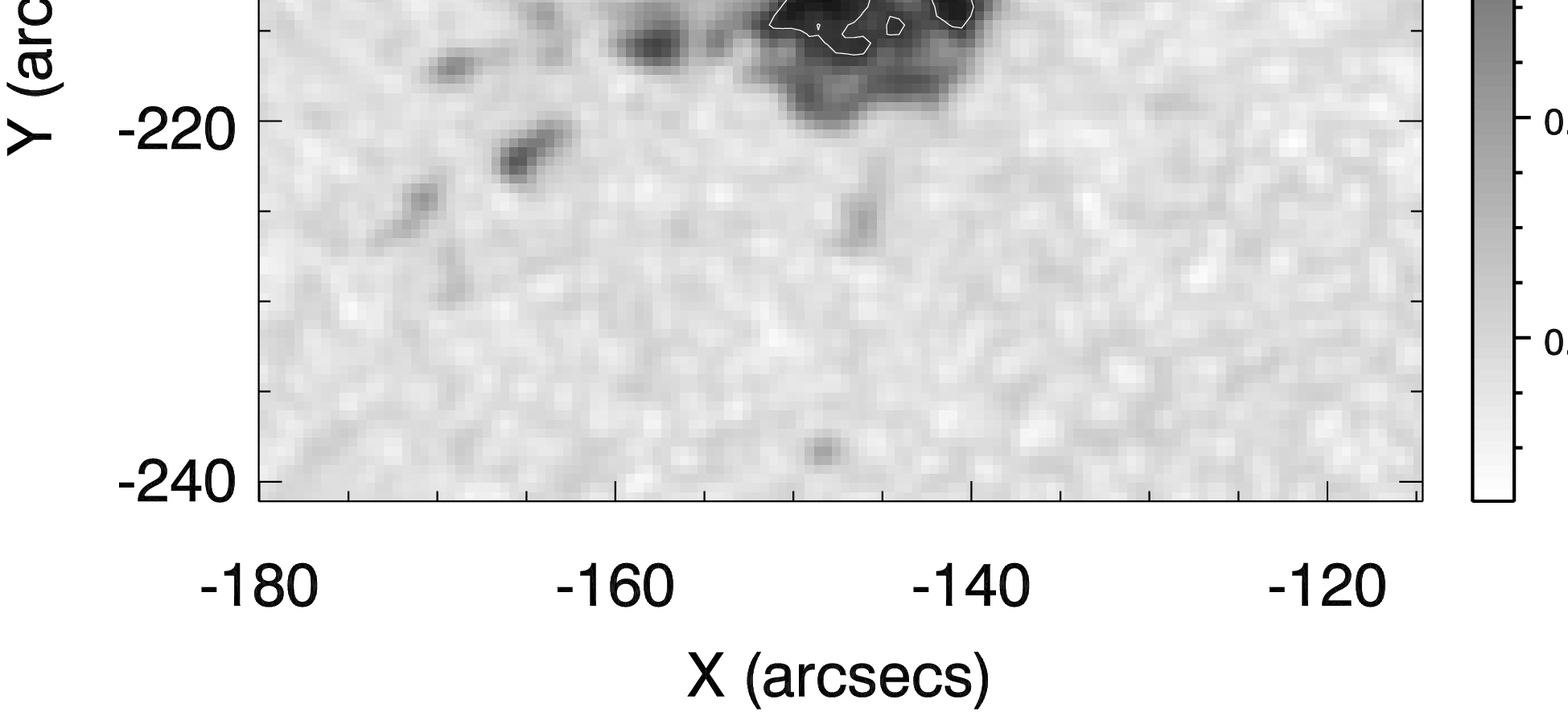}
	\includegraphics[scale=0.245, clip, trim=70 200 76 230]{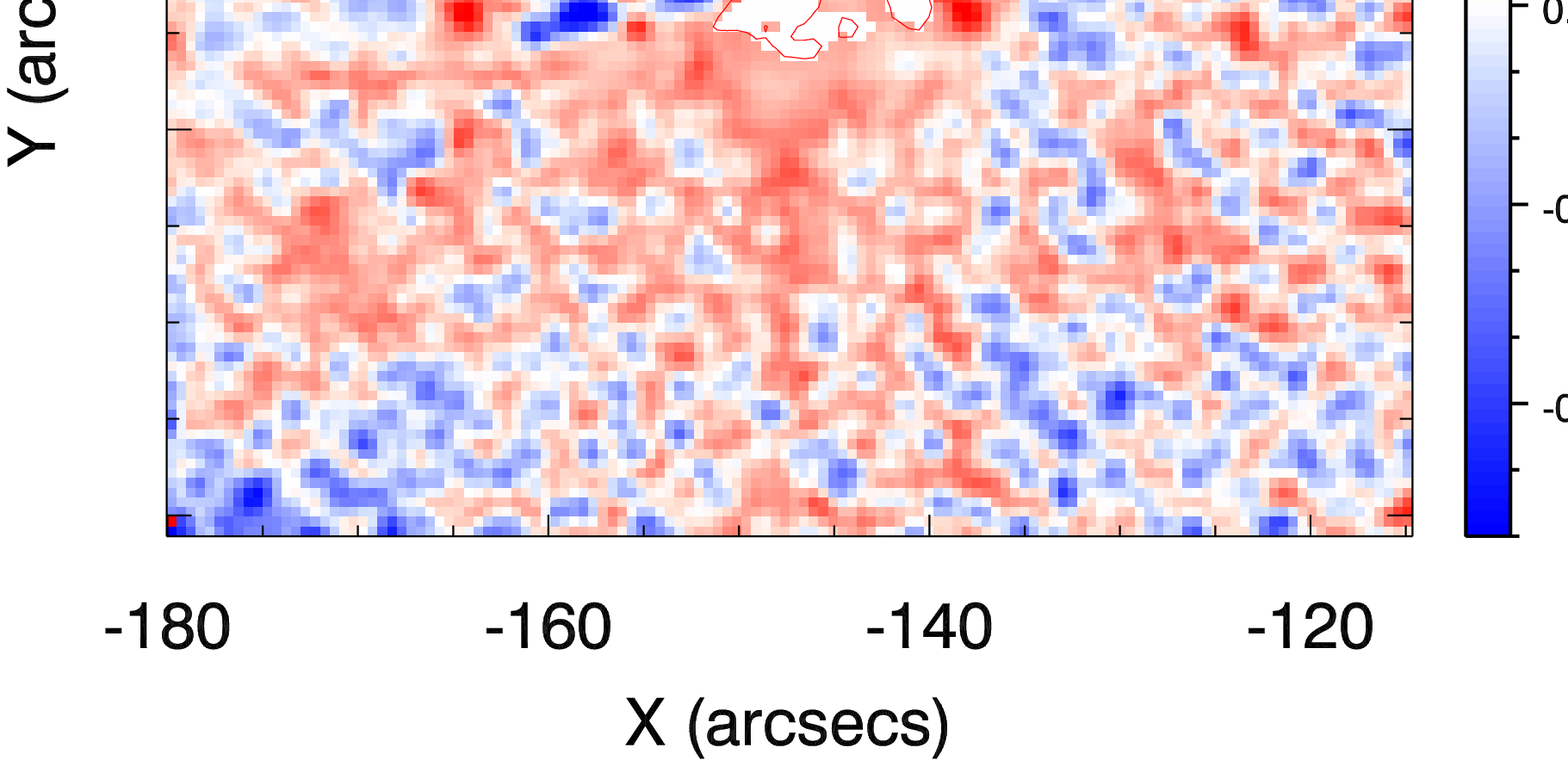}
	\includegraphics[scale=0.245, clip, trim=70 200 76 230]{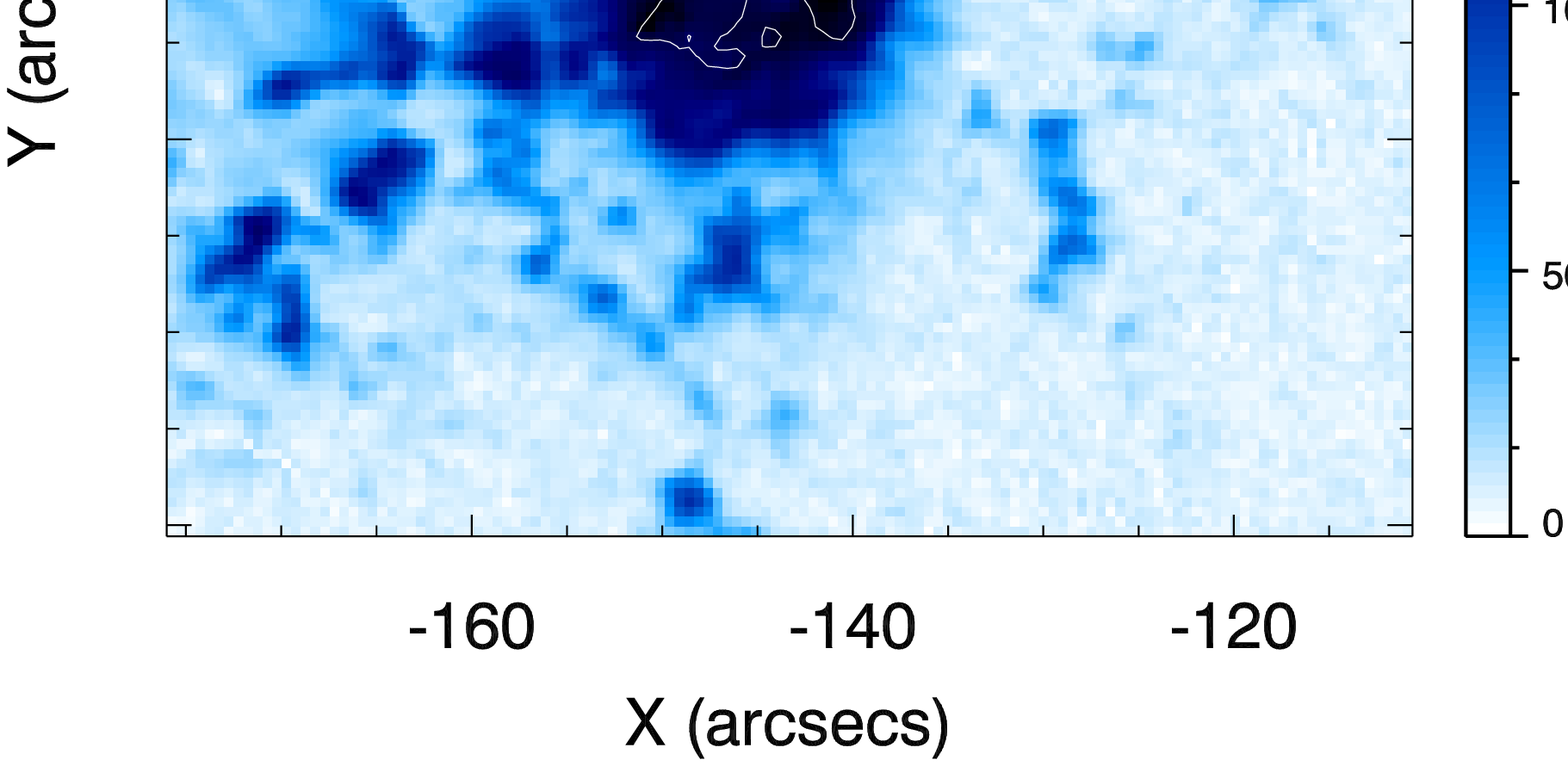}
	\includegraphics[scale=0.245, clip, trim=70 200 76 230]{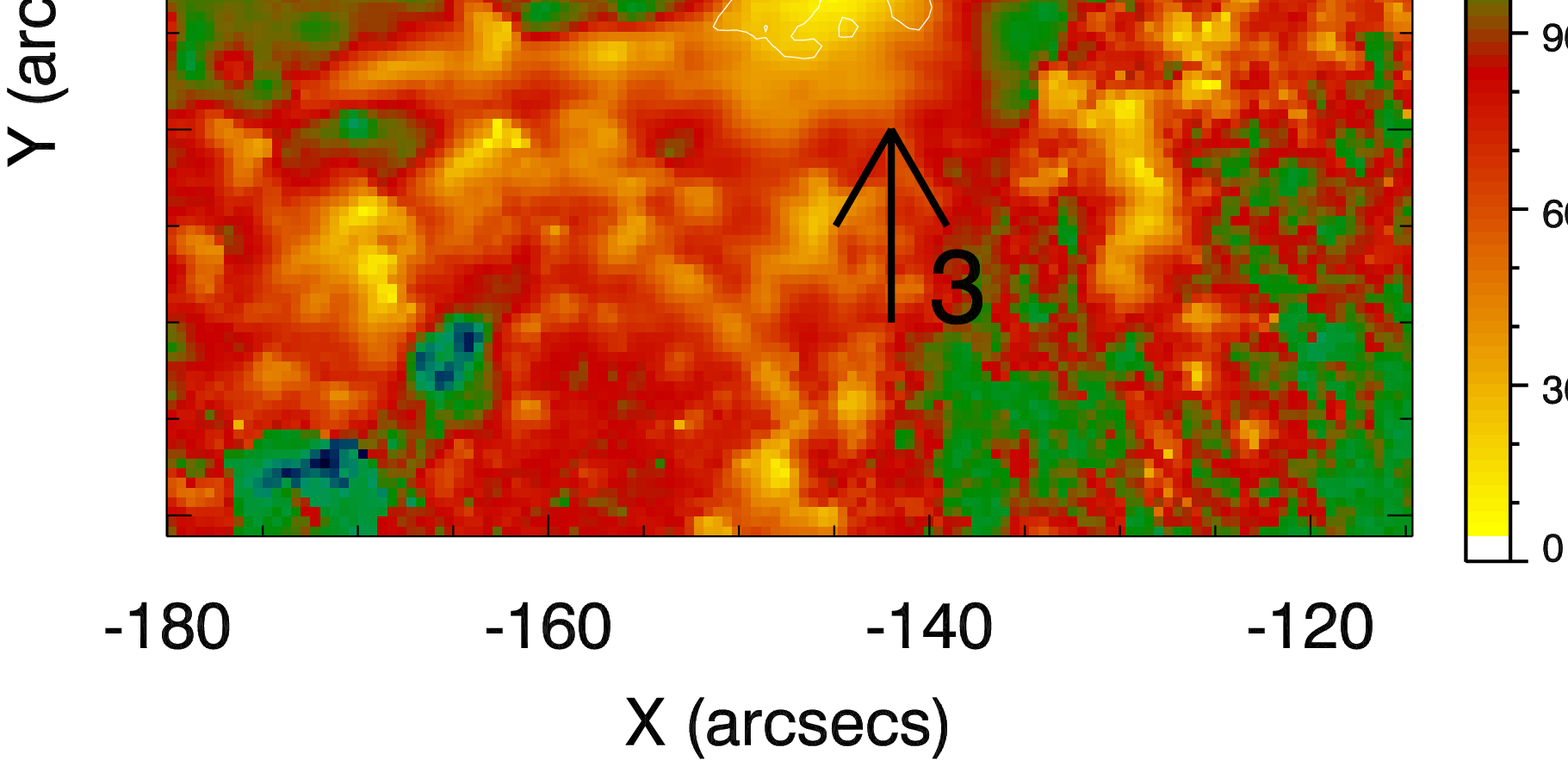}\\
	\includegraphics[scale=0.245, clip, trim=0  200 76 230]{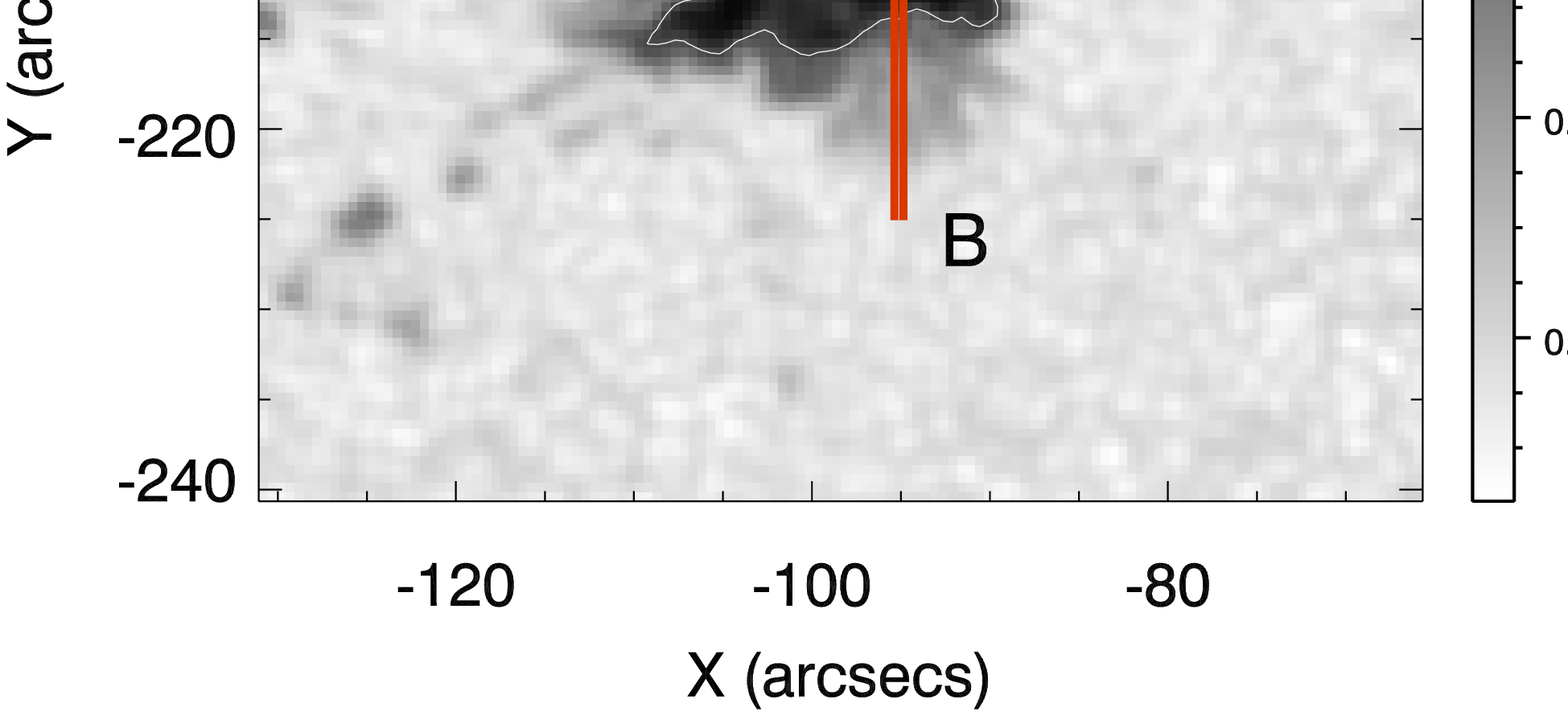}
	\includegraphics[scale=0.245, clip, trim=70 200 76 230]{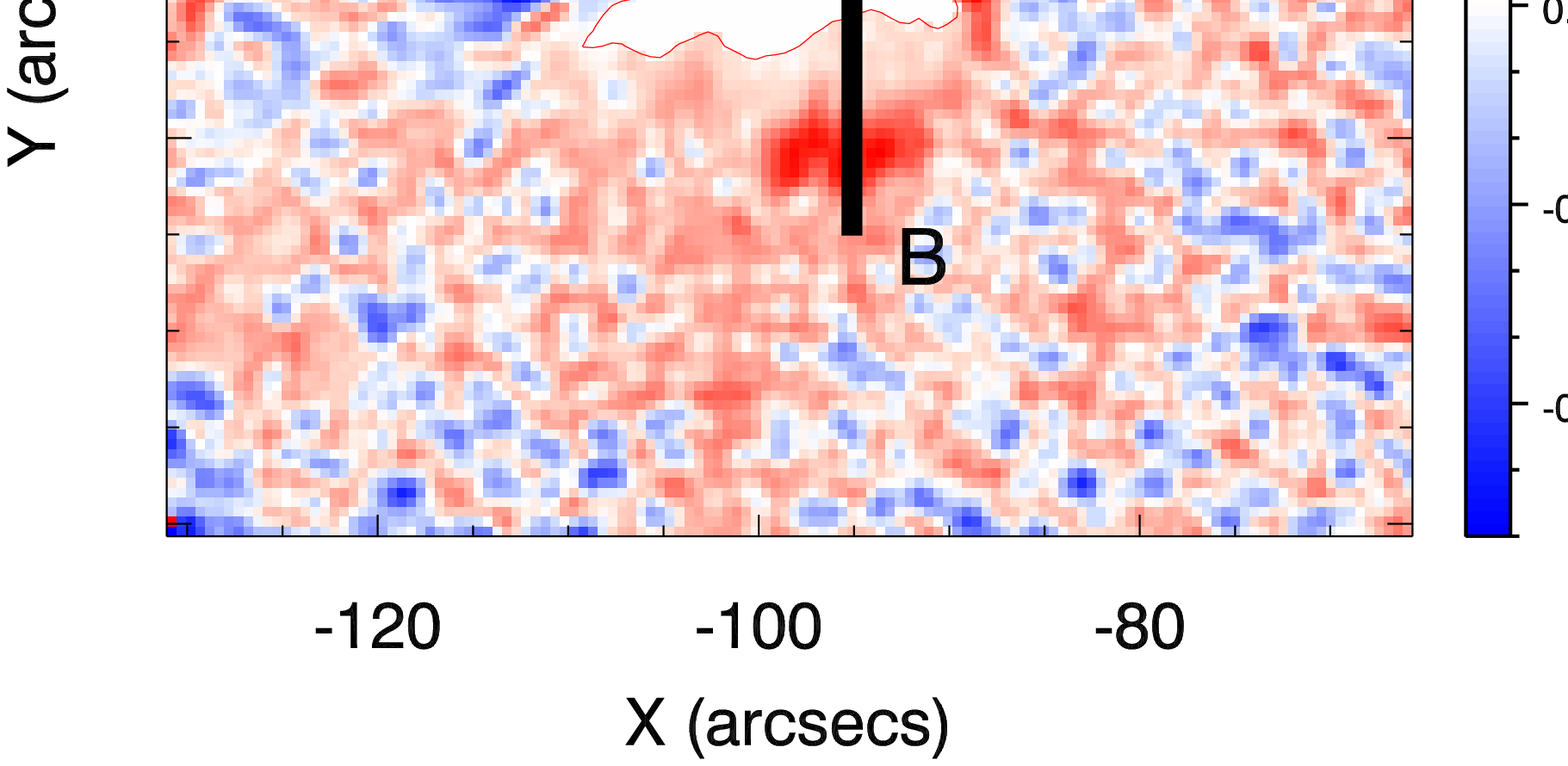}
	\includegraphics[scale=0.245, clip, trim=70 200 76 230]{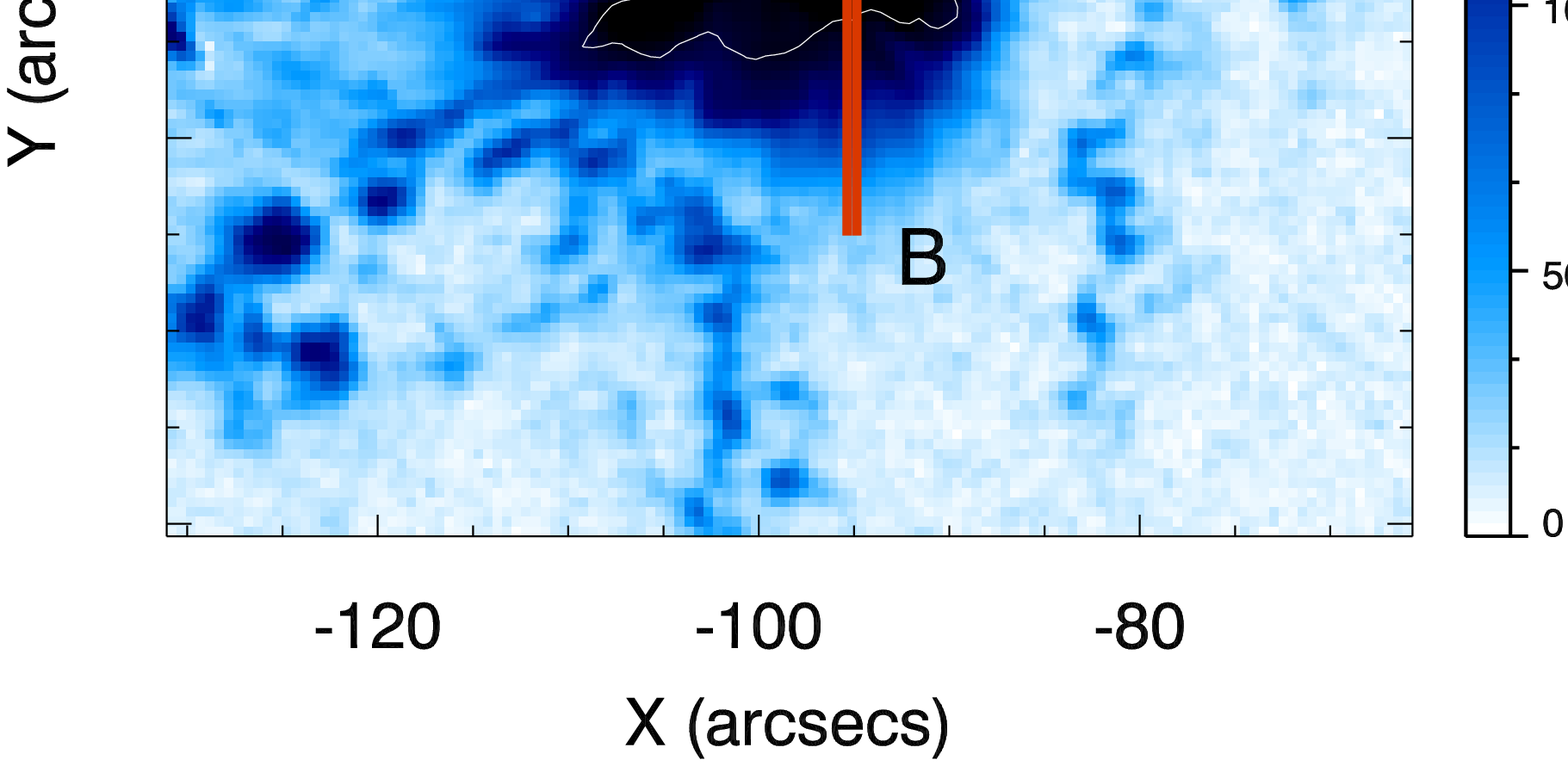}
	\includegraphics[scale=0.245, clip, trim=70 200 76 230]{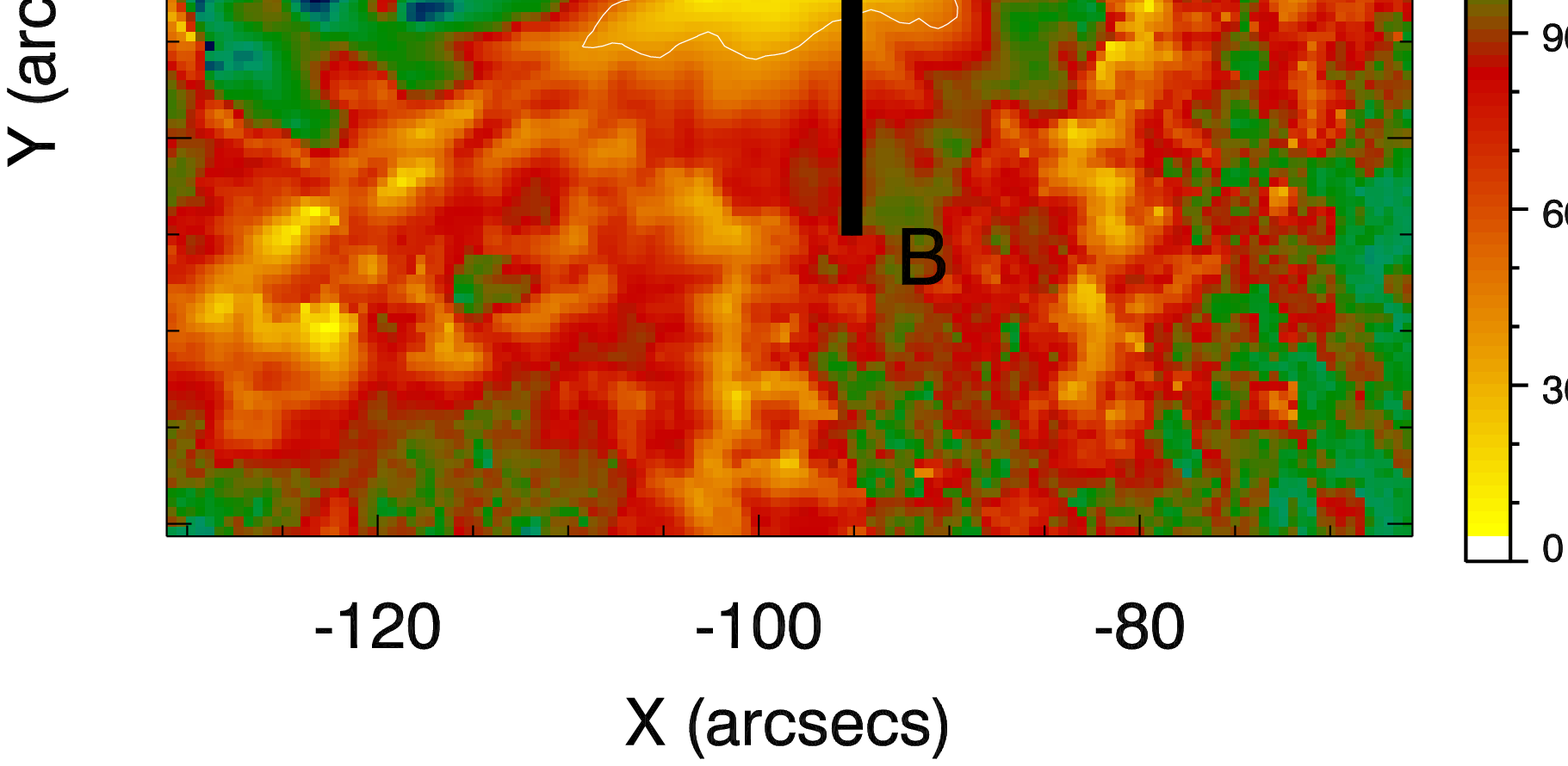}\\
	\includegraphics[scale=0.245, clip, trim=0  200 76 230]{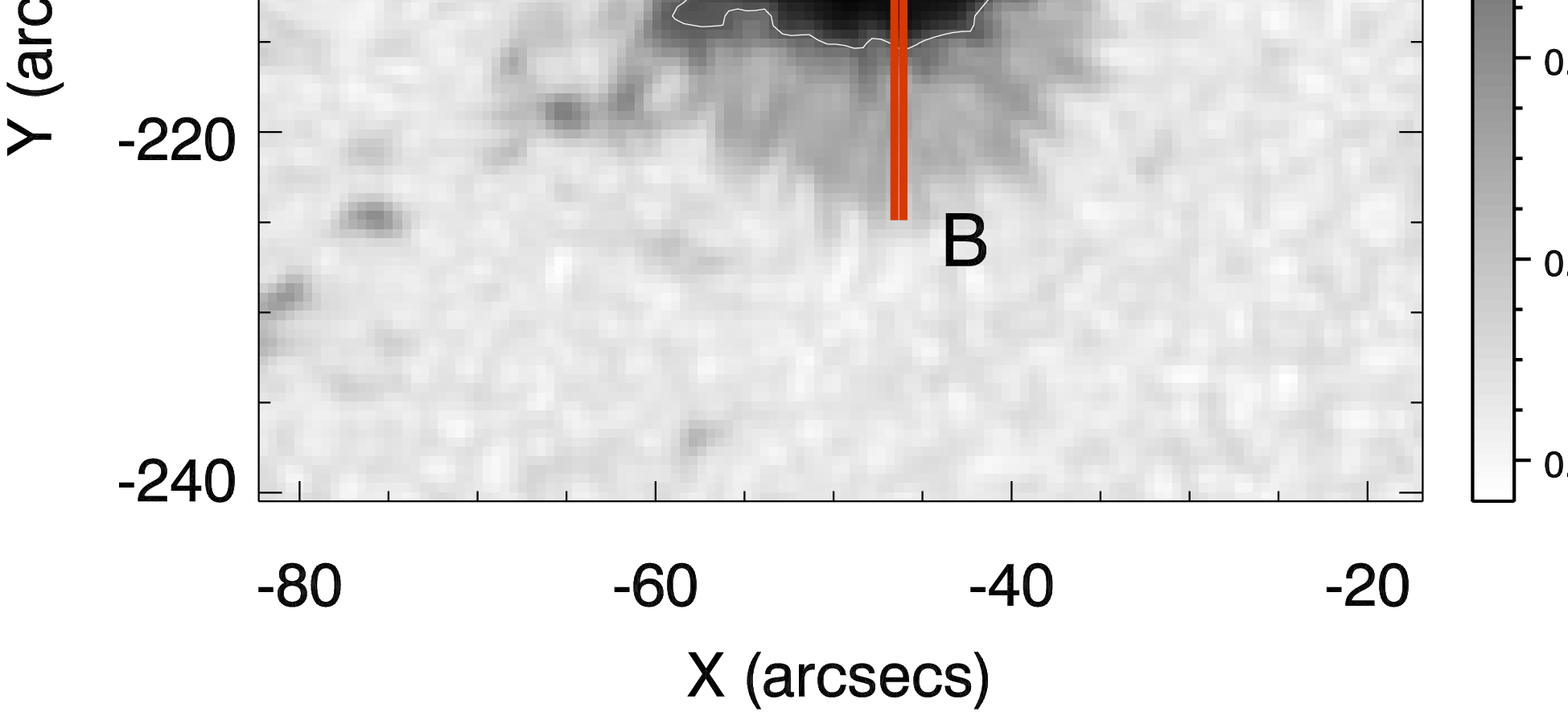}
	\includegraphics[scale=0.245, clip, trim=70 200 76 230]{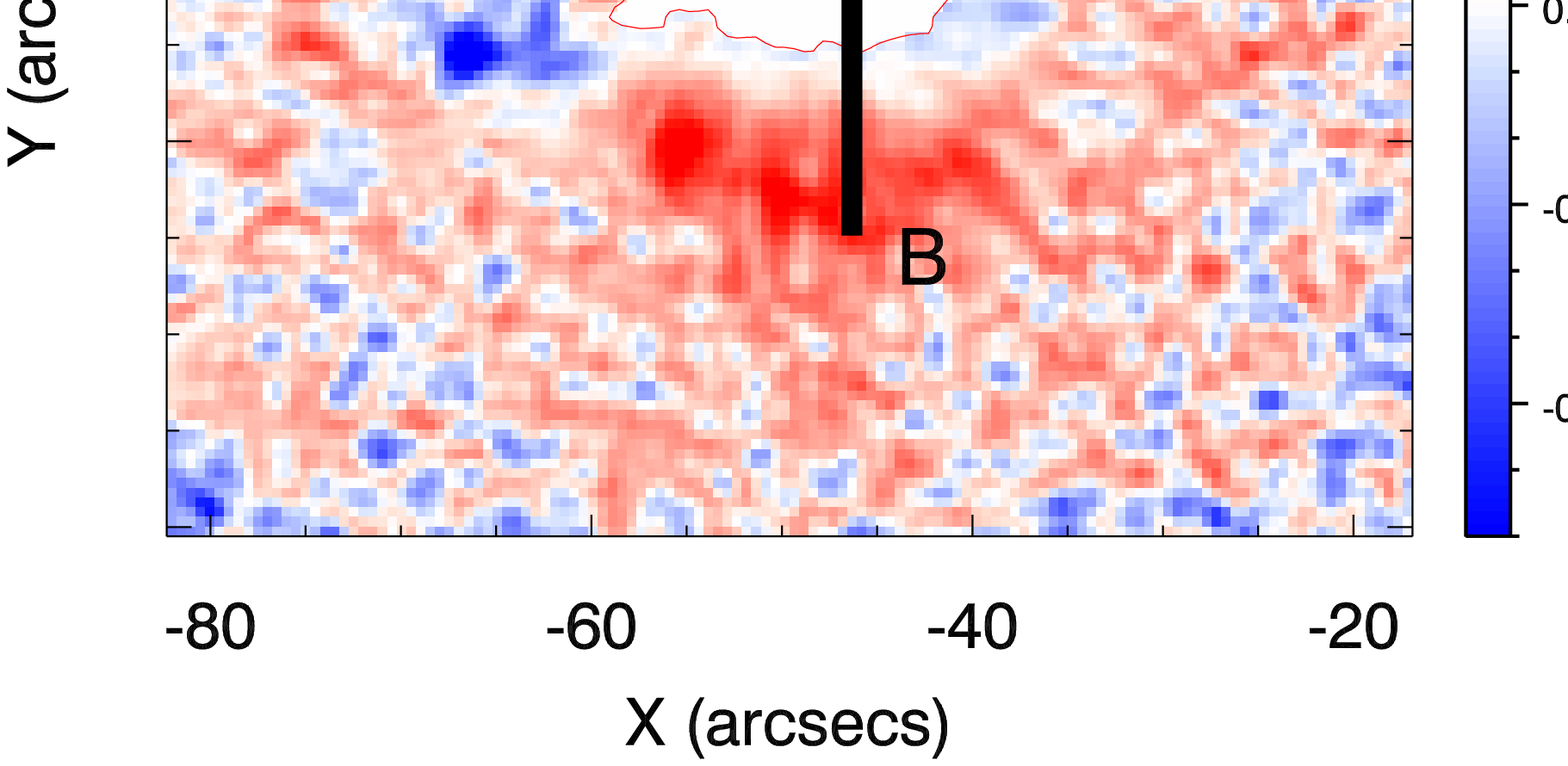}
	\includegraphics[scale=0.245, clip, trim=70 200 76 230]{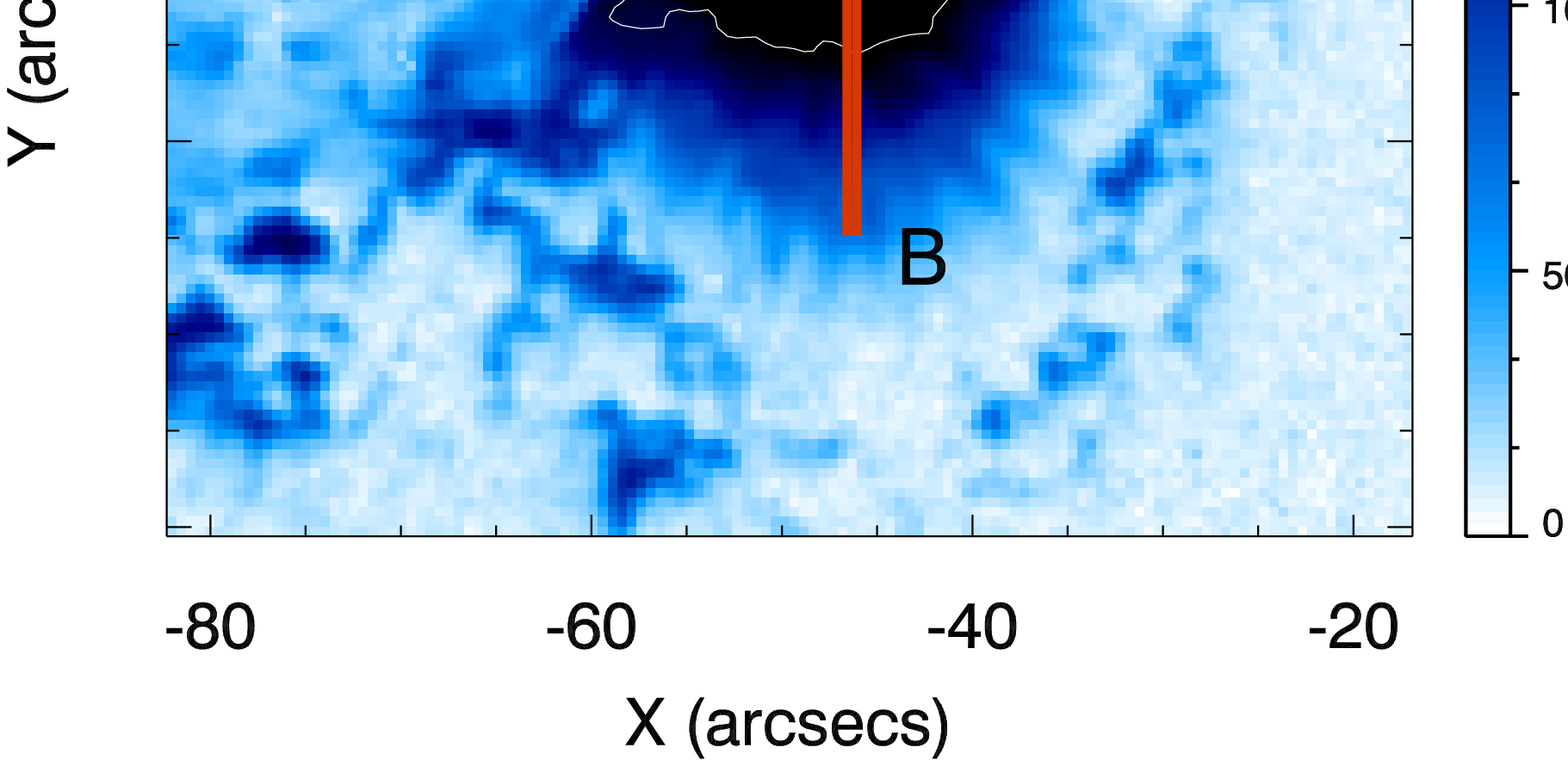}
	\includegraphics[scale=0.245, clip, trim=70 200 76 230]{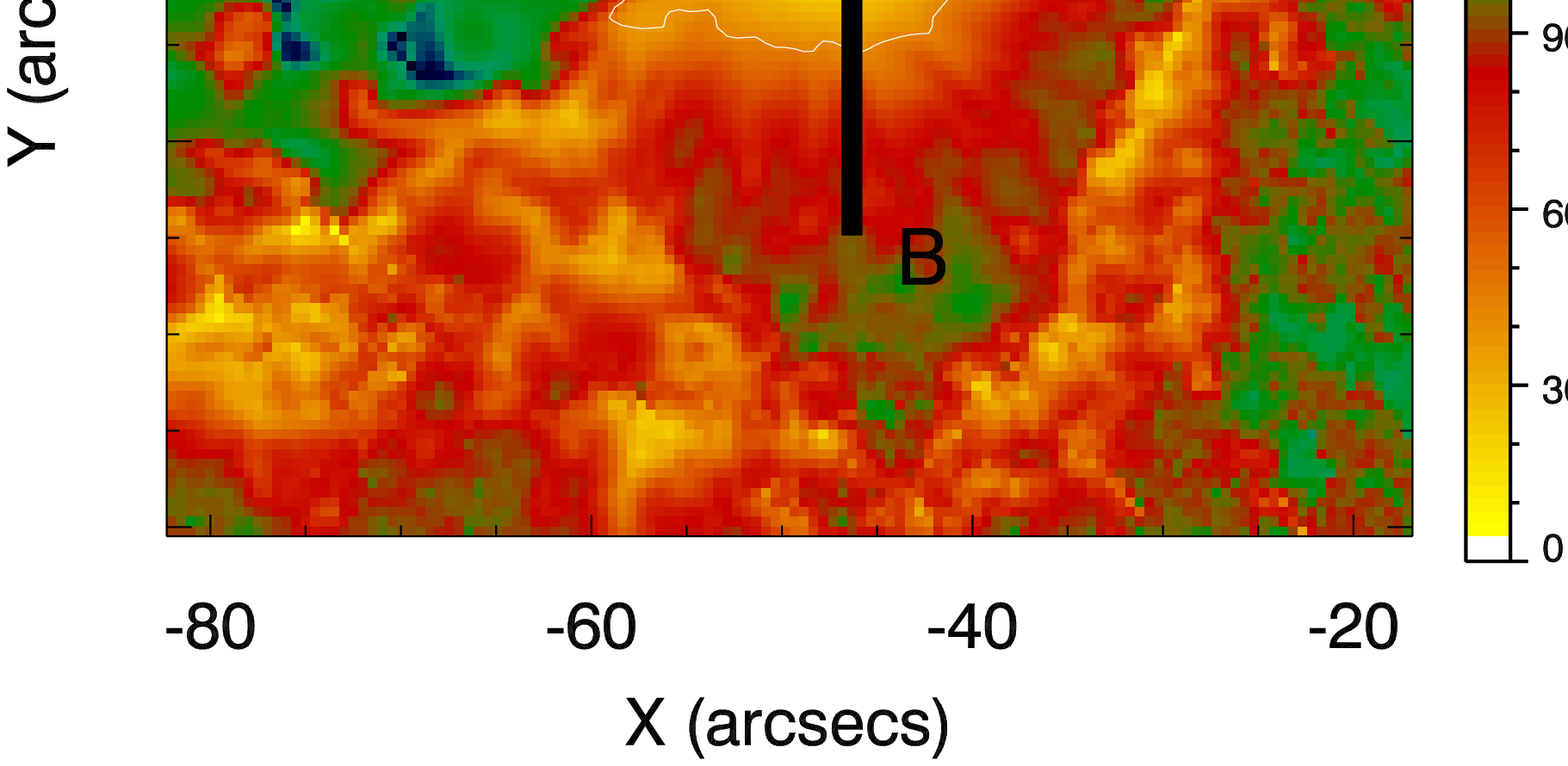}\\
	\includegraphics[scale=0.245, clip, trim=0  200 76 230]{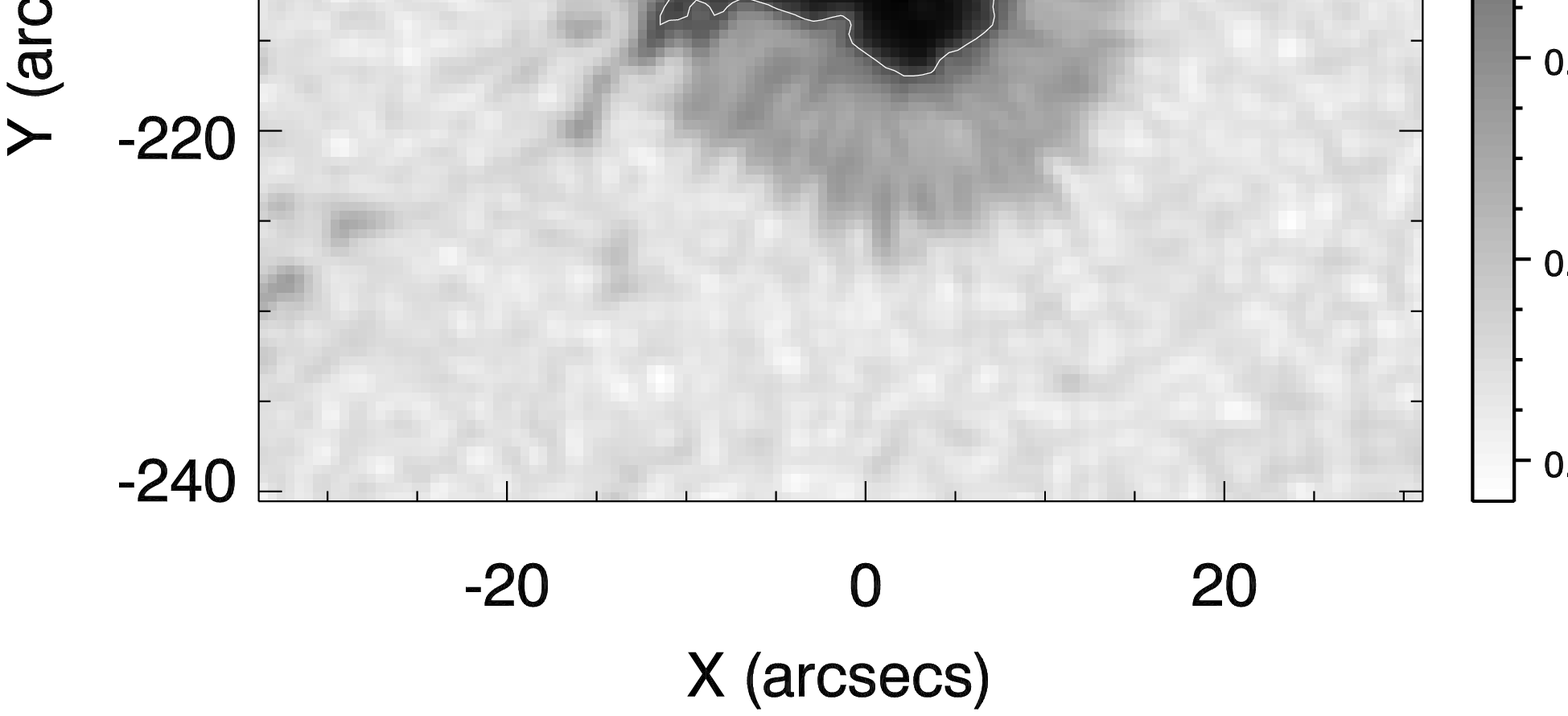}
	\includegraphics[scale=0.245, clip, trim=70 200 76 230]{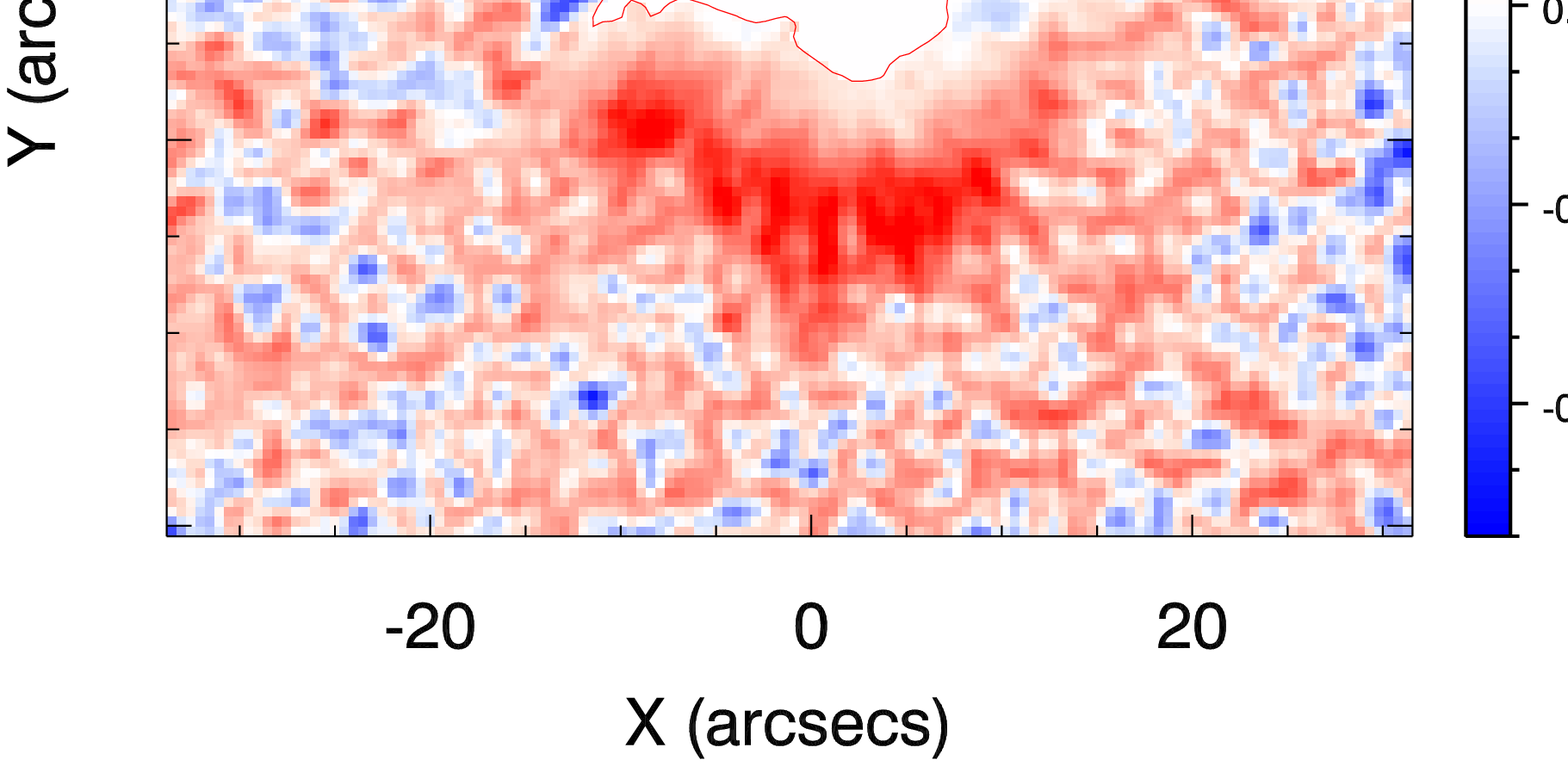}
	\includegraphics[scale=0.245, clip, trim=70 200 76 230]{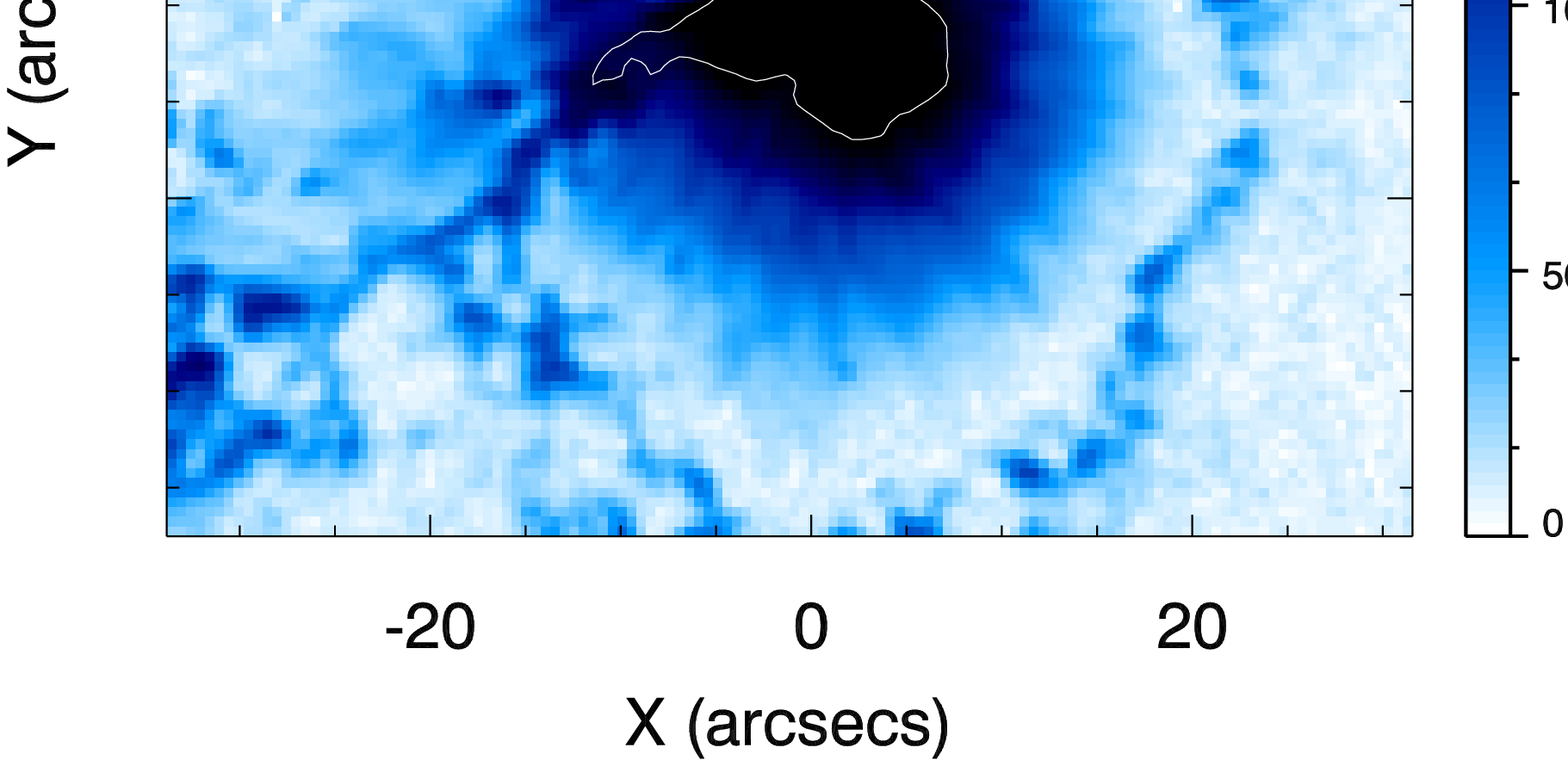}
	\includegraphics[scale=0.245, clip, trim=70 200 76 230]{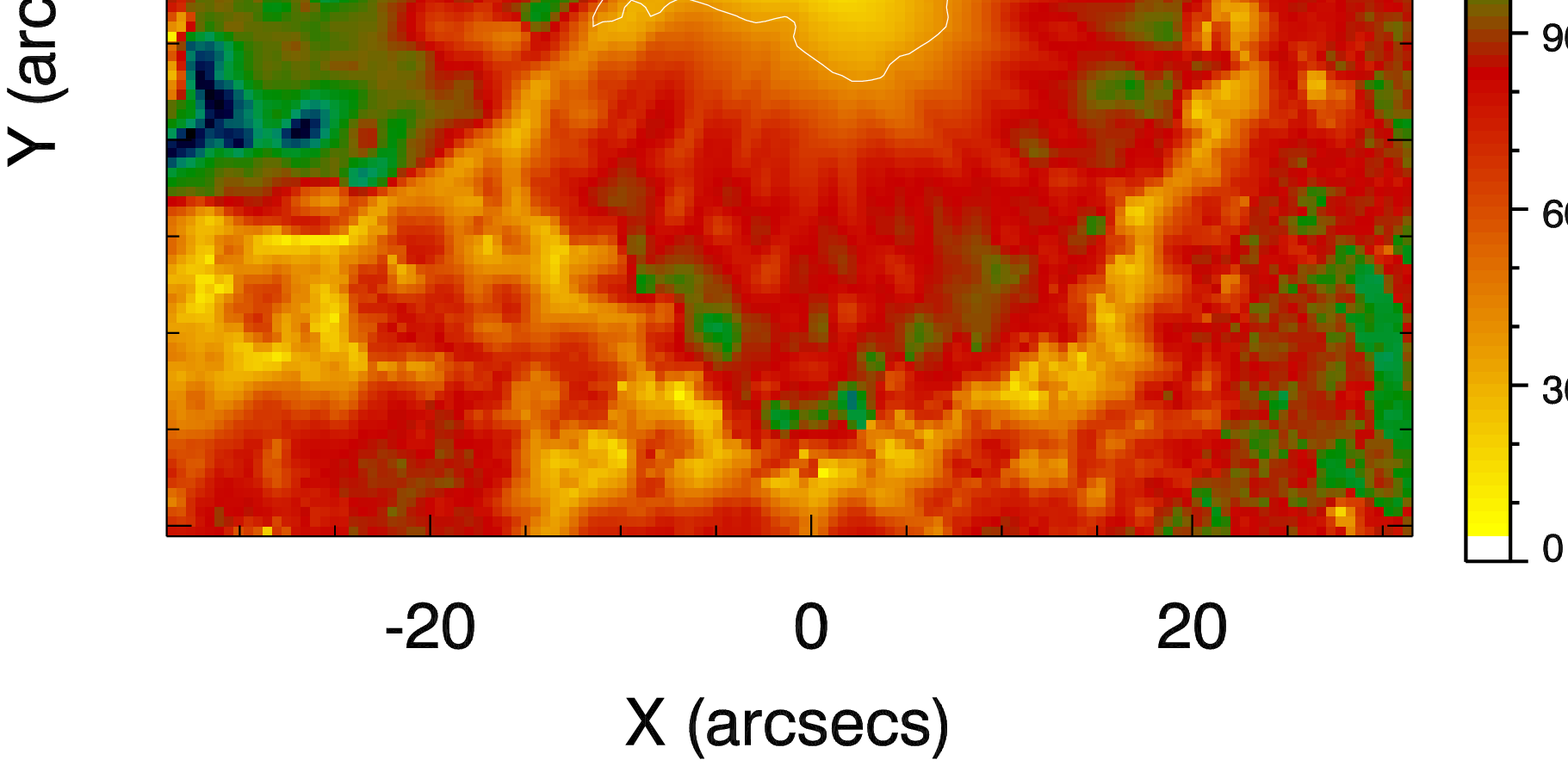}\\
	\includegraphics[scale=0.245, clip, trim=0 200 76 230]{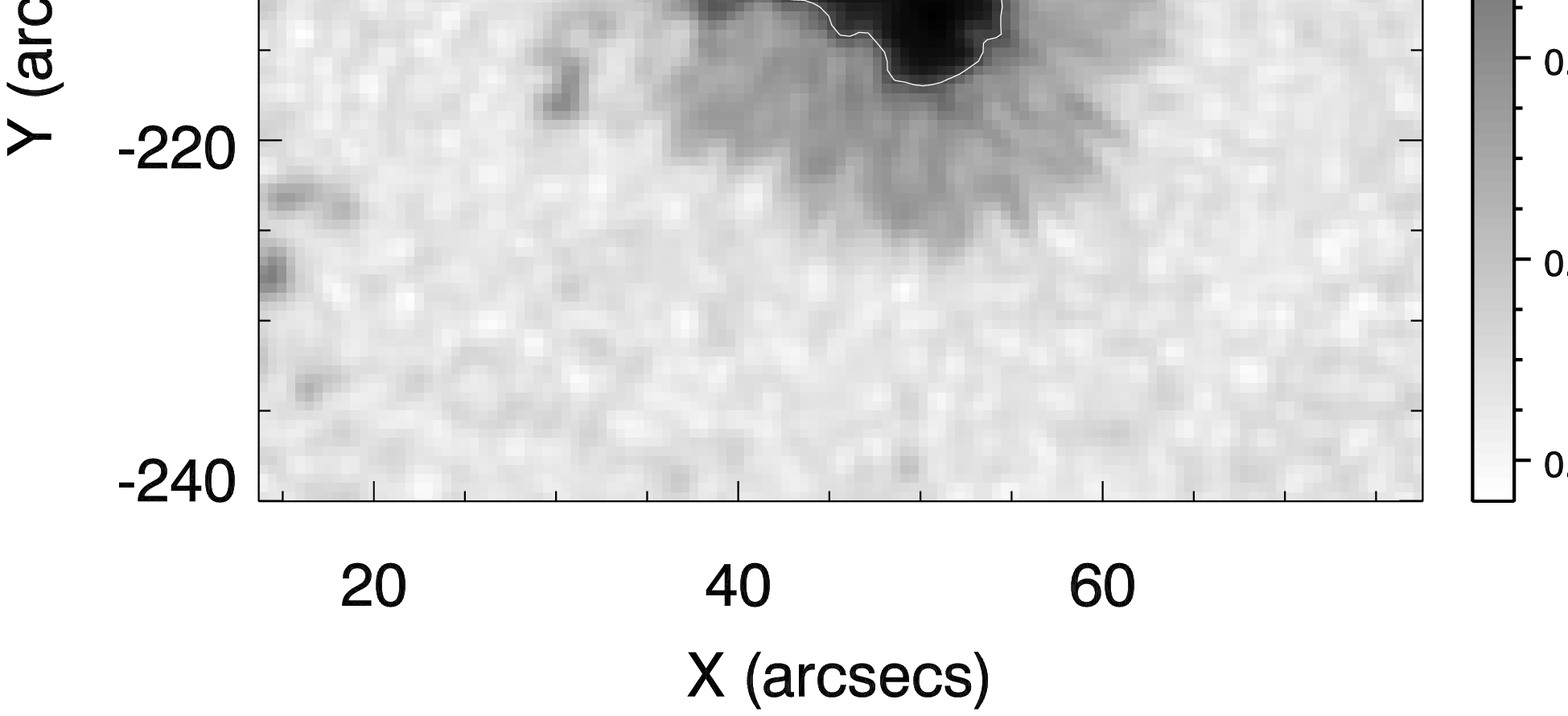}
	\includegraphics[scale=0.245, clip, trim=70 200 76 230]{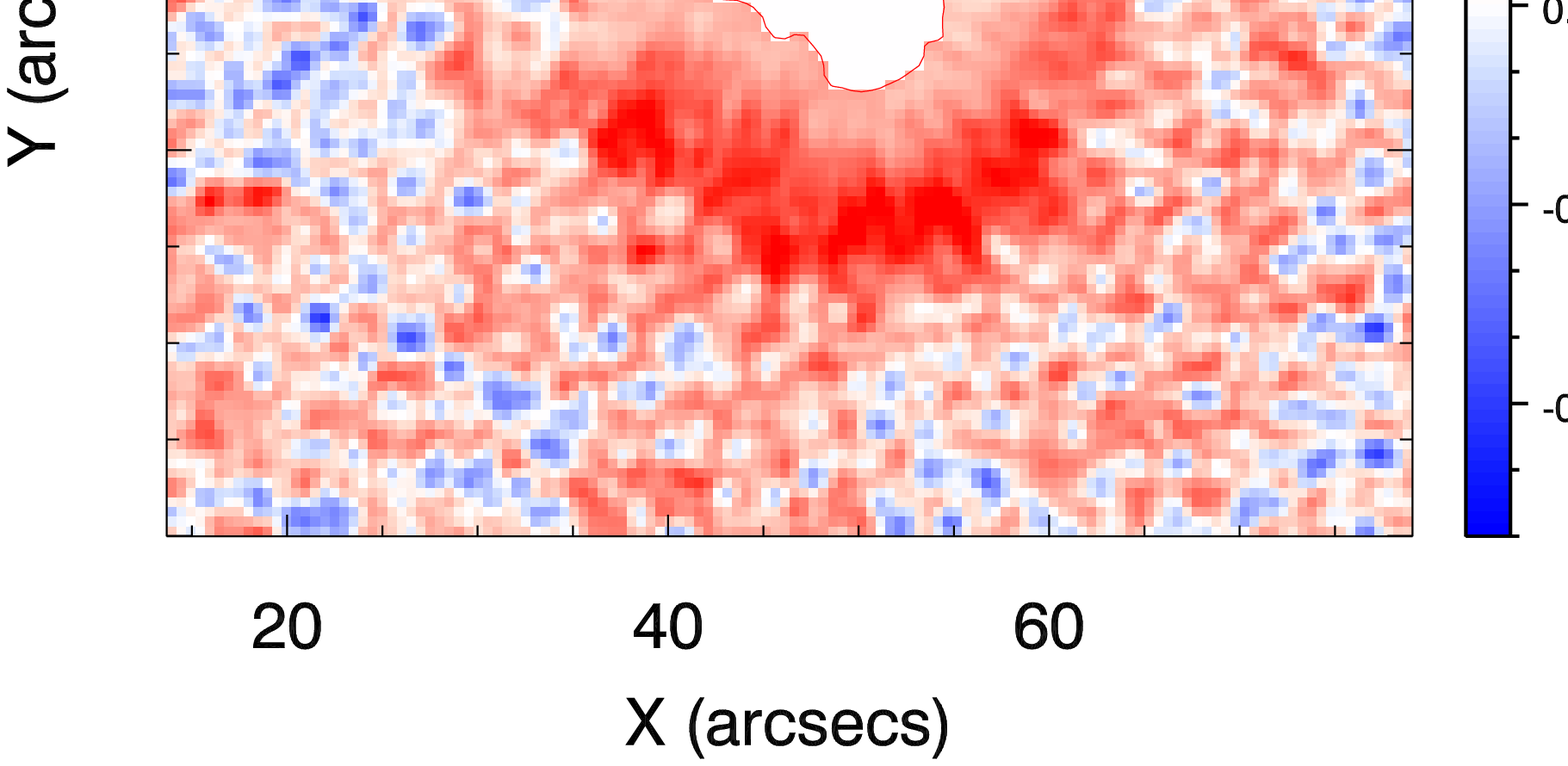}
	\includegraphics[scale=0.245, clip, trim=70 200 76 230]{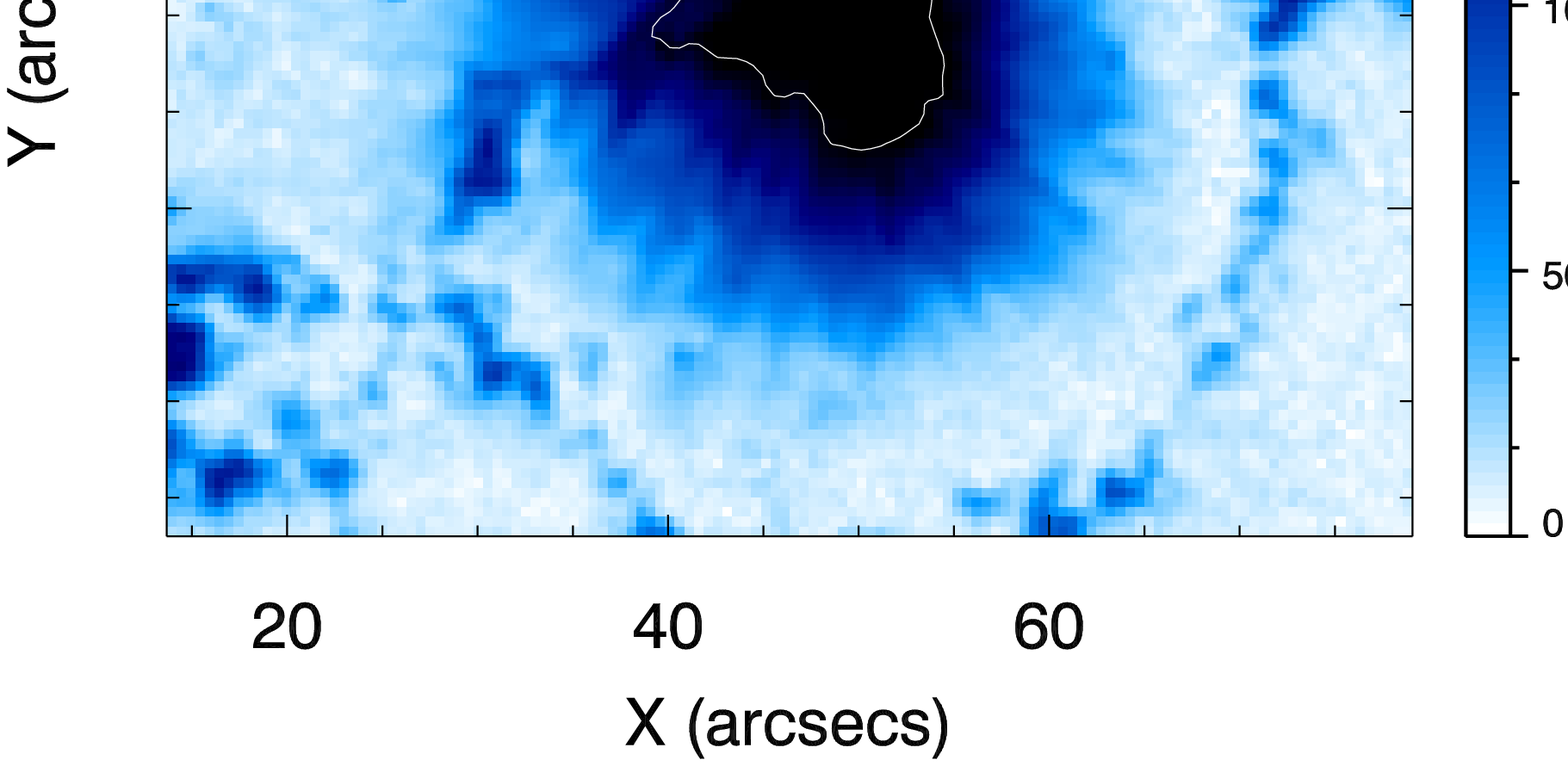}
	\includegraphics[scale=0.245, clip, trim=70 200 76 230]{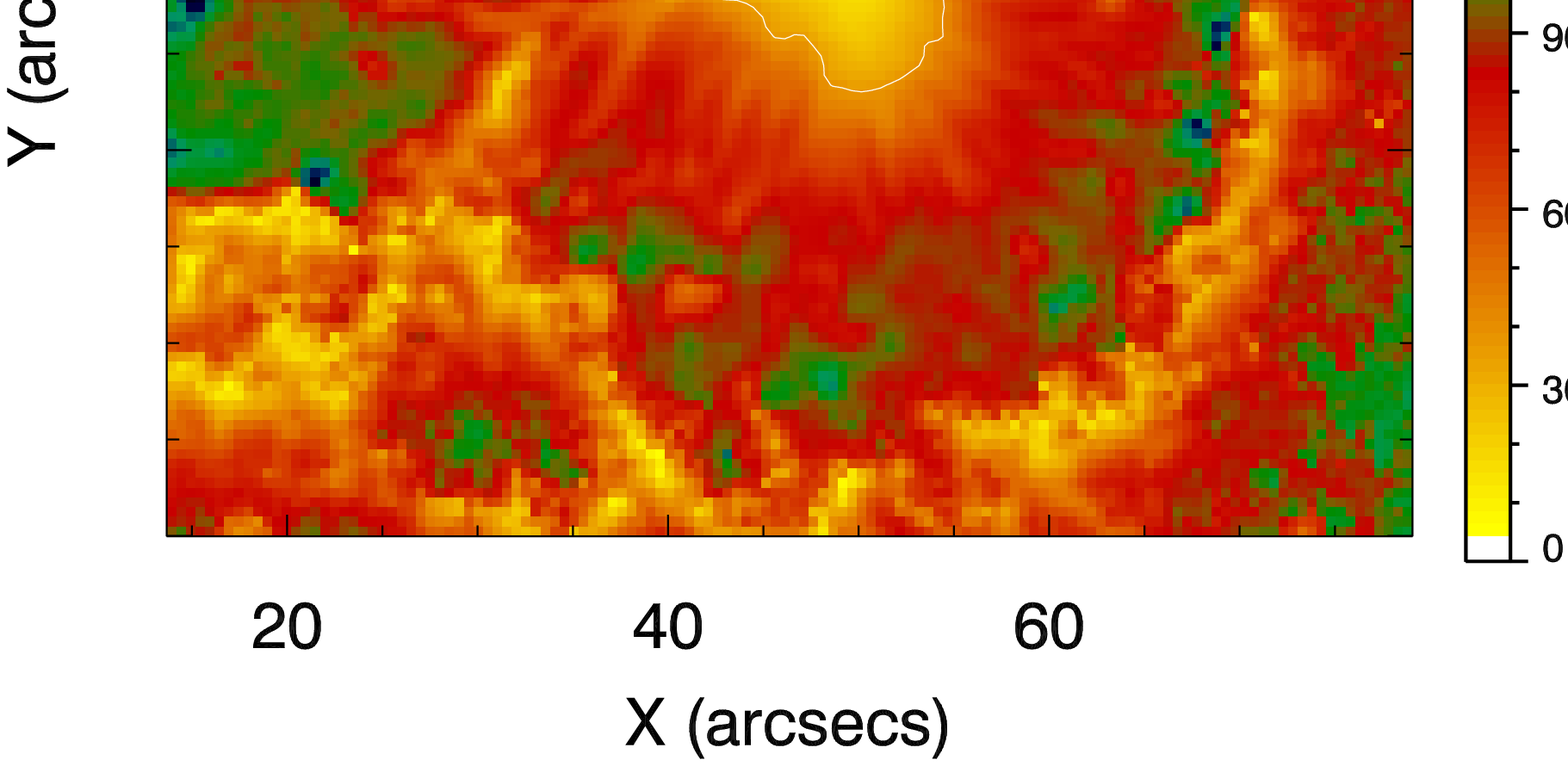}\\
	\includegraphics[scale=0.245, clip, trim=0 200 76 230]{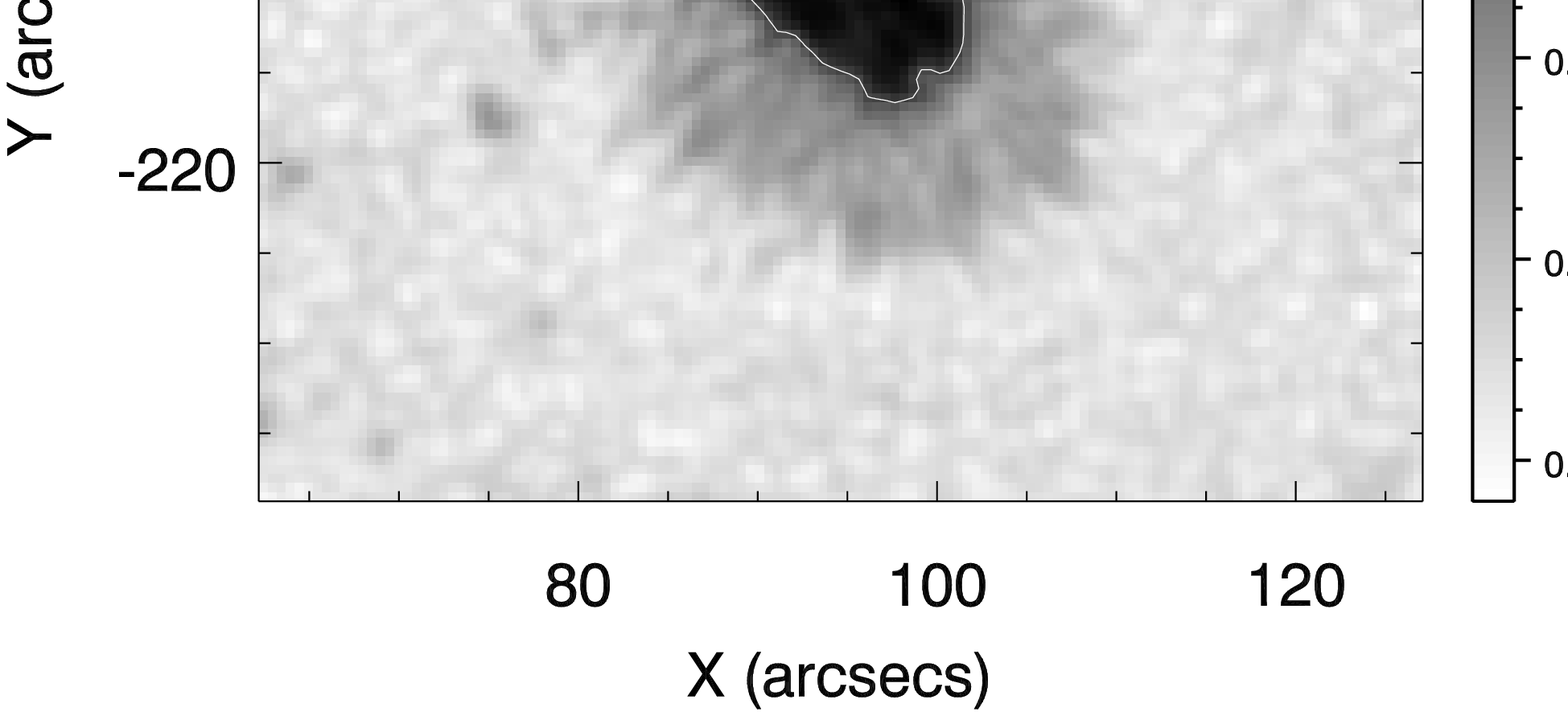}
	\includegraphics[scale=0.245, clip, trim=70 200 76 230]{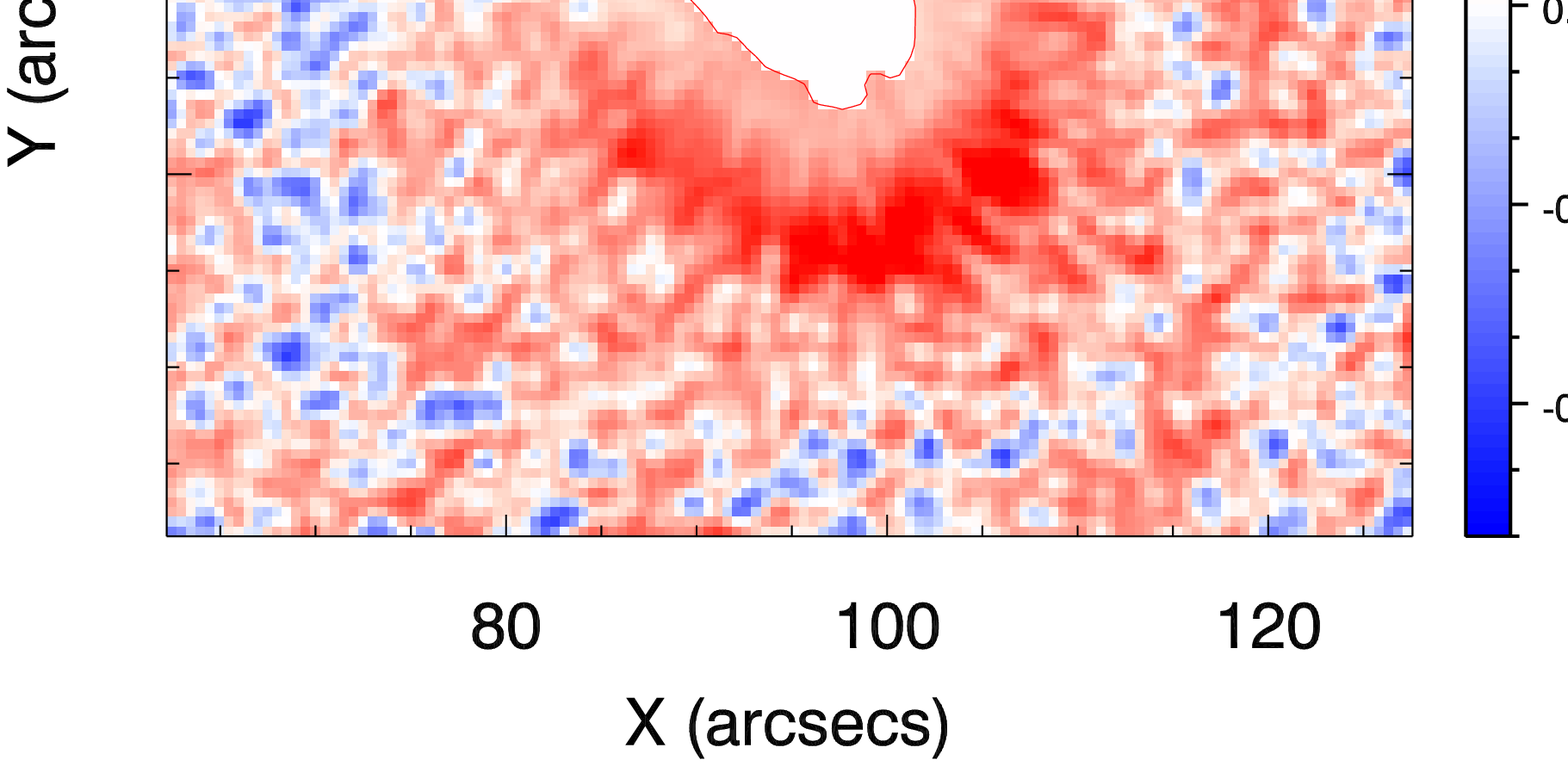}
	\includegraphics[scale=0.245, clip, trim=70 200 76 230]{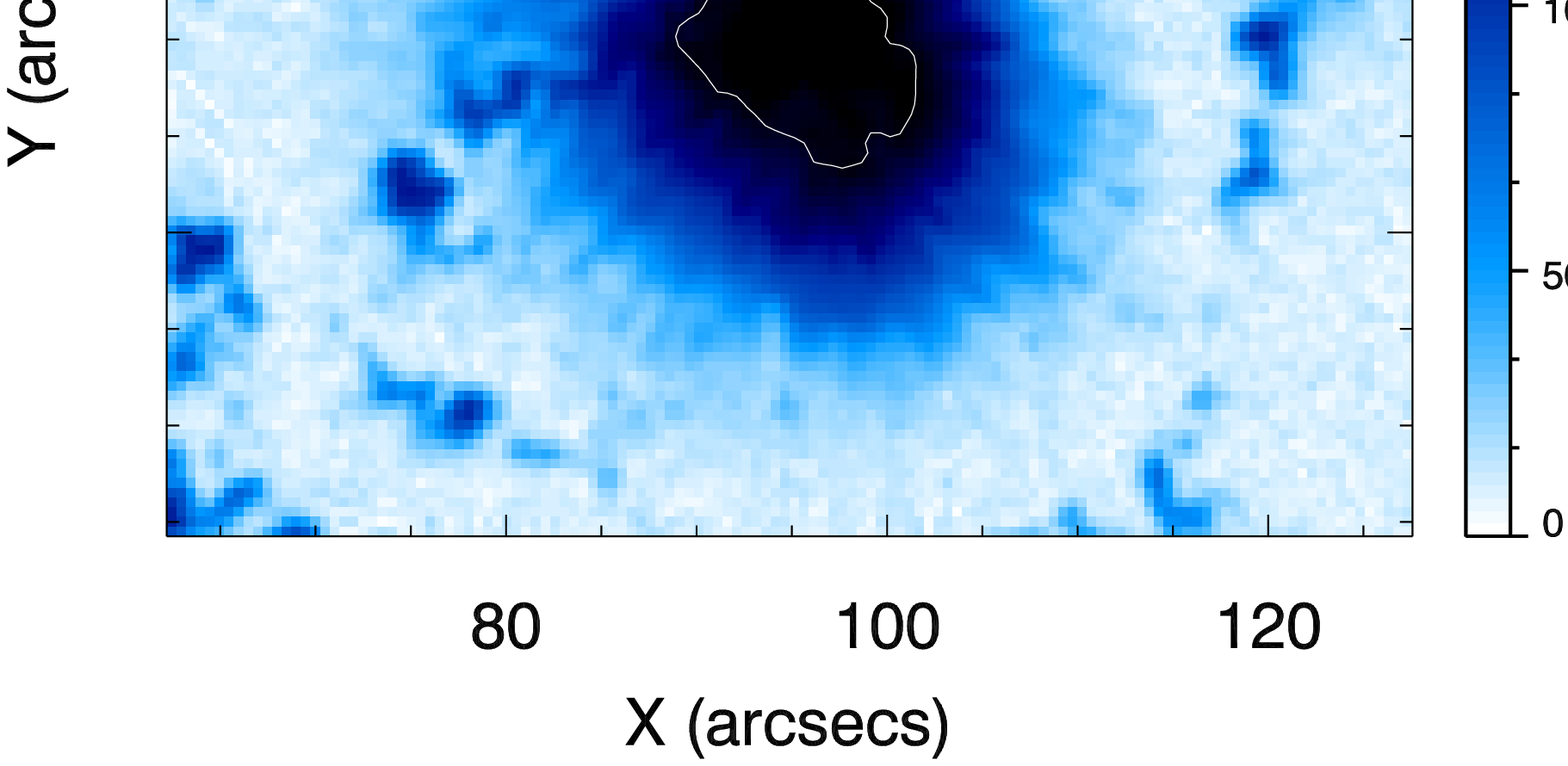}
	\includegraphics[scale=0.245, clip, trim=70 200 76 230]{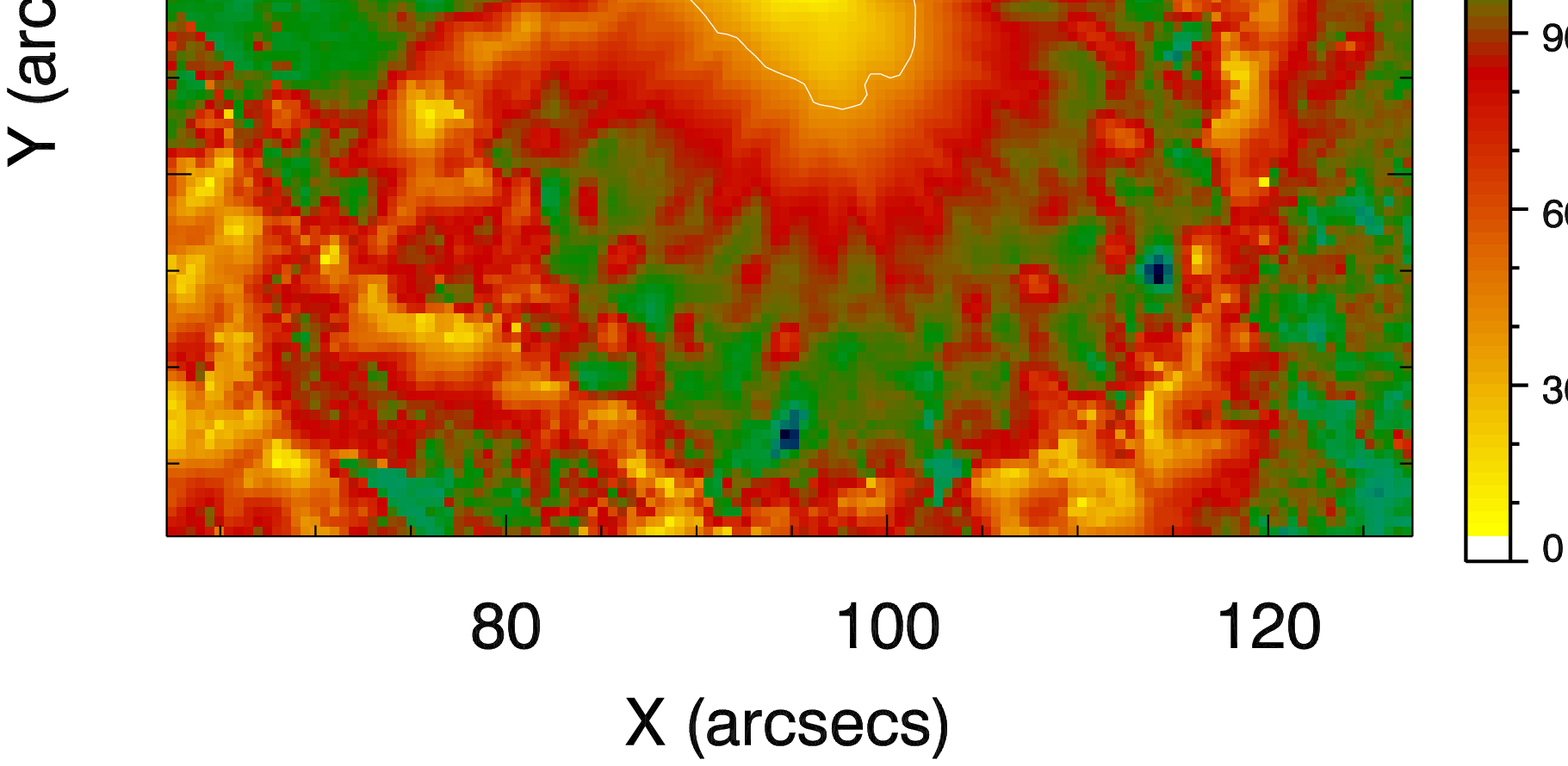}\\
	\includegraphics[scale=0.245, clip, trim=0  155 76 550]{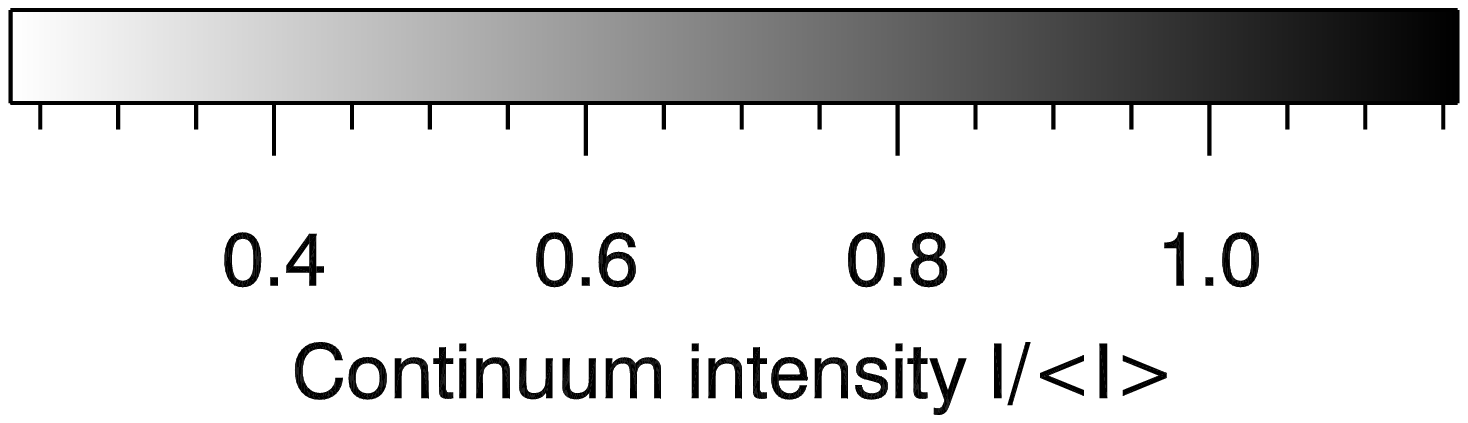}
	\includegraphics[scale=0.245, clip, trim=70 155 76 550]{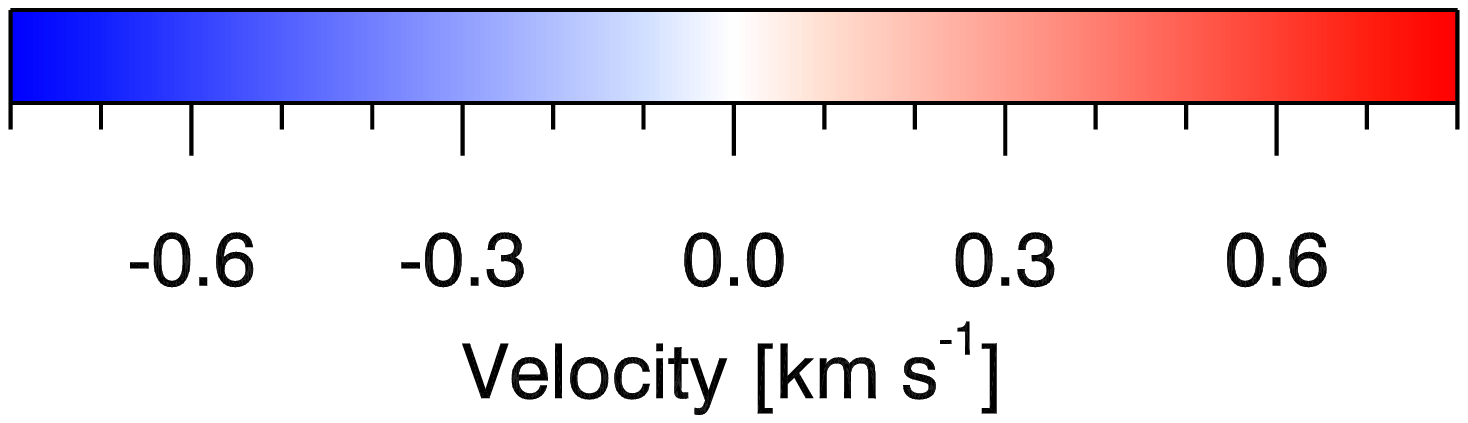}
	\includegraphics[scale=0.245, clip, trim=70 155 66 550]{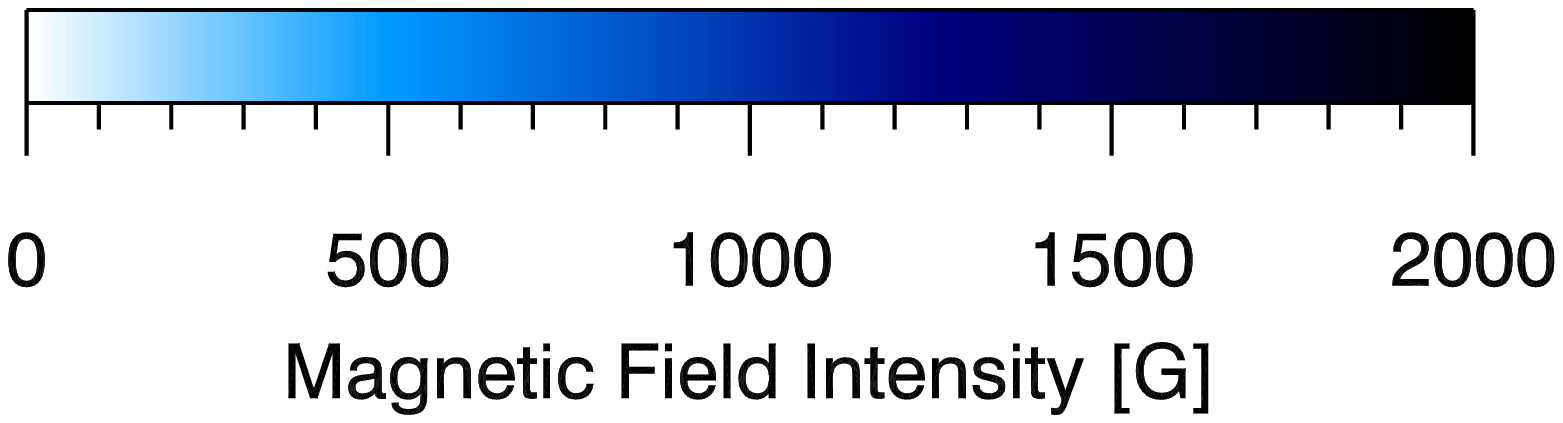}
	\includegraphics[scale=0.245, clip, trim=82 155 72 550]{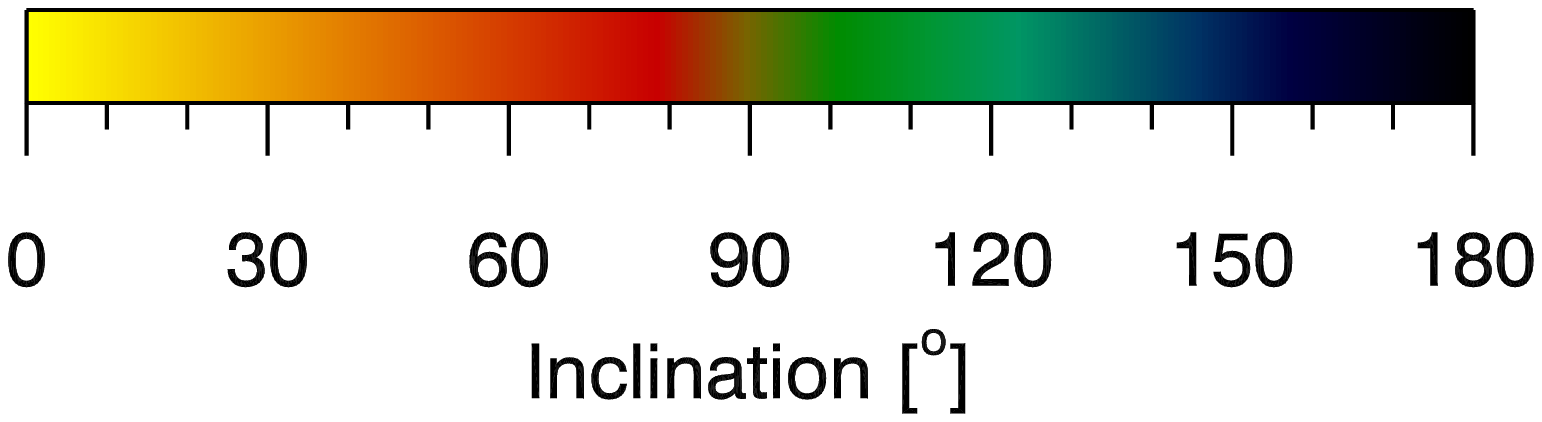}
	\caption{From left to right: Maps of the intensity, LOS velocity, magnetic field strength and inclination angle at different times from 2012 May 28 at 14:58 UT (\textit{top row}) to 2012 May 29 at 14:58 UT \textit{bottom row}) as deduced by SDO/HMI data acquired at 617.3 nm. The black contour in the inclination map indicates the edge of the pore or umbra as seen in the continuum intensity image. Positive and negative velocities correspond to downflows and upflows, respectively. The arrows in the first inclination map indicate the sectors described in the text. The 2-pixel wide segment A in each image of the second and third row is used for the analysis shown in Figure 5, 6 and 7. The arrow points to the disc center. (An animation of this figure is available on-line.) \label{fig3}}
\end{figure*}

Figure 6 presents the evolution of the inclination angle and strength of the magnetic field along the segment A shown in Figure 3. These plots reveal that the magnetic field strength (\textit{bottom panels} of Figure 6) changes significantly ( about 400 G) between 6\arcsec\/ and 13\arcsec\/ from the inner edge of the segment. The inclination angle of the magnetic field in the region between 3\arcsec\/ and 7\farcs5 (\textit{top panels} of Figure 6) reaches values up to 80$^{\circ}$, indicating positive polarity. We also notice that at 7\farcs5 (corresponding to the outer edge of the penumbra), it varies from 80$^{\circ}$ to 70$^{\circ}$, becoming more vertical. However, note that this variation is within the uncertainty of the inclination determined from SDO/HMI. Finally, from 8\arcsec\/ to 13\arcsec\/ beyond the outer penumbral boundary, the magnetic field changes sign, in fact at 19:00 UT it is larger than 90$^{\circ}$ while at 24:00 UT it has values smaller than 90$^{\circ}$. We notice that the area where the inclination changes sign does not correspond to the penumbra but it belongs to the moat region. Also, the field strength along the same cut at the location of the forming penumbra increased mainly only some minutes after the Evershed-like flow had already been established.

\begin{figure*}[t]
	\centering
	\includegraphics[scale=0.40, clip, trim=0  200 0 230]{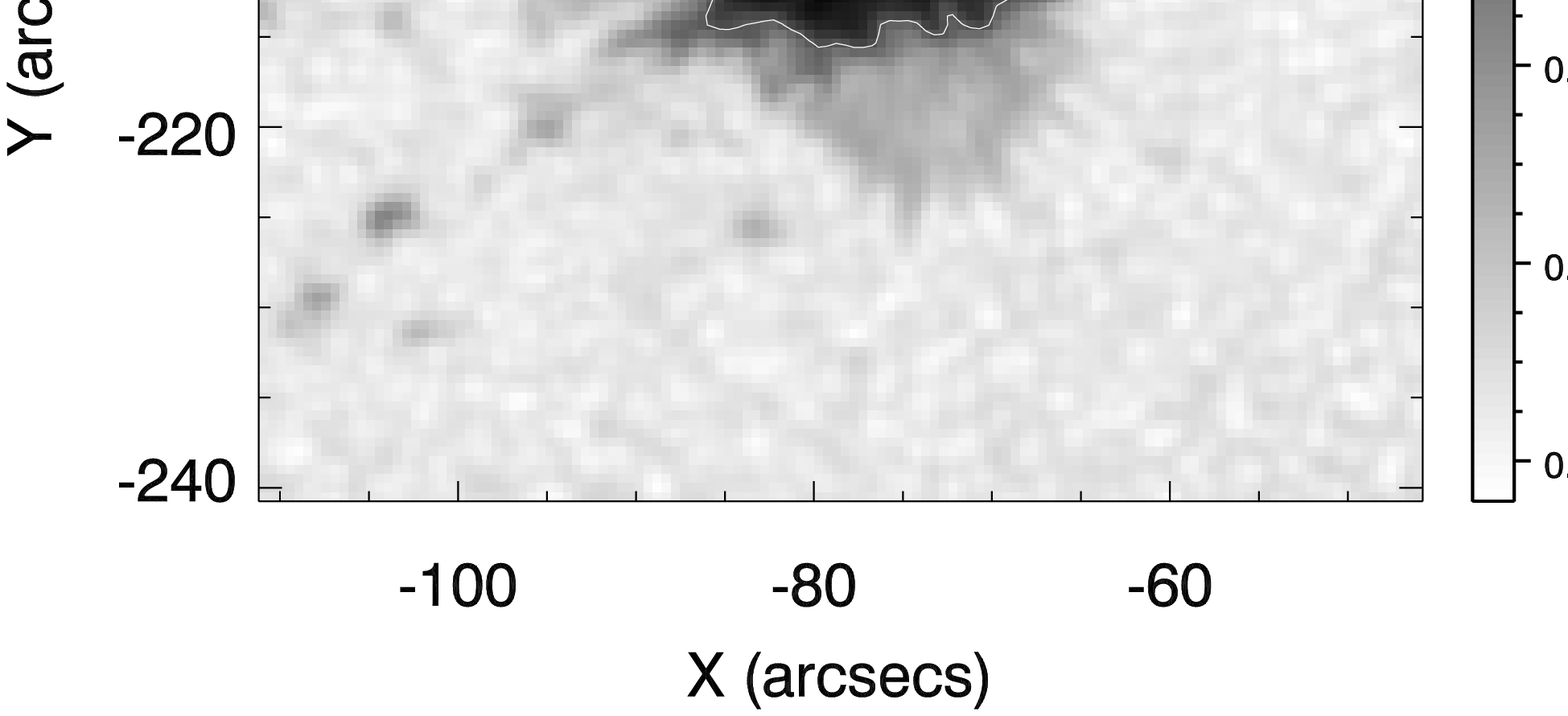}
	\includegraphics[scale=0.40, clip, trim=84 200 0 230]{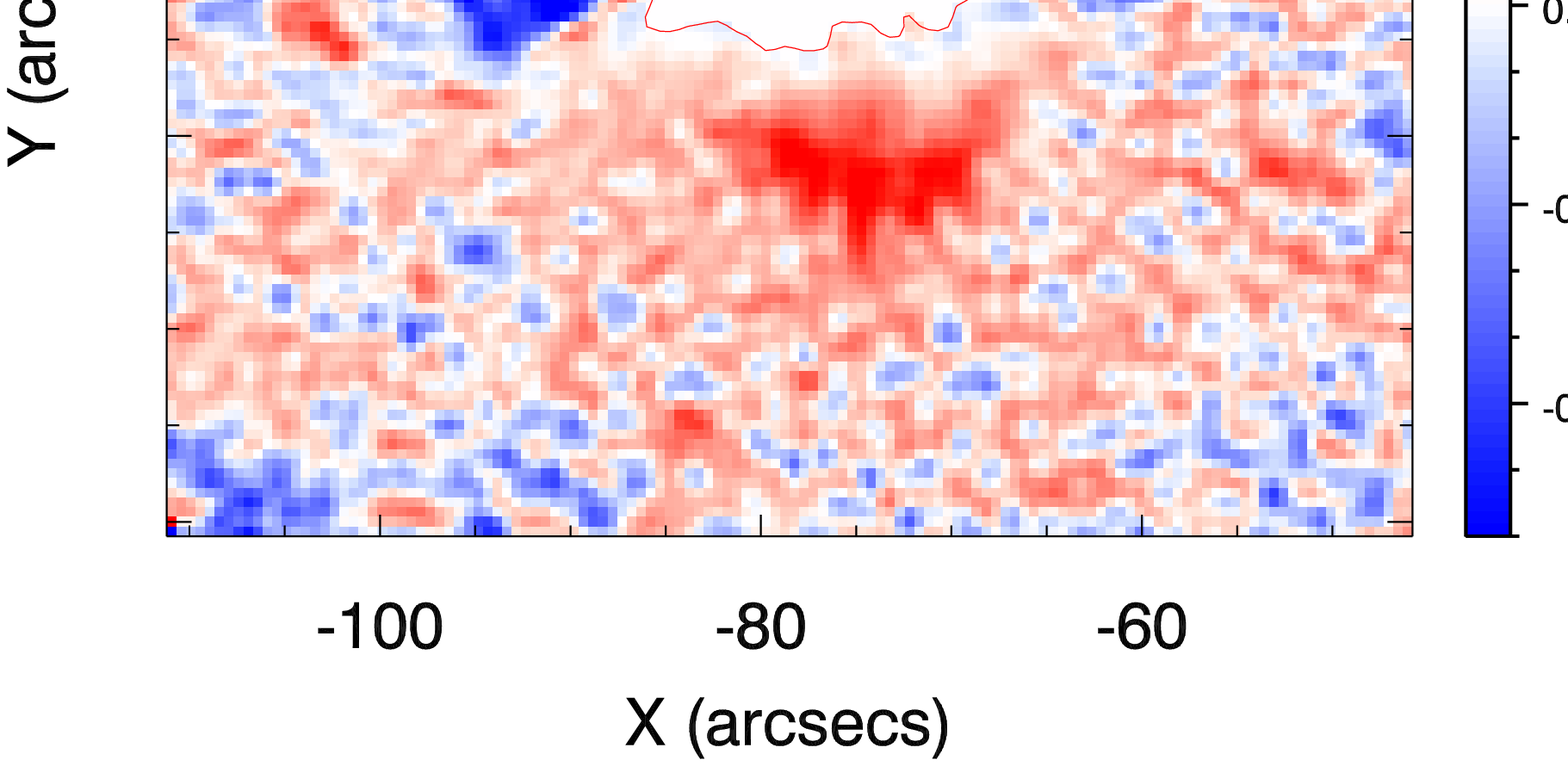}\\
	\includegraphics[scale=0.40, clip, trim=0  200 0 230]{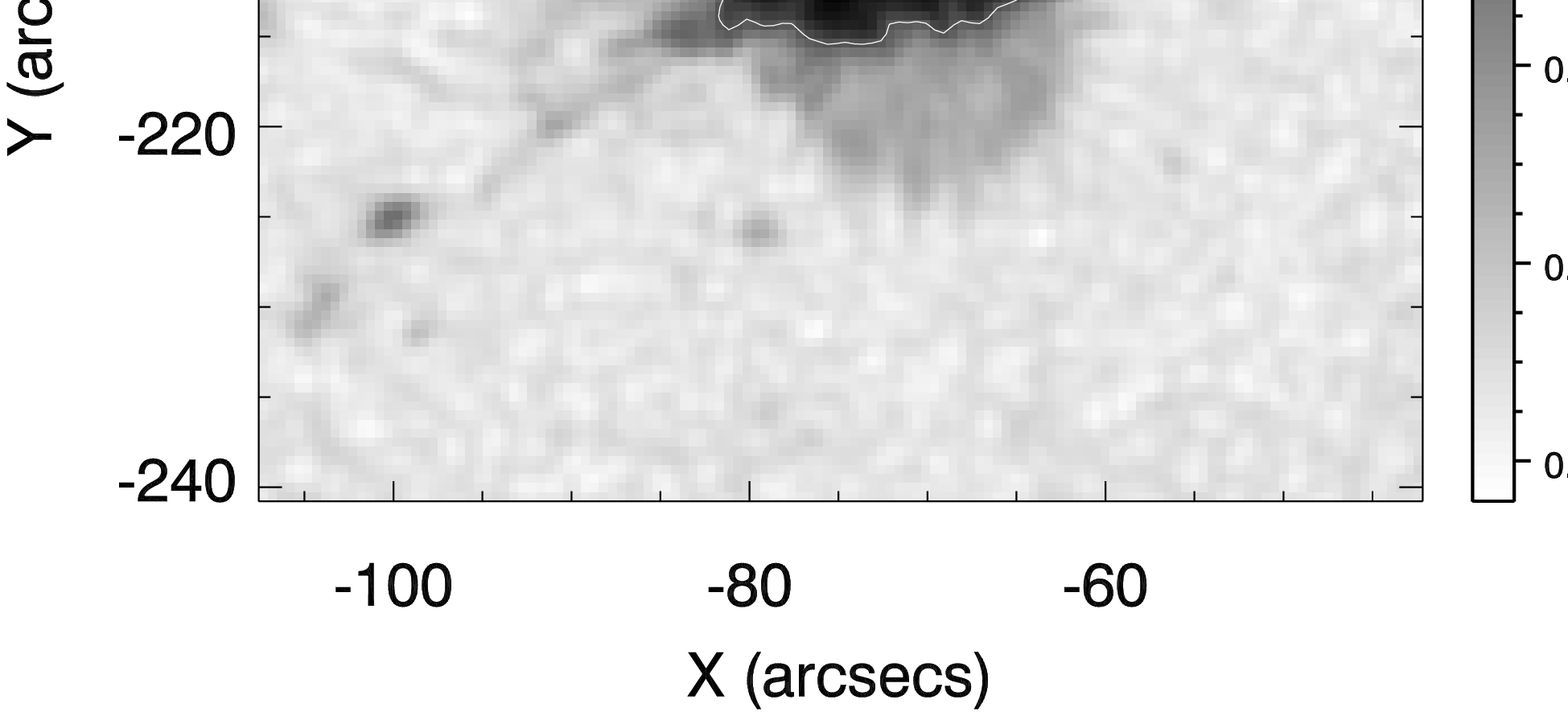}
	\includegraphics[scale=0.40, clip, trim=84 200 0 230]{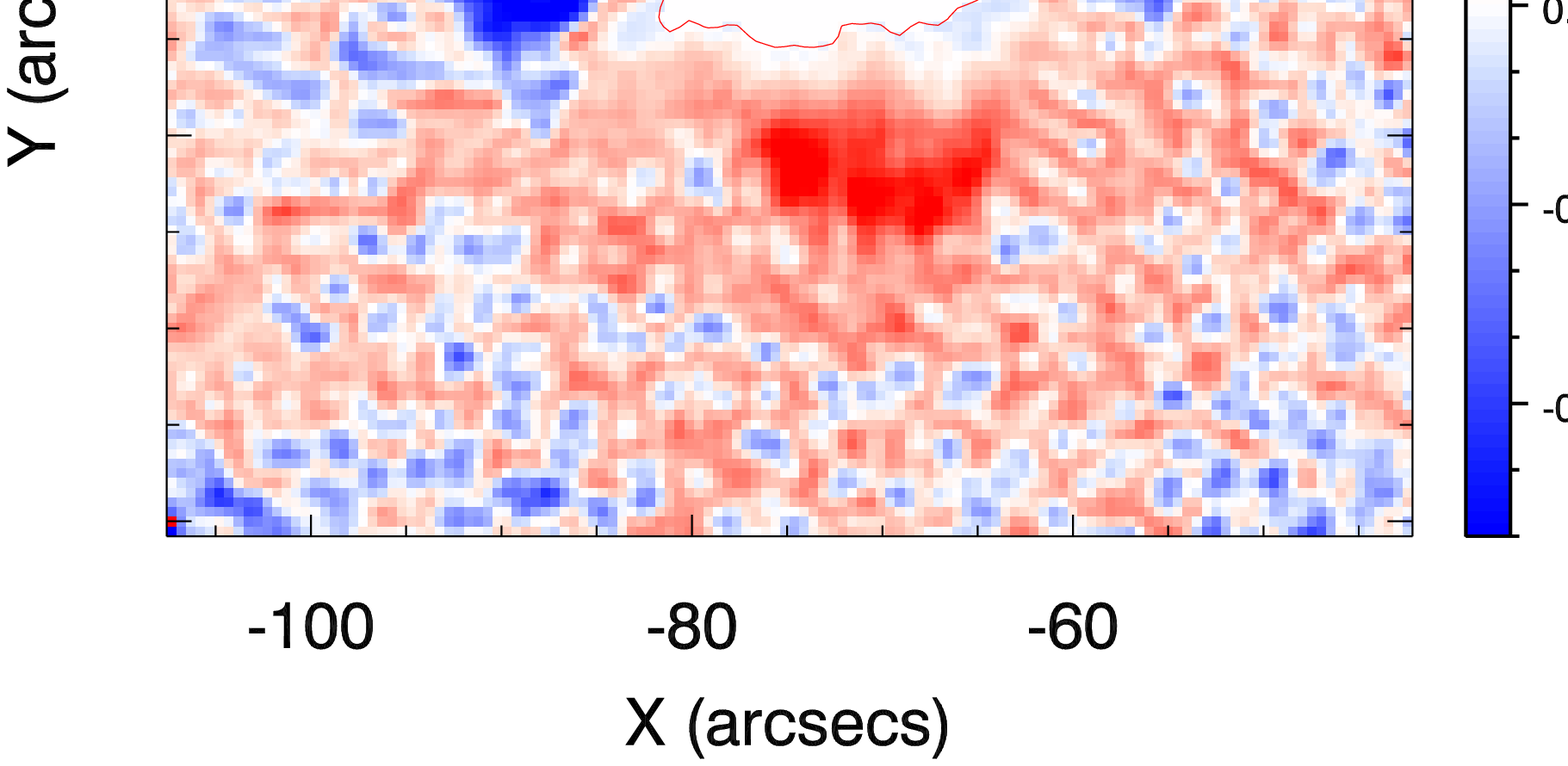}\\
	\includegraphics[scale=0.40, clip, trim=0  180 0 230]{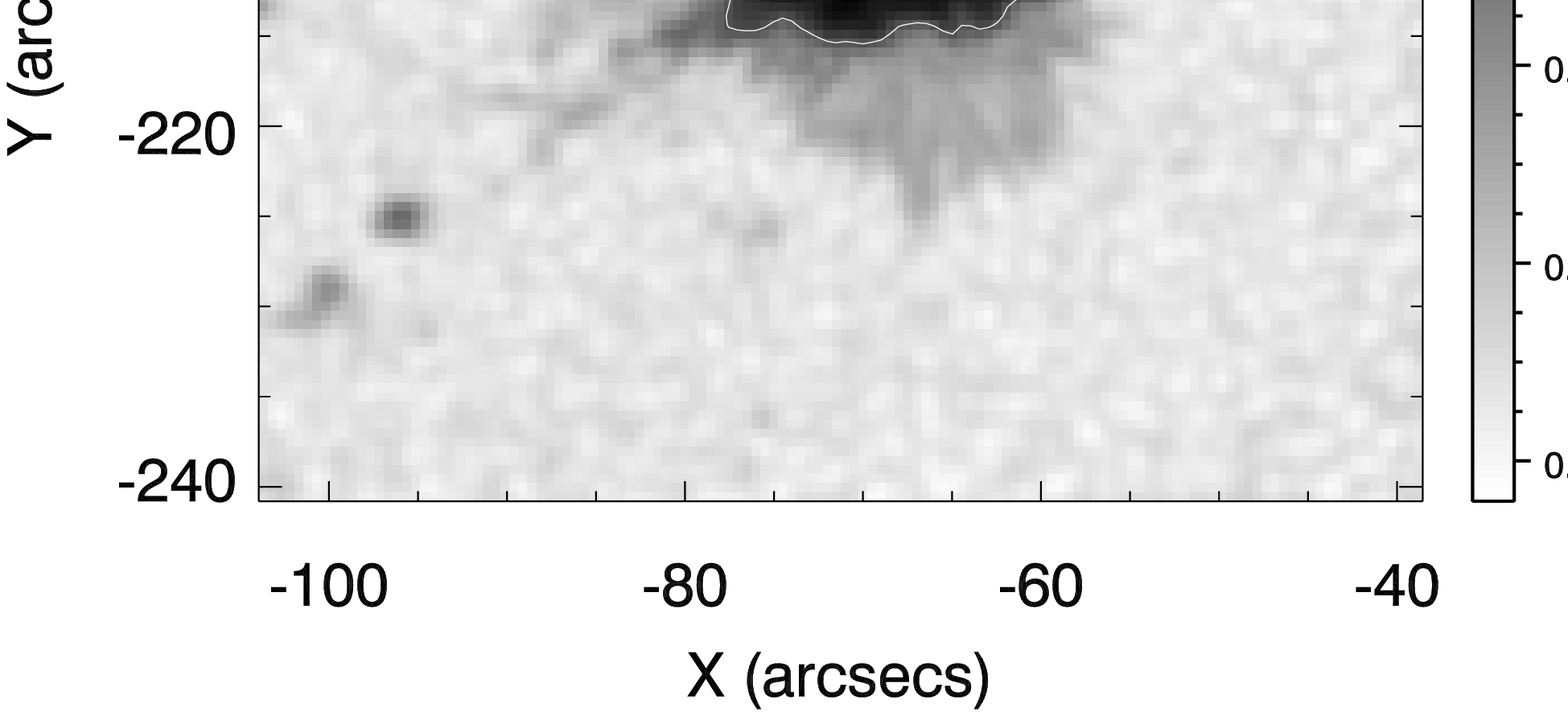}
	\includegraphics[scale=0.40, clip, trim=84 180 0 230]{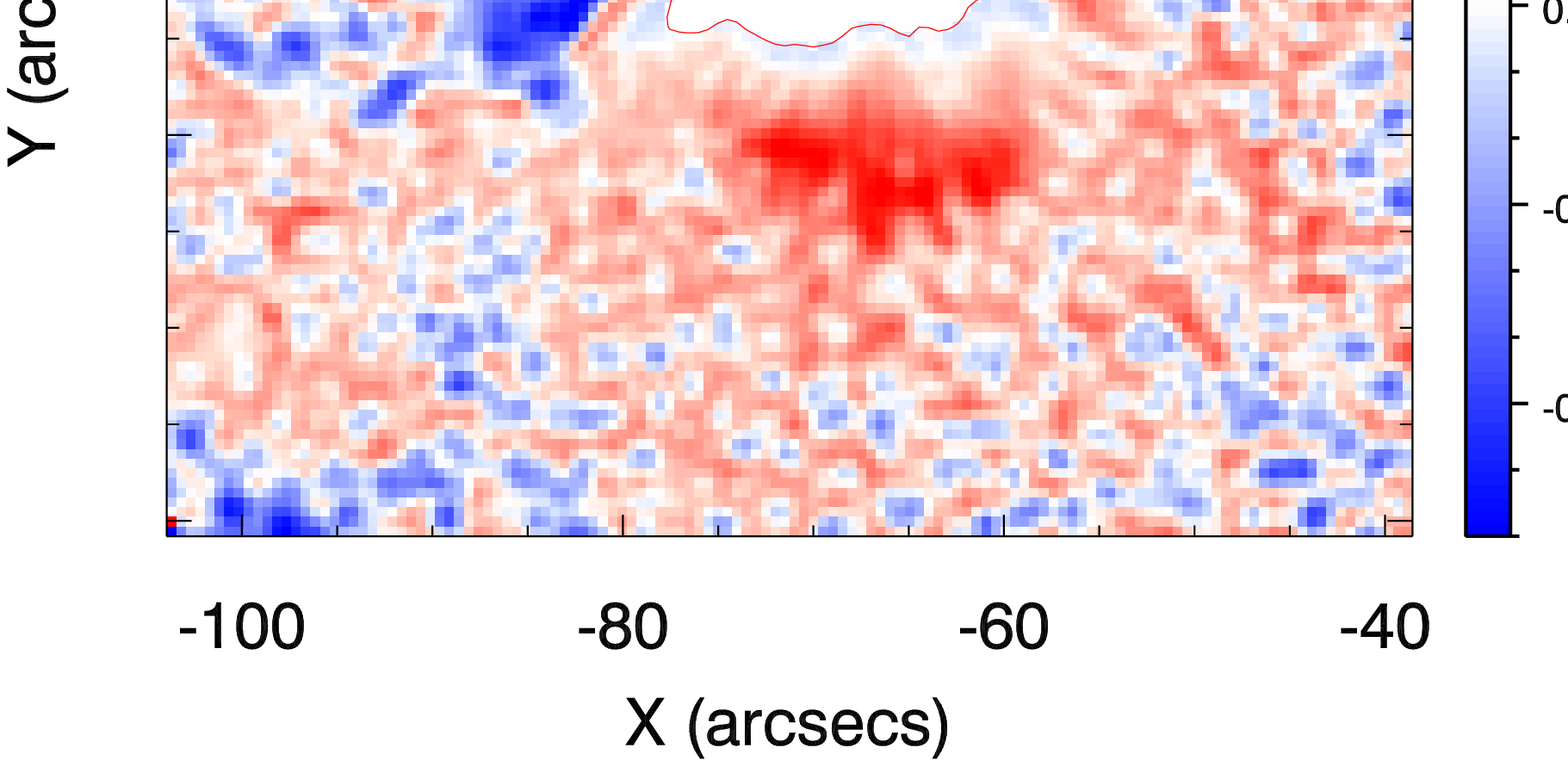}\\
\caption{Maps of the continuum intensity and LOS velocity from 2012 May 28 at 21:12 UT to 2012 May 28 at 21:58 UT as deduced by SDO/HMI. The arrow points to the disc center. \label{fig4}}
\end{figure*}

In order to highlight the variations of the continuum intensity, the magnetic field strength and inclination temporal, we report in Figure 7 the differences between the values of these quantities measured on May 28 at 24:00 UT and at 19:00 UT. The \textit{left panel} of Figure 7 indicates that the continuum intensity decreases by about 20\% of the quiet Sun value in 5 hours between 6\arcsec\/ and 11\arcsec\/ from the inner edge of the segment and it increases by about 27\% between 2\arcsec\/ and 6\arcsec\/. This increase can be ascribed to the shrinking of the pore along this particular segment (to become the umbra of the forming sunspot). We note that the intensity of the magnetic field increases by about 500 G in 5 hours between 6\arcsec\/ and 9\arcsec\/ from the inner edge of the segment (see the \textit{middle panel} of Figure 7). The decrease of the magnetic field between 4\arcsec\/ and 6\arcsec\/ can be attributed to the shrinking of the pore field and to the consequent inward migration of the UP boundary (Fig. 5)(see \citealp{Jur15}). Finally, a variation in the inclination angle of the magnetic field up to 20$^\circ$ can be detected, with the vertical component of the field changing sign (Figure 7, \textit{right panel}).

The same analysis has been performed along the segment B (see the second row of Figure 3). In this region the penumbra formed a few hours before than region marked by the segment A, with significant changes in the magnetic field strength but slight changes in the inclination. In particular, along the region indicated by the segment B the onset of the Eversed flow occurs in about 3 hours, i.e., from 15:00 UT to 19:00 UT.

\begin{figure*}[htbp]
\centering
\includegraphics[height=10.0cm, width=8.0cm, clip, trim=0 20 0 30]{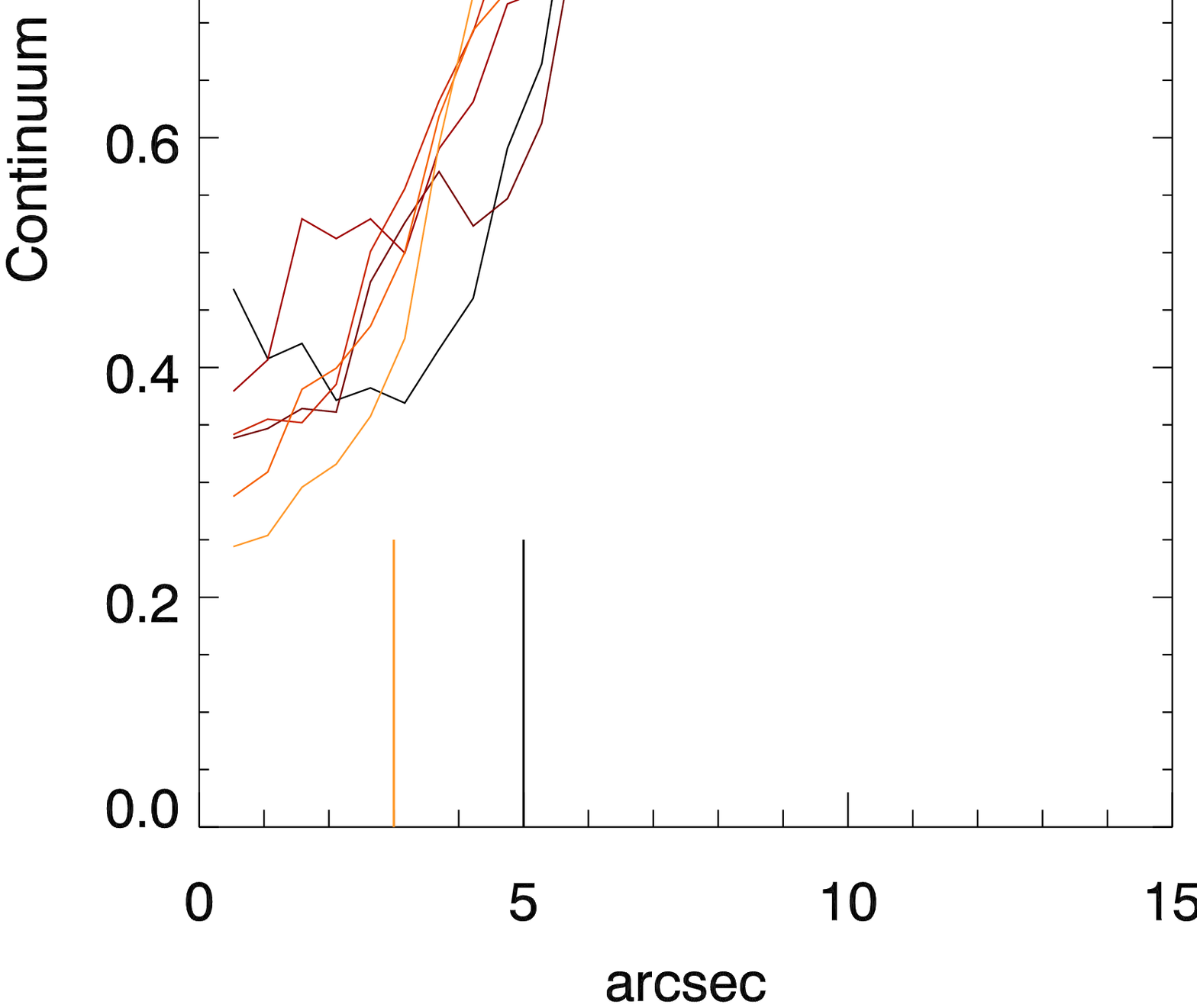}
\includegraphics[height=10.0cm, width=8.0cm, clip, trim=0 20 0 30]{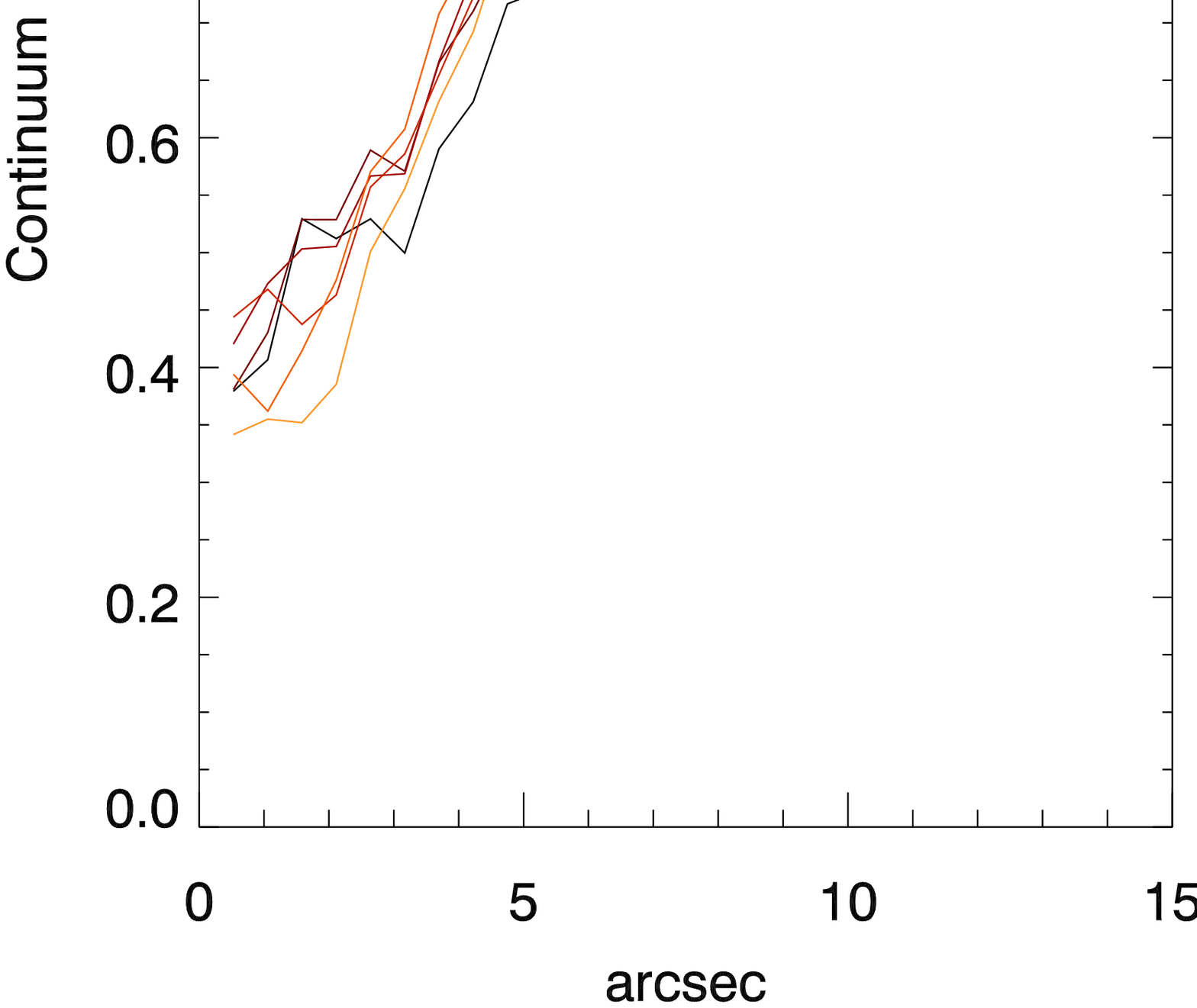}\\
\includegraphics[height=10.0cm, width=8.0cm, clip, trim=0 20 0 0]{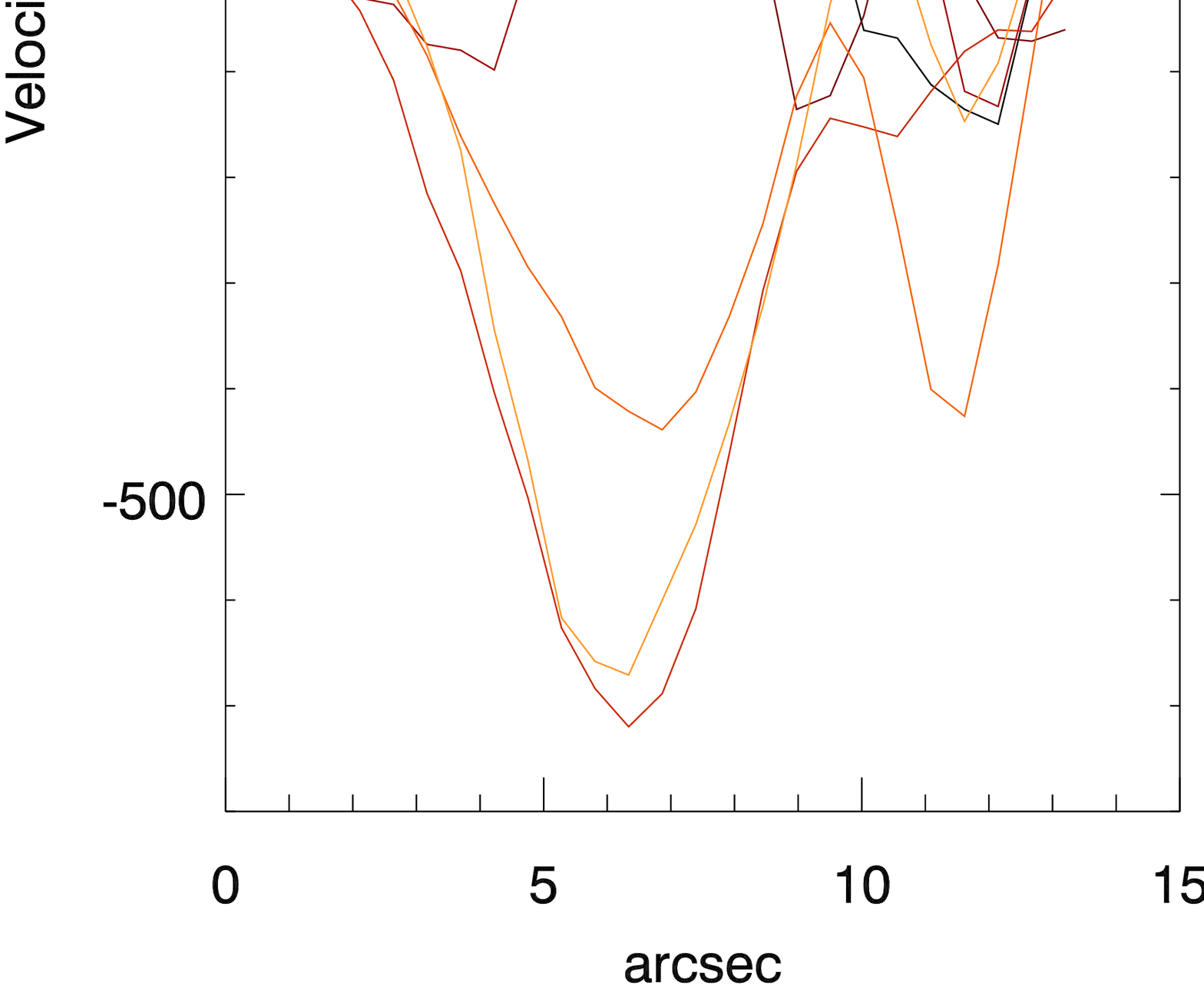}
\includegraphics[height=10.0cm, width=8.0cm, clip, trim=0 20 0 0]{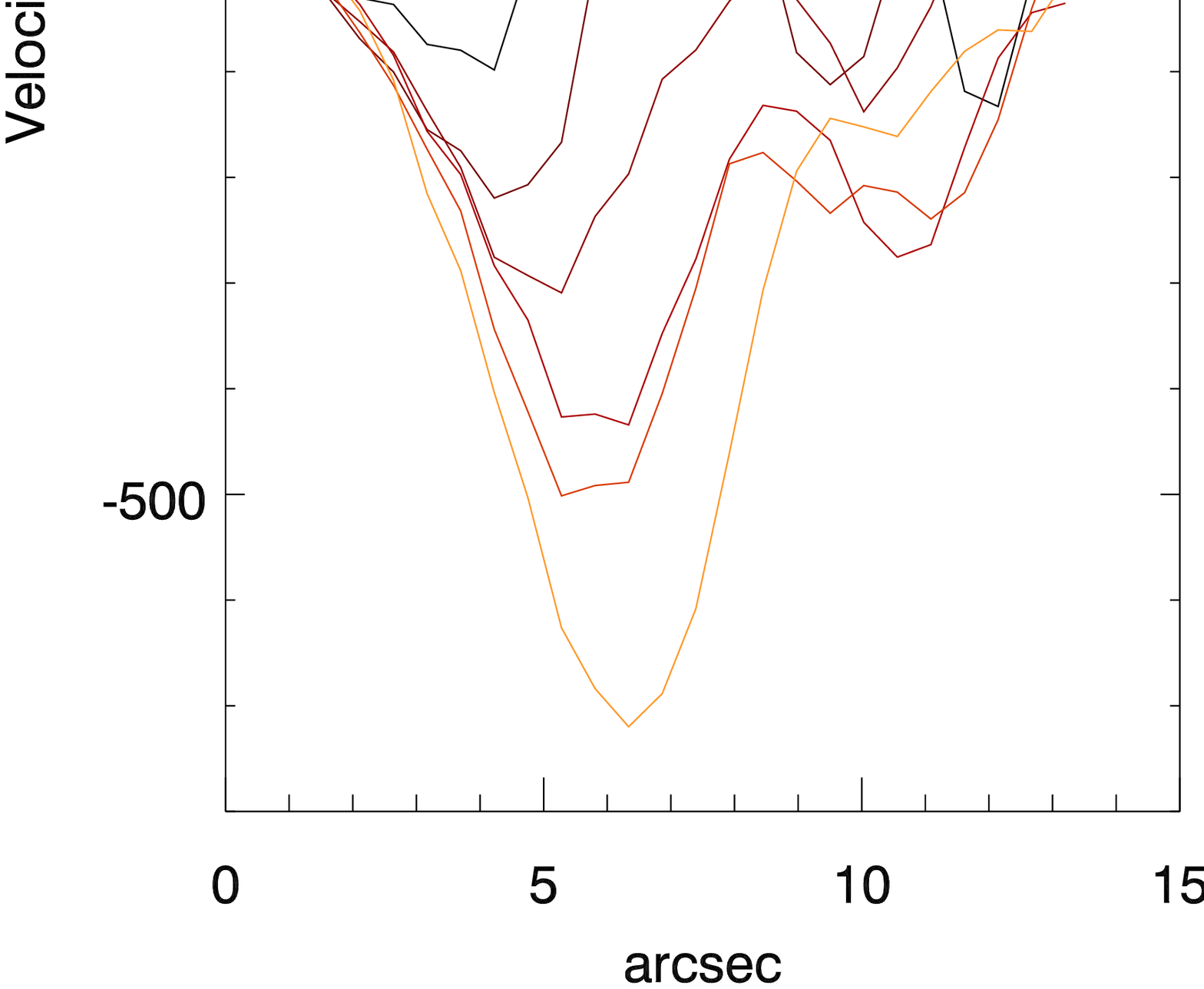}
\caption{Variation of the continuum intensity (\textit{top panels}) and of the LOS velocity (\textit{bottom panels}) along the segment A in the western part of the pore indicated in Figure 3 (\textit{second row}) and Figure 4. The origin of the horizontal axis denotes the end of the segment within the umbra. The figure is based on SDO/HMI data. The left and right panels cover intervals of 5 hours and 1 hour, (when the largest changes in LOS velocity and continuum intensity occur), respectively. In the \textit{top left panel} we report the positions of the umbra-quiet Sun boundary at 19:00 UT and the positions of the umbra-penumbra (UP) boundary at 24:00 UT using vertical bars at coordinates 3\arcsec\/ and 5\arcsec\/, respectively. \label{fig5}}
\end{figure*}

\begin{figure*}[htbp]
\centering
\includegraphics[height=10.0cm, width=8.0cm, clip, trim=0 20 0 30]{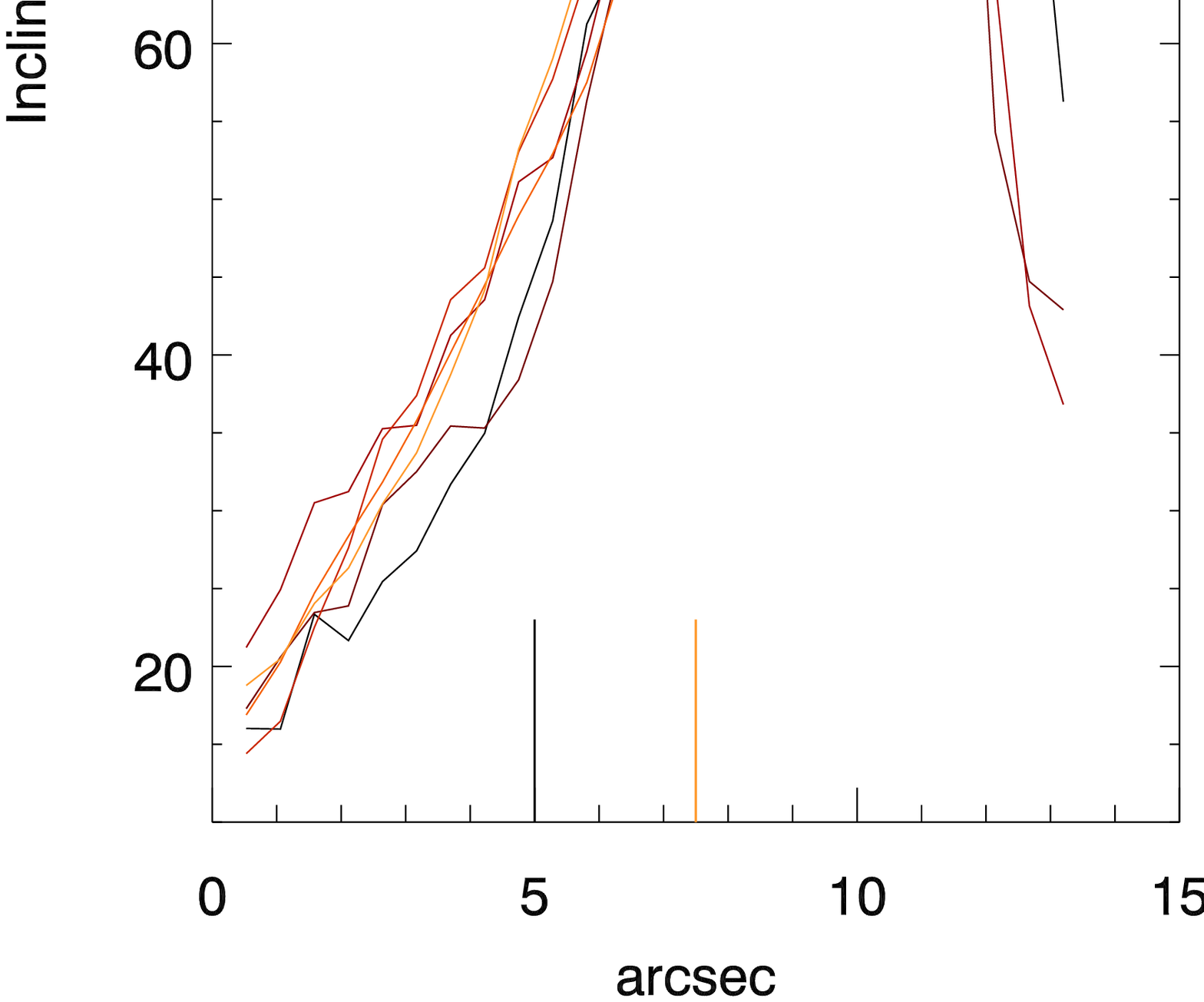}
\includegraphics[height=10.0cm, width=8.0cm, clip, trim=0 20 0 30]{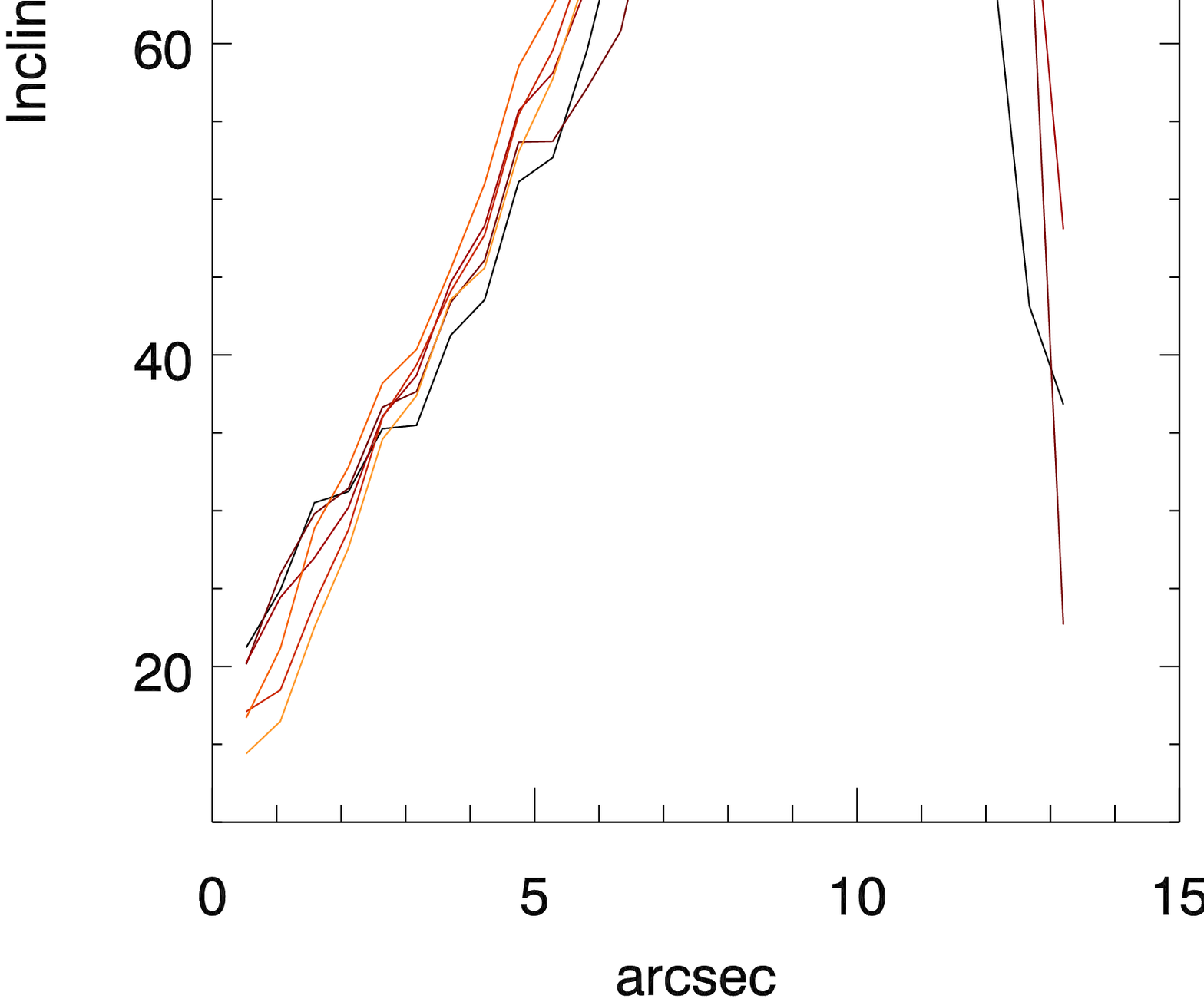}
\includegraphics[height=10.0cm, width=8.0cm, clip, trim=0 20 0 0]{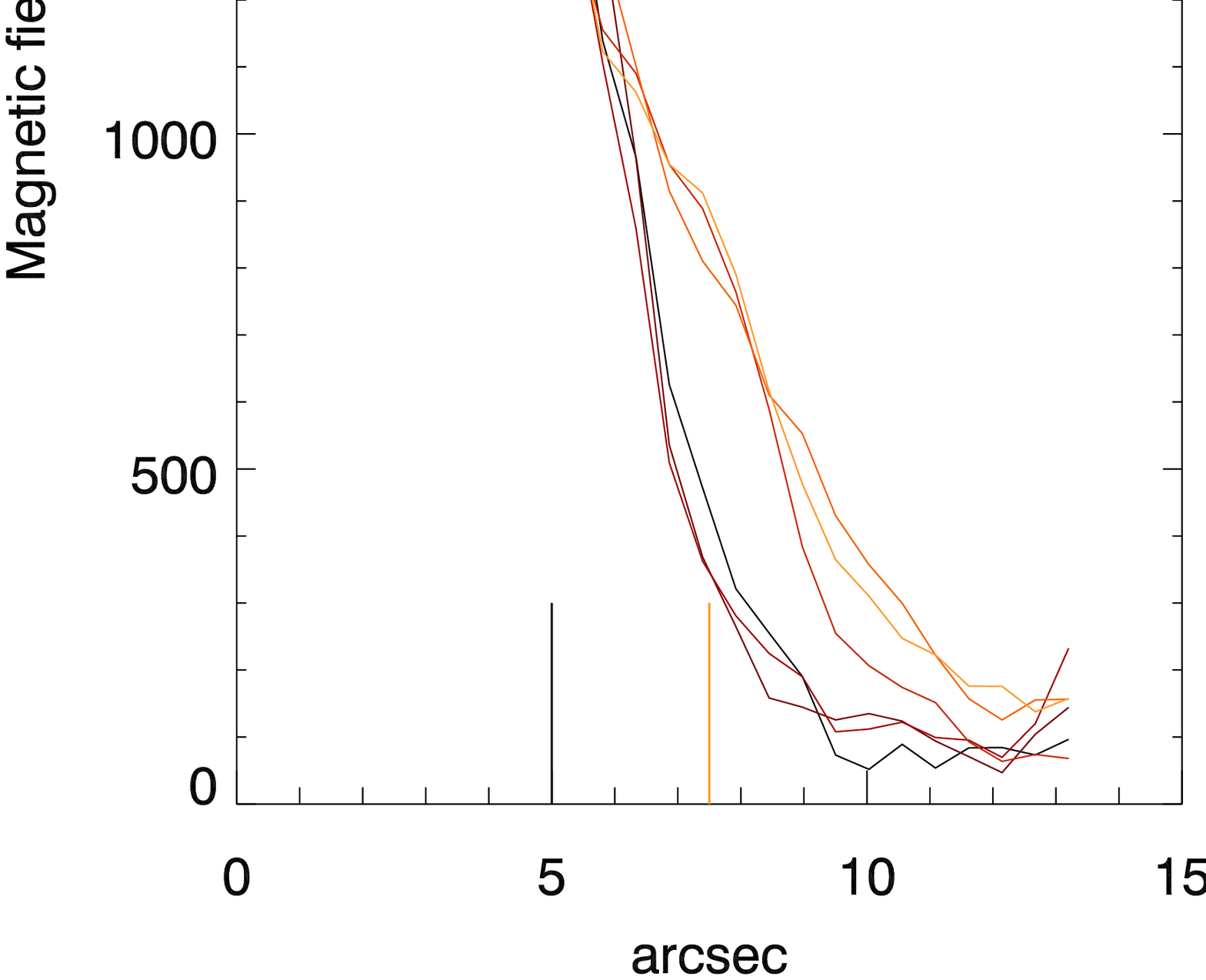}
\includegraphics[height=10.0cm, width=8.0cm, clip, trim=0 20 0 0]{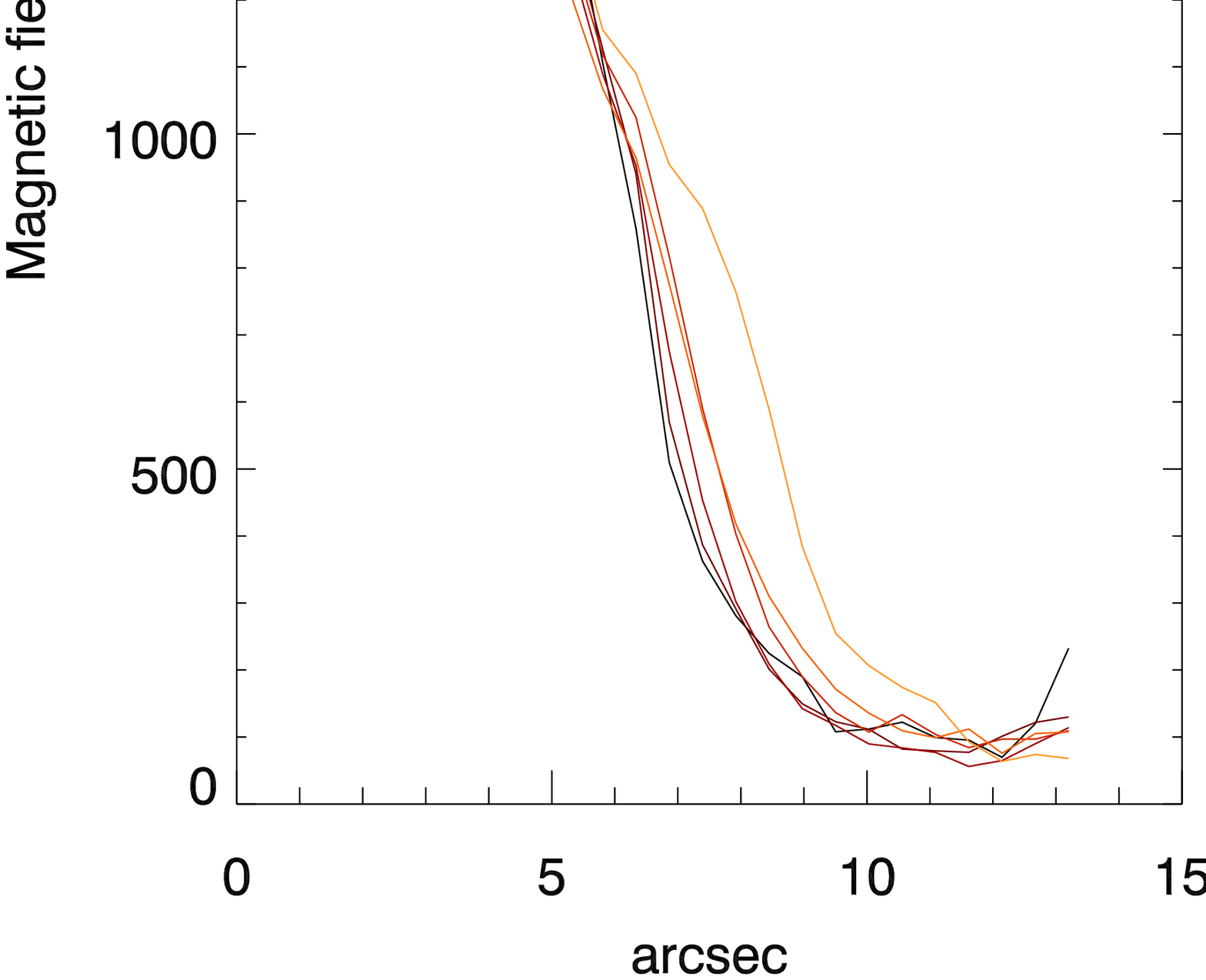}
\caption{Variation of inclination angle (\textit{top panels}) and strength of the magnetic field (\textit{bottom panels}) along the segment in the western part of the pore indicated in Figure 3 (\textit{second row}) and Figure 4. The origin of the horizontal axis denotes the end of the segment within the umbra. The figure is based on SDO/HMI data. The left and right panels cover intervals of 5 hours and 1 hour, (when the largest changes in LOS velocity and continuum intensity occur), respectively. In the \textit{left panels} we report the positions of the umbra-quiet Sun boundary at 19:00 UT and the positions of the outer edge of the penumbra at 24:00 UT using vertical bars at coordinates 5\arcsec\/ and 7.5\arcsec\/, respectively. \label{fig6}}
\end{figure*}

\begin{figure*}[htbp]
\centering
\includegraphics[scale=0.3, clip, trim=35 20 20 20]{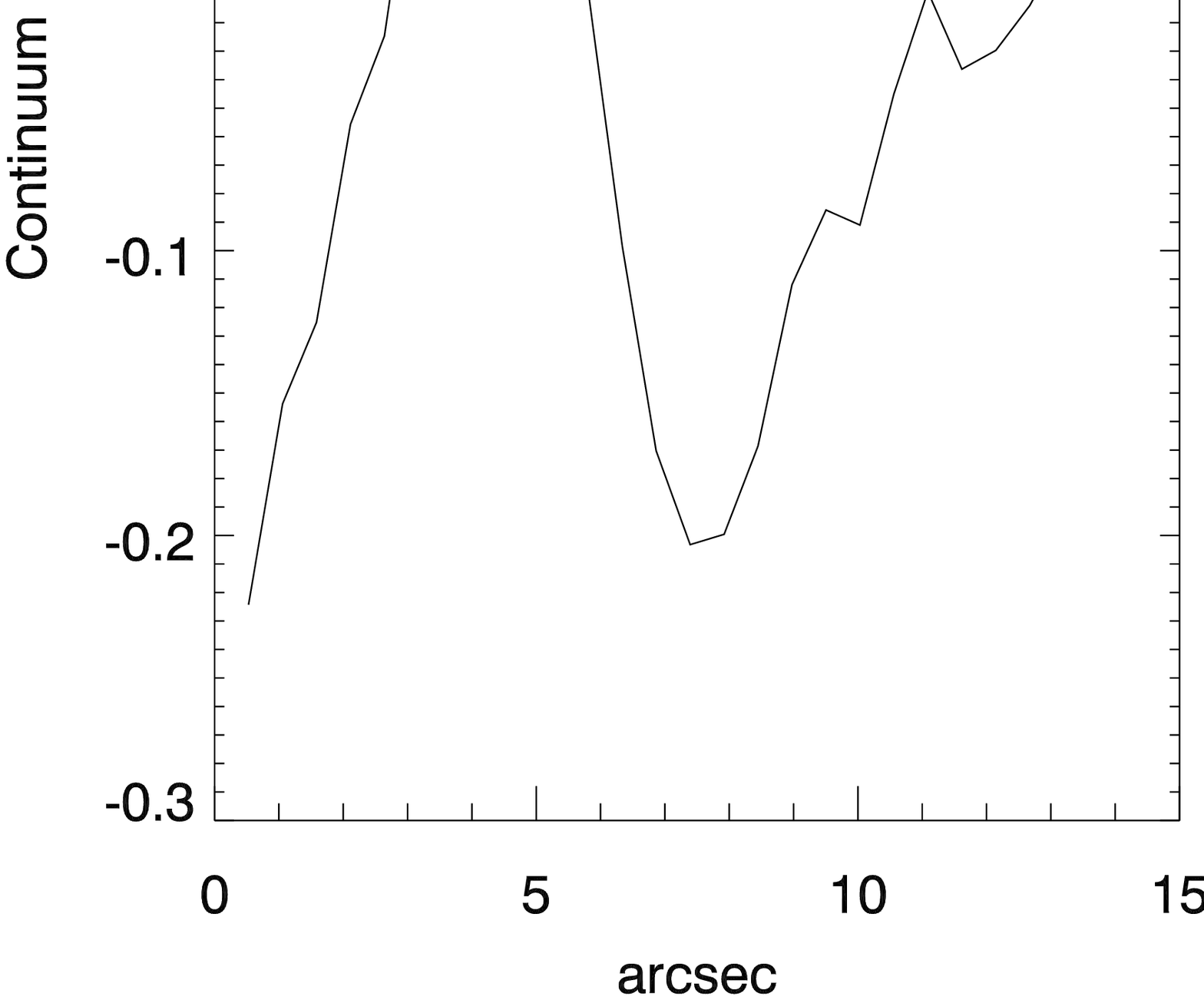}%
\includegraphics[scale=0.3, clip, trim=35 20 20 20]{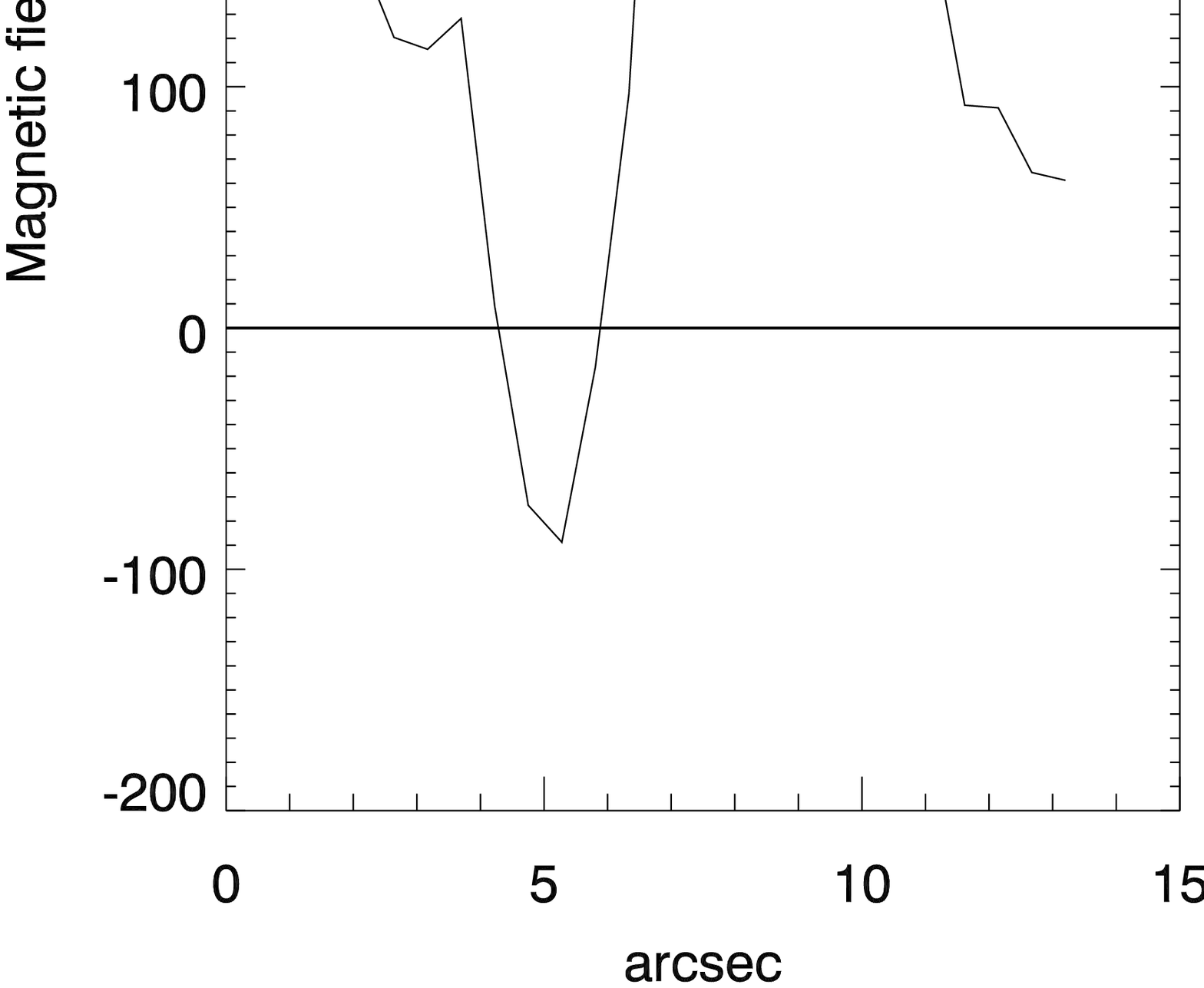}%
\includegraphics[scale=0.3, clip, trim=35 20 20 20]{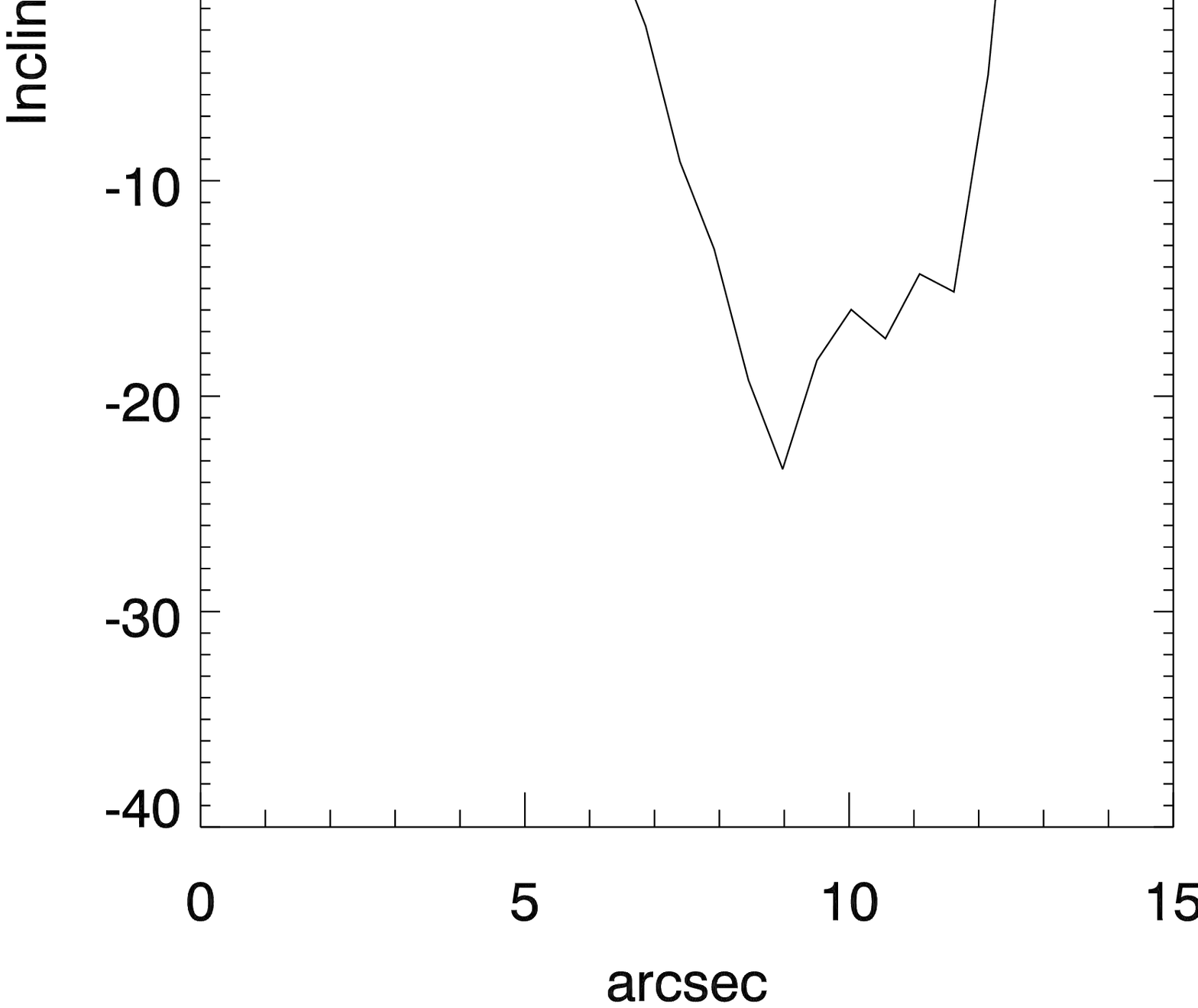}
\caption{Differences between the continuum intensity, magnetic field strength, and inclination angle measured on May 28 at 24:00 UT and at 19:00 UT (SDO/HMI dataset). \label{fig7}}
\end{figure*}

\section{Discussion and Conclusions}

In this paper we presented results concerning the formation of the penumbra of a sunspot and the associated onset of the Evershed flow. We studied the magnetic field and the LOS velocity field of the preceding sunspot of the AR NOAA 11490, whose penumbra formed in about 10 hours. We found that the LOS component of the velocity field compatible with the Evershed flow appeared in parallel with the formation of the penumbra. The velocity field changed sign in $1-3$ hours along a cut passing from the pore into the nearby quiet Sun and the Evershed flow has been established. This occured at the same time as the continuum intensity was lowered from quiet-Sun values to typical penumbral values. Interestingly, the field strength along the same cut at the location of the forming penumbra increased mainly only some minutes after the Evershed-like flow had already been established.

Before the formation of the penumbra, the photospheric magnetic field configuration of the pore that later turned into the sunspot showed the presence of an annular zone just outside its boundary, characterized by a magnetic field strength larger than 1000 G, having an (upside down) ballerina skirt structure of the magnetic field on a large azimuthal scale. During this phase, in the inner part of this annular zone we observed redshifts of the spectral lines of about 500 m s$^{-1}$. If we assume that the flow follows the field and that the magnetic field lines connect the pore with the other photospheric structures of opposite polarity, then the inferred direction of flow was opposite to that expected for plasma motion related to the Evershed radial outflow.

This flow, however, changed its direction at the same time as the local part of the penumbra formed, from 21:00 UT to 22:00 UT, when a plasma blueshift with a maximum velocity of about 700 m s$^{-1}$ was observed in the northern part of the penumbra. We found that the change in velocity preceded the change in magnetic field. In fact, the velocity had already changed sign (at 21:24 UT) while the magnetic field had still a very low value in the penumbral region (it increased to a higher value at 22:00 UT). Moreover, the magnetic field changes inclination from 80$^{\circ}$ to roughly 70$^{\circ}$ becoming slightly more vertical at the outer edge of the penumbra (at 7\farcs5). This could indicate that a nearly horizontal canopy-like field is converted into a more penumbra-like field, i.e., one that is more inclined on average. Before the penumbra forms the field is mostly a nearly horizontal canopy field, i.e. it does not pass through the solar surface at these locations. However, after the formation of the penumbra at least some of the field (that emerging in the spines within the penumbra) passes through the solar surface at a considerable angle to the horizontal. This aspect needs to be verified by future, high-resolution observations as well as by detailed sunspot modelling. Furthermore, beyond the outer penumbral boundary in the moat region the magnetic field changes sign from a flat opposite polarity field, to having the same polarity as the sunspot.

The results obtained here may provide us some insight into where the penumbral field comes from. A hint is given by the ring of redshifted material surrounding the pore. This may be the material flowing up through the solar surface in the outer opposite polarity magnetic footpoints, now flowing down again as was proposed by \citet{Rom14}. 

We present the following scenario for the formation of the penumbra and the start of the Evershed flow to explain the observations presented here and in earlier papers. The canopy field of the initial pore gets weaker at greater distances from the boundary of the pore. At some distance the field is sufficiently weak that convective flows can drag field lines down into the photosphere forming small U-loops whose inner footpoint has the opposite magnetic polarity to the pore. Such a magnetic structure is found around our forming spot in Figure 3. This footpoint is at the same time now the outer footpoint of an inverted U-loop connecting it to the pore. Such a pulling down of a canopy field has been demonstrated with the help of numerical simulations by \citet{Piet10} and has also been proposed to explain the formation of bipolar moving magnetic features around sunspots \citep{Zha03,Zha07}. Since the external footpoint of the inverted U-loop is brighter (hotter) than the pore, has little magnetic flux and has a comparatively weak  field, a siphon flow directed towards the pore is set up (which may be driven by either the temperature or the magnetic field strength difference between the pore and the the external footpoint). This seemingly inverse Evershed flow is compatible with the findings of \citet{Rom14} and of this paper. 

With time more and more flux is dragged down, increasing the flux in the external footpoint of the inverted U-loop. This loop is kept flat and low-lying by the overlying canopy, as proposed by Romano et al. (2014) in the cartoon shown in their Figure 4. A similar action of the canopy was indeed found by \citet{Gugl14} during the formation of penumbral-like structures. As its flux increases, at some point the field lines reach the solar surface along the complete length of the loop. The region darkens as the magnetic field inhibits convection, but the darkening stops, i.e., the brightness reaches a new equilibrium at a lower value, as magneto-convection starts. At the same time the Evershed flow is set up as part of the magneto-convective process. This flow is directed radially outward, as within the penumbral filaments harbouring this flow both the brightness and the magnetic field are distributed such as to accelerate the gas away from the umbra \citep{Tiw13}.

This scenario is compatible not only with the present observations, including the ring of opposite polarity features surrounding the forming sunspot (and the ring of same polarity flux surrounding this ring), but also nicely explains why the process does not work on the side of the spot where flux is still emerging. There the field of the outer footpoint of the  inverted U-loop gets cancelled by the emerging flux. Also, because the formation starts in the canopy near the final outer boundary of the penumbra (assuming that this is the place where the canopy field becomes sufficiently weak to be dragged down by convective and other flows) the scenario leads to a natural explanation of why the penumbra first leaves a mark in the low chromosphere/upper photosphere before becoming visible at the solar surface.

The analyzed IBIS/DST observations available for the time, before and after penumbra formation, together with the complementary SDO/HMI data have guided us in coming up with a new scenario for the formation of the penumbra, including the start of the Evershed effect. These and future such observations are likely to set useful constraints on quantitative models describing the beginning of the Evershed flow in sunspot penumbrae. For example, we found that a flow qualitatively compatible with a radial outflow starts in a short time, in our case in less than one hour (see Figure 4, \textit{top right panel}). It would be interesting if future numerical simulations of penumbra formation could reproduce this fast evolution of the plasma flow, accompanied or followed shortly afterwards by an increase of the total magnetic field and a slight change of its inclination.

In the near future we plan to perform new observing campaigns with high performance instruments, such as IBIS or CRISP, in order to obtain other high-quality data sets where the evolution of the annular zone may be observed for a longer time. Also, studying more sunspots will help to determine how universal the observational results obtained here are. In this context, the next generation solar telescopes with larger aperture such as the GREGOR telescope \citep{Schm12}, the Daniel K. Inouye Solar Telescope \citep[DKIST, formerly the Advanced Technology Solar Telescope (ATST),][]{Kei10}, and the European Solar Telescope \citep[EST,][]{Col10}, are expected to provide more information on the processes underlying the formation of the penumbra and the beginning of the Evershed flow. Additional observations will provide further tests of the proposed scenario and should enable us to further refine and extend it.

\acknowledgments

The authors wish to thank the DST staff for its support during the observing campaigns. The research leading to these results has received funding from the European Commission's Seventh Framework Programme under the grant agreement SOLARNET (project n$^{o}$ 312495). This work was also supported by the Istituto Nazionale di Astrofisica (PRIN-INAF-2014), by the University of Catania (PRIN MIUR 2015) and by Space Weather Italian COmmunity (SWICO) Research Program. This work was partly supported by the BK21 plus program through the National Research Foundation (NRF) funded by the Ministry of Education of Korea.



{\it Facilities:} \facility{DST (IBIS)}, \facility{SDO (HMI)}





\clearpage

\end{document}